\documentclass[12pt]{article}
\pdfoutput=1
\usepackage[latin1]{inputenc}
\usepackage{amsmath}
\usepackage{amsfonts}
\usepackage{amssymb}
\usepackage{pstricks}
\usepackage{amsthm}
\usepackage{mathrsfs}
\usepackage{amssymb}
\usepackage{cancel}
\usepackage{slashed}
\usepackage{graphicx}
\usepackage[font=small,labelfont=bf]{caption}
\usepackage{float}
\usepackage{csquotes}
\usepackage[bottom]{footmisc} 
\usepackage{setspace}
\usepackage{bbding}
\usepackage{cite}
\usepackage{framed}
\usepackage{color}
\usepackage{wrapfig}
\definecolor{shadecolor}{RGB}{224,238,238}
\newcommand{\nn}{\nonumber}

\newcommand{\fb}{$\text{fb}^{-1}$ }

\makeatletter
\renewcommand{\theequation}{\thesection.\arabic{equation}}
\@addtoreset{equation}{section}
\makeatother

\def\lsim{\;\raise0.3ex\hbox{$<$\kern-0.75em\raise-1.1ex\hbox{$\sim$}}\;}
\def\gsim{\;\raise0.3ex\hbox{$>$\kern-0.75em\raise-1.1ex\hbox{$\sim$}}\;}
\def\beq{\begin{equation}}   \def\eeq{\end{equation}}
\def\ba{\begin{array}}       \def\ea{\end{array}}
\def\bea{\begin{eqnarray}}   \def\eea{\end{eqnarray}}
\def\nn{\nonumber}
\def\nl{\newline}

\def\noi{\noindent}

\theoremstyle{definition} 

\begin{document}

\begin{titlepage}
\begin{flushright}
LPT Orsay 17-33 \\
\end{flushright}


\begin{center}

\begin{doublespace}

\vspace{1cm}
{\Large\bf Simultaneous Search for Extra Light and Heavy Higgs Bosons
via Cascade Decays} 
\vspace{2cm}

{\bf{Ulrich Ellwanger$^{a}$ and Mat\'ias Rodr\'iguez-V\'azquez$^a$}}\\
\vspace{1cm}
{\it  $^a$ Laboratoire de Physique Th\'eorique, UMR 8627, CNRS,
Universit\'e de Paris-Sud, Universit\'e
Paris-Saclay, 91405 Orsay, France}
\end{doublespace}
\end{center}
\vspace*{2cm}

\begin{abstract}
Models with extended Higgs sectors can contain several additional Higgs
states, heavier or lighter than the SM Higgs boson. The couplings of
lighter extra states to SM particles can be strongly reduced, leading to
small cross sections for their direct production. Heavier extra states
can have larger couplings to SM particles and, moreover, have large
branching fractions into lighter extra states, notably into a SM-like
Higgs boson accompagnied by another Higgs state which can be lighter or heavier
than 125~GeV.
Motivated by corresponding scenarios in the NMSSM we study the prospects for
the discovery or exclusion of cascade decays $ggF \to H_3 \to H_2 + H_1$ in the
$b\bar{b}b\bar{b}$, $b\bar{b}\tau\tau$ and $b\bar{b}\gamma\gamma$ final states
where either $H_1$ or $H_2$ can be SM-like. Significant regions of the NMSSM
parameter space can be tested by these searches. These are, however, not
confined to models of the NMSSM type.

\end{abstract}

\end{titlepage}
\newpage
\section{Introduction}
\label{sec:intro}

Extended Higgs sectors are frequent properties of models beyond the
Standard Model (BSM). Such extra states can have very small couplings
to quarks, leptons and SM gauge fields. For instance, for singlets under
the SM gauge symmetries such renormalizable couplings are disallowed by
gauge invariance. The direct production cross sections for these states
are then strongly suppressed in all channels. On the other hand,
couplings of singlets to SU(2) Higgs doublets of the SM- or BSM-type are
possible and typically present in BSM models. This allows for the
discovery of such states in cascade decays of heavy BSM SU(2) Higgs doublets,
provided the production cross sections of the latter are large enough.

The final states after BSM-Higgs to BSM-Higgs + SM-Higgs cascades
typically correspond to the ones
in searches for resonant SM-Higgs ($H_{125}$) pair production: mainly 
$b\bar{b}b\bar{b}$, $b\bar{b}\tau\tau$ and $b\bar{b}\gamma\gamma$.
Corresponding searches have been performed at the LHC by ATLAS
\cite{ATL-CONF-2014-005,Aad:2014yja,Aad:2015uka,Aad:2015xja,ATL-CONF-2016-049,
TheATLAScollaboration:2016ibb} and by CMS 
\cite{Khachatryan:2016sey,CMS:2014eda,Khachatryan:2014jya,Khachatryan:2015yea,
Khachatryan:2015tha,Khachatryan:2016cfa,CMS:2016tlj,CMS:2016pwo,
CMS-PAS-HIG-16-013,CMS-PAS-HIG-16-029,CMS-PAS-HIG-16-032,Sirunyan:2017djm,
CMS-PAS-HIG-17-006,Sirunyan:2017tqo,CMS-PAS-HIG-17-008}. However, one
of the SM-like Higgs bosons would now be replaced by a lighter or heavier
BSM-Higgs boson. One can argue that the cross sections for such processes
can be more promising than for resonant SM-Higgs pair production:

\noi a) A sizeable gluon-gluon-fusion ($ggF$) production cross section of a heavy scalar or
pseudoscalar $\Phi$, i.e. a sizeable coupling of $\Phi$
to top quarks, requires $\Phi$ to possess a sizeable SU(2)-doublet component.
However, since $H_{125}$ is also a SU(2)-doublet,
trilinear couplings $\Phi-H_{125}-H_{125}$ (with $\Phi$ a pure doublet) violate
the SU(2) symmetry and must be proportional to a SU(2) symmetry breaking vev;
the latter is limited from above by the $Z/W$ masses. This limits the possible
partial width for $\Phi \to H_{125}+H_{125}$, whereas the concurrent decay
$\Phi\to t\bar{t}$ is always possible
if $\Phi$ can be produced in $ggF$ and is heavier than $2\, m_{top}$.\\
\noi b) In the case $ggF\to \Phi\to H_{125} + H'$ with $\Phi$ a pure doublet,
the trilinear coupling $\Phi-H_{125}-H'$ can be SU(2) invariant if $H'$ is a
singlet. In models with extended Higgs sectors including both an extra doublet
and a singlet, such a coupling can thus be much larger than the $Z/W$ masses
leading to sizeable $\Phi\to H_{125}+H'$ partial widths.

In Two-Higgs-Doublet-Models of type II such as the Minimal Supersymmetric
Standard Model (MSSM) the production cross sections for extra CP-even ($H$)
and CP-odd ($A$) Higgs doublets are not suppressed, and are dominated by
$ggF$ for $\tan\beta$ not too large
\cite{Djouadi:2005gj,Djouadi:2013vqa,Djouadi:2015jea}.
$H$ or $A$ can thus play the r\^ole of $\Phi$ above.
The Next-to-Minimal Supersymmetric Standard Model (NMSSM)
\cite{Maniatis:2009re,Ellwanger:2009dp} contains additional CP-even ($H_S$)
and CP-odd ($A_S$) singlet-like states with masses typically
below $M_H \sim M_A$. One finds that
the $BR(H\to H_S + H_{125})$ and $BR(A\to A_S+ H_{125})$ can be up to $\sim 50\%$
\cite{Kang:2013rj,King:2014xwa,Carena:2015moc,Ellwanger:2015uaz,Costa:2015llh,
Baum:2017gbj}, for
the reasons given above and detailed in the next section.

In the NMSSM this offers the possibility to produce otherwise practically invisible mostly
singlet-like states $H_S/A_S$ in cascade decays of $H/A$
\cite{Kang:2013rj,King:2014xwa,Carena:2015moc,Ellwanger:2015uaz,Costa:2015llh,Baum:2017gbj}.
It is the aim of the present paper to study the prospects
for discovery or exclusion of, simultaneously, $H/A$ and $H_S/A_S$ states in
$ggF \to H \to H_S + H_{125}$ or $ggF \to A \to A_S + H_{125}$
in the final states $b\bar{b}b\bar{b}$,
$b\bar{b}\tau\tau$ and $b\bar{b}\gamma\gamma$. Supersymmetry plays no r\^ole
here, accordingly our results are applicable to any models with similarly
extended Higgs sectors; see, e.g., \cite{vonBuddenbrock:2016rmr}.

We will adopt various strategies from the searches for resonant SM
Higgs pair production by ATLAS
\cite{ATL-CONF-2014-005,Aad:2014yja,Aad:2015uka,Aad:2015xja,ATL-CONF-2016-049,
TheATLAScollaboration:2016ibb} and by CMS 
\cite{Khachatryan:2016sey,CMS:2014eda,Khachatryan:2014jya,Khachatryan:2015yea,Khachatryan:2015tha,
Khachatryan:2016cfa,CMS:2016tlj,CMS:2016pwo,
CMS-PAS-HIG-16-013,CMS-PAS-HIG-16-029,CMS-PAS-HIG-16-032,Sirunyan:2017djm,
CMS-PAS-HIG-17-006,Sirunyan:2017tqo,CMS-PAS-HIG-17-008}. Moreover,
 for $M_{H_S}$ near 125~GeV we can
compare our backgrounds and expected 95\%~CL upper limits on the
cross sections times branching fractions to the ones obtained in these
publications.

On the other hand, the analyses presented here are complicated by the fact
that the masses $M_{H_S}/M_{A_S}$ are not known {\it a priori}.
An important aspect of optimal search strategies are $M_{H_S}/M_{A_S}$ dependent
selection criteria (cuts) on events, hence different analyses should
be performed, varying the assumptions on $M_{H_S}/M_{A_S}$. Only at
the end of each of these analyses a search for a resonance-like bump in
the total invariant mass of the $H_S/A_S$ plus $H_{125}$ decay products, which
should correspond to $M_H/M_A$, is proposed.

In the next section we discuss shortly the Higgs sector of the NMSSM and
the couplings relevant for the processes considered here.
In section 3 we present features of our signal simulations. In section~4 we discuss
the optimal search strategy for the $b\bar{b}b\bar{b}$ final state, and compare
expected 95\%~CL upper limits and $5\,\sigma$ discovery limits on the
cross sections times branching fractions to the ones possible in the NMSSM.
Sections 5 and 6 are devoted to the $b\bar{b}\tau\tau$ and $b\bar{b}\gamma\gamma$
final states. All these search strategies and results are identical for
$ggF \to H \to H_S + H_{125}$ and $ggF \to A \to A_S + H_{125}$, for
notational simplicity we will refer to $H \to H_S + H_{125}$ only.
In section 7 we conclude with a summary and an outlook.

\section{The neutral Higgs sector of the NMSSM}
\label{sec:thehiggssector}
In this section we discuss briefly some properties of the Higgs sector
of the  CP-conserving $\mathbb{Z}_3$-invariant NMSSM. It consists in two
SU(2) doublets $H_u$, $H_d$ and a complex singlet $S$.
The superpotential of the Higgs sector reads
\beq
W_{\text{Higgs}}=\lambda\hat{S}\hat{H}_u\cdot \hat{H}_d+\frac{\kappa^3}{3}\hat{S}^3
\eeq
where $\lambda$ and $\kappa$ are dimensionless Yukawa couplings, and $\hat{H}_u$,
$\hat{H}_d$ and $\hat{S}$ denote chiral superfields. Once the real
component of the superfield $\hat S$ develops a vacuum expectation value (vev) $s$,
the first term in the superpotential generates an effective $\mu$ term
\beq
\mu= \lambda s\; .
\eeq
The vev $v_u$ of $H_u$ generates up-type quark masses, the vev $v_d$ of $H_d$
generates masses for down-type quarks and leptons, and both vevs contribute to
the $Z$ and $W^\pm$ masses. Their ratio is $\tan\beta = v_u/v_d$.

Decays of a heavy Higgs state into two lighter Higgs states occur in the presence
of corres\-ponding trilinear Higgs couplings. Most of the trilinear Higgs couplings in the
$\mathbb{Z}_3$-invariant NMSSM originate from quartic terms in the Higgs
potential (see \cite{Maniatis:2009re,Ellwanger:2009dp}) proportional to two
powers of $\lambda$, $\kappa$ or the electroweak gauge couplings, once the
(neutral) Higgs fields are expanded around their vevs and decomposed into
their real and imaginary parts:
\beq
H^0_u= v_u + \frac{1}{\sqrt{2}}(H^0_{u,r} + i H^0_{u,i}), \quad
H^0_d= v_d + \frac{1}{\sqrt{2}}(H^0_{d,r} + i H^0_{d,i}), \quad
S=s+\frac{1}{\sqrt{2}}(S_r +i S_i)\; .
\eeq
Hence the trilinear couplings are proportional to the vevs $v_u$, $v_d$ or $s$.
Whereas $v_u$, $v_d$ are limited from above by $M_Z^2 = \frac{g_1^2+g_2^2}{2}(v_u^2+v_d^2)$,
a large vev $s$ can generate
a trilinear coupling $\sim H_u\cdot H_d S$. Another source for such a
coupling is a trilinear Higgs-dependent soft SUSY breaking term
\beq
\label{eq:Alambda}
\lambda A_\lambda H_u\cdot H_d S + h.c.
\eeq
where the dimensionful parameter $A_\lambda$ can be much larger than Higgs
vevs. 

In order to obtain its impact on trilinear couplings among Higgs mass
eigenstates, the mass matrices have to be diagonalized. In the CP-even sector,
where one deals with a $3\times 3$ mass matrix, a first step in this direction
is a rotation in the SU(2) doublet sector into the so-called Higgs basis
\beq
\label{eq:rotat}
H^0_{u,r} = \sin\beta h' - \cos\beta H',\qquad H^0_{d,r} = \cos\beta h' + \sin\beta H'
\eeq
where the vev of $H'$ is zero, and the vev $\left<h'\right>=\sqrt{v_u^2+v_d^2}$
is equal to the one of the SM Higgs boson.
In fact, in most of the phenomenological acceptable regions of the parameter space
of the NMSSM (near the alignment limit \cite{Carena:2015moc}) the eigenstates of the full
$3\times 3$ CP-even mass matrix are not very different from $h'$, $H'$ and $S_r$,
and will be denoted by $H_{125}$ ($\sim h'$, approximately SM-like), $H$ ($\sim H'$,
approximately MSSM-like) and $H_S$ ($\sim S_r$, approximately singlet-like) in
the following. 

The corresponding rotation of the imaginary components $H^0_{u,i}$ and $H^0_{d,i}$
(with $\beta\to -\beta$) diagonalizes their $2\times 2$ mass submatrix exactly and generates the Goldstone
boson together with the MSSM-like pseudoscalar $A'$. The latter still mixes with
the singlet-like $S_i$, but typically both differ little from the mass eigenstates
$A$ and $A_S$.

Performing the rotation (\ref{eq:rotat}) in (\ref{eq:Alambda}) and using the
previous approximations in the CP-even and CP-odd sectors, one obtains the trilinear couplings
\beq
\frac{\lambda A_\lambda}{\sqrt{2}}\left(\frac{\tan^2\beta-1}{\tan^2\beta+1} 
H_{125} \left(H H_S - A A_S\right) + \dots\right)
\eeq
where the omitted terms are suppressed by $\tan\beta$. Hence, for not too small $\tan\beta\to 1$,
trilinear couplings $g_{H_{125} H H_S}$ and $g_{H_{125} A A_S}$ are generated
which have no analog in the MSSM, and are larger than all other trilinear Higgs couplings
if $\lambda A_\lambda$ is large.

On the other hand the masses $M_{H/A}$ of the nearly degenerate mostly MSSM-like states $H/A$
are approximatively given by
\beq
M_{H/A}^2 \sim \mu\left(A_\lambda+\frac{\kappa}{\lambda}\mu\right)
\frac{1+\tan^2\beta}{\tan\beta}\;,
\eeq
which limits $A_\lambda$ from above for fixed $M_{H/A}$, $\tan\beta$,
small $|\kappa/\lambda|$ and $|\mu| \gsim 100$~GeV (as
required by the lower LEP bound on higgsino-like charginos). Accordingly the
trilinear couplings $g_{H_{125} H H_S}$ and $g_{H_{125} A A_S}$ can be
larger for larger $M_{H/A}^2$.

The production cross sections for the mostly 
MSSM-like states $H/A$ is dominated by $ggF$ \cite{Djouadi:2005gj,Djouadi:2013vqa,Djouadi:2015jea};
at $\sqrt{s}=13-14$~TeV and for $\tan\beta \sim 2-3$ (typical in the NMSSM) they are
$\gsim {\cal O}(1\,\text{pb})$ up to $M_{H/A}\sim 600$~GeV. The trilinear couplings 
$g_{H_{125} H H_S}$ and $g_{H_{125} A A_S}$
induce the decays $H \to H_{125}+ H_S$ and $A \to H_{125}+ A_S$ if kinematically allowed. The branching fractions
$BR(H/A \to H_{125}+ H_S/A_S)$ can be as large as~$\sim 50\%$, in contrast to the
decay $H \to H_{125} +H_{125}$.

The singlet-like states $\Phi_S = H_S/A_S$
 have small couplings to quarks, leptons and gauge fields induced by mixings
with $h'$, $H'$ and $A'$. Hence the production cross sections for $\Phi_S$
are typically small, and their discovery may have to rely on
$H/A \to H_{125}+ \Phi_S$ decays. Via the couplings induced by mixing, $\Phi_S$ can decay
into the same channels as $H_{125}$ and $H/A$. For $M_{\Phi_S} > 2m_{top}$, decays
into $t\bar{t}$ are dominant, whereas decays $\Phi_S\to b\bar{b}$ dominate typically for
$M_{\Phi_S} < 2m_{top}$. (The branching ratio for $H_S\to W^+ W^-$ can also be
sizeable \cite{Carena:2015moc}.)
For $M_{H_S} > 250$~GeV, decays $H_S\to H_{125}+ H_{125}$
are possible, leading to double-resonant tri-Higgs production (not considered here).
Decays $\Phi_S\to \tau^+ +\tau^-$ are practically always possible. In the regions in
the NMSSM parameter space with all present constraints
on the signal rates of $H_{125}$ being satisfied the $BR(\Phi_S\to \gamma + \gamma)$ is
in the $0.1-0.3\%$ range, making this decay observable as well. Henceforth we will
consider resonant $b\bar{b}b\bar{b}$, $b\bar{b}\tau^+ \tau^-$ and $b\bar{b}\gamma\gamma$
final states originating from $\Phi_S\to b\bar{b}$, $\Phi_S\to \tau^+ \tau^-$ and
$\Phi_S\to \gamma\gamma$ decays.

Of interest will be the product of cross sections times branching fractions
$\sigma(ggF\to H/A)\times BR(H/A \to H_{125}+ H_S/A_S
 \to b\bar{b}b\bar{b},\; b\bar{b}\tau^+ \tau^-\ \text{and}\ b\bar{b}\gamma\gamma)$ for
various masses $M_H$, $M_{H_S}$ and $M_{A_S}$, for realistic regions in the parameter space
of the NMSSM. To this end we have performed scans  using the public code
\texttt{NMSSMTools\_5.1.0} \cite{Ellwanger:2004xm,Ellwanger:2005dv} including
the radiative corrections from \cite{Degrassi:2009yq}. All phenomenological constraints,
including the absence of Landau singularities below the GUT scale and,
notably, constraints from Higgs searches in various channels at LEP and LHC
are applied. 

These include searches for scalar and pseudoscalar Higgs production at
LEP (including unconventional Higgs decays), constraints from $B$-physics,
constraints on the mass of $H_{125}$ ($\pm 3$~GeV to account for theoretical
uncertainties) and on its signal rates from the combined ATLAS and CMS run I results
which disallow too large $H_{125}-H_S$ mixings,
constraints from searches for $ggF\to H_S\to \gamma\gamma$ for $M_{H_S}=65-122$~GeV,
and searches for $H/A$ in the $H/A\to \tau\tau$ channel
with $H/A$ produced in association with $b$-quarks and via $ggF$.
We note that
constraints from these latter searches in the $M_A-\tan\beta$ plane are weak for
$\tan\beta \approx 2-3$, typical in the NMSSM, and further alleviated
if $H/A$ have large branching fractions into the final states considered here.

The results of these scans for $\sigma(ggF\to H/A)\times BR(H/A \to H_{125}+ H_S/A_S
 \to b\bar{b}b\bar{b},\; b\bar{b}\tau^+ \tau^-$ and $b\bar{b}\gamma\gamma)$
will be compared to the sensitivities in different final states in the following
sections. The $ggF$ production cross sections for $H/A$
have been obtained from the CERN Yellow Report web page \cite{BSMAt13TeV}
at NNLO+NNLL, after an appropriate rescaling of the $H/A$-gluon-gluon
coupling provided by \texttt{NMSSMTools\_5.1.0}. Also the branching ratios
of $H/A$ and $H_S/A_S$ are taken from \texttt{NMSSMTools\_5.1.0}.
In the Figures showing the 95\%~CL exclusion limits and $5\,\sigma$
discovery cross sections, viable values for the cross sections times branching fractions in
the parameter space of the NMSSM will be indicated as light shaded blue regions
for $ggF\to H \to H_{125}+ H_S$, and as light shaded red regions for
$ggF\to A \to H_{125}+ A_S$.
For simplicity we will use the notation $ggF\to H \to H_{125}+ H_S$ for the
search strategies; the same search strategies apply to $ggF\to A \to H_{125}+ A_S$.

\section{Simulation of Signal Samples}
\label{sec:sigsim}

Signal events for the production of $H$ in $ggF$ are generated by \texttt{MadGraph5\_aMC@NLO}
\cite{Alwall:2014hca} with matrix elements at NLO taken from \texttt{aMC\_SusHi}
\cite{Harlander:2012pb,Liebler:2015bka,Mantler:2015vba} using the NNPDF2.3NLO PDF
set \cite{Ball:2013hta} in the 4-flavour scheme. Renormalization and factorization
scales are chosen as $H_T/2$ on an event-wise basis. \texttt{Pythia6.4}
\cite{Sjostrand:2006za} is used for the $H \to H_{125}+ H_S$ decays, the
$H_{125}$ and $H_S$ decays and the subsequent showering and hadronization.
The total widths of $H$ are below $M_H/50$ in all cases, below $M_H/100$ for
$M_H < 500$~GeV, hence the narrow width approximation is well satisfied.

Separate signal events have been generated for each pair $(M_H, M_{H_S})$.
For $M_H$ we chose $M_H=425,\ 500,\ 625,\ 750\ \text{and}\ 1000$~GeV.
We varied $M_{H_S}$ in steps of 10~GeV in the range $25-225$~GeV,
and in steps of 20~GeV above 225~GeV up to the kinematic boundary
$M_H-125$~GeV (except for $M_H=1000$~GeV where $M_{H_S}$ was varied in
steps of 25/50~GeV).

For each pair $(M_H, M_{H_S})$ we generated 150k unweighted events, more than the expected
number of events at 3000~fb$^{-1}$. Accordingly the statistical fluctuations
from the Monte Carlo (MC) generation are small compared to the expected
statistical fluctuations from the data; the latter will be taken into account.

The output is given to the detector simulation \texttt{Delphes 3}
\cite{deFavereau:2013fsa}. Jets are clustered with \texttt{FastJet 3.0.1}
\cite{hep-ph/0512210} using the anti-$k_T$ algorithm with $\Delta R = 0.4$.
For $b$-tagging the ATLAS card is used in \texttt{Delphes 3}. 

The $p_T$ dependence of the $b$-tagging and mistagging efficiencies is
chosen in the ATLAS card following the parametrizations given in
\cite{ATL-PHYS-PUB-2015-022}. The default value of the parameters in
\texttt{Delphes 3.3.2} correspond to a working point $\varepsilon_b = 70\%$
as, e.g., in the ATLAS search for Higgs pair production in the $4b$
final state at 13~TeV in \cite{ATL-CONF-2016-049}. We will
employ the same settings except for the $b\bar{b}\gamma\gamma$ final state.

\section{Search Strategy for the $b\bar{b}b\bar{b}$ Final State}

Searches for resonant SM Higgs pair production in the $b\bar{b}b\bar{b}$ final state
have been performed before by ATLAS at 8~TeV \cite{ATL-CONF-2014-005,Aad:2015uka,
Aad:2015xja} and at 13~TeV \cite{ATL-CONF-2016-049}, and
by CMS at 8~TeV \cite{CMS:2014eda,Khachatryan:2015yea,Khachatryan:2016cfa} and at 13~TeV
\cite{CMS:2016tlj,CMS:2016pwo}.

Searches for $ggF \to H \to H_S + H_{125} \to b\bar{b}b\bar{b}$ are complicated
by the presence of two unknown masses of $H$ and $H_S$. A naive approach would be
to require one $b\bar{b}$ pair with a mass near $125$~GeV, and to look for simultaneous
excesses in the plane of invariant masses of the other $b\bar{b}$ pair and the total
$4b$ invariant mass. However, this approach does not allow to optimize cuts as
function of different masses of $H$ and $H_S$. An at least $\sim 20\%$ gain in efficiency
can be obtained as follows:

\noi a) Choose a tentative value for $M_{H_S}$, and optimise the cuts
and the choice of $b\bar{b}$ pairs as function of this value;

\noi b) Search subsequently for an excess in the total $4b$ invariant mass
(suitably corrected, see below).

Subsequently we describe first our simulation of signal samples and 
the strategies for the analysis. In the following subsections we discuss
the background simulation and validation, and finally the results for
the expected 95\%~CL upper limits and $5\,\sigma$ discovery limits on the
cross sections times branching fractions
as function of $M_H$ and $M_{H_S}$. The latter are compared
to possible production cross sections times branching fractions in the NMSSM.

\subsection{Analyses of Signal Samples}
\label{sec:bbbbsignals}

After the simulation of signal samples as described in section~\ref{sec:sigsim},
at least four $b$-tagged jets with
$p_T > 30$~GeV and $|\eta| < 2.5$ are required.
Four $b$-tagged jets can be paired in six different ways. The two invariant
masses of two $b\bar{b}$ pairs are tentatively denoted by $M_{b\bar{b}}(H_{125})$ and
$M_{b\bar{b}}(H_S)$. The subsequent procedure depends on the chosen value for
$M_{H_S}$, and has to be repeated for each choice.

\begin{figure}[!t]
\centering
\includegraphics[scale=0.55]{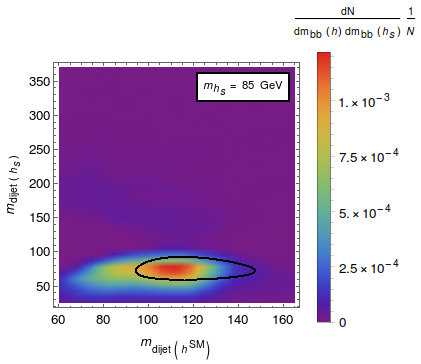}\hfill
\includegraphics[scale=0.55]{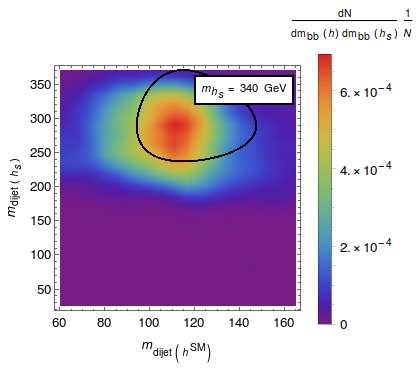}
\caption{Dijet mass distributions $M_{b\bar{b}}(H_{125})$ and $M_{b\bar{b}}(H_S)$
for two different benchmark points $M_{H_S}=85$~GeV and $M_{H_S}=340$~GeV. The black
contours indicate the signal regions defined by $\chi < 2$.}
\label{fig:dijetmasses}
\end{figure}

An event is kept only if a pairing exists for which $M_{b\bar{b}}(H_{125})$ is
sufficiently close to 125~GeV, and $M_{b\bar{b}}(H_S)$ is sufficiently close to
$M_{H_S}$. In practice, the measured invariant masses of $b\bar{b}$ pairs are
often somewhat smaller than the mass of the decaying Higgs boson. Therefore
``sufficiently close to'' should better be replaced by ``slightly below''
as in \cite{ATL-CONF-2016-049,CMS:2016tlj}. Generalizing the conditions applied in
\cite{ATL-CONF-2016-049,CMS:2016tlj}, an event is kept only if a pairing exists for which 
\beq
\label{eq:chi}
\chi = \sqrt{\left(
\frac{M_{b\bar{b}}(H_{125})-115\ \text{GeV}}{0.1\; M_{b\bar{b}}(H_{125})}\right)^2 +
\left(\frac{M_{b\bar{b}}(H_S)-0.85\; M_{H_S}}{0.1\; M_{b\bar{b}}(H_S)}\right)^2}
< 2\; .
\eeq
If different pairings within a given event satisfy (\ref{eq:chi}), the combination
that minimizes $\chi$ is chosen.
In Figs.~\ref{fig:dijetmasses} we show, for $M_H=500$~GeV, the distributions of the dijet masses
$M_{b\bar{b}}(H_{125})$ and $M_{b\bar{b}}(H_S)$ for the pairing minimizing $\chi$,
for two different benchmark points $M_{H_S}=85$~GeV and $M_{H_S}=340$~GeV in the case where
$M_{H_S}$ for the analysis was chosen correctly. The black
contours indicate the signal regions defined by $\chi < 2$.
\footnote
{An alternative pairing algorithm based on the angular distances of the
constituent $b$-jets was studied. We found that the resulting sensitivities are
mostly similar to those obtained using the mass-based algorithm described above when the
jets are sufficiently collimated. However, as $M_{H_S}$ increases, the jets become
too back-to-back, making the angular-pairing contraproductive. For this reason, we used the
mass-based algorithm for the reconstruction of the Higgs candidates.}

In the case of heavy resonances with masses above $\sim 1$~TeV, the two b-jets from
a single Higgs boson tend to merge into a single fat jet. Accordingly a
``boosted'' analysis based on single $\Delta R=1.0$ jets per Higgs boson was applied
in \cite{ATL-CONF-2016-049} for searches for such heavy resonances. We found that
for such heavy $H$ states the production cross sections become too small for
reasonable sensitivities, and limit ourselves to $M_H \leq 1$~TeV subsequently.
We had tried to vary the jet reconstruction parameter $\Delta R$ without
observing, however, a major impact on the possible sensitivities.
The next step in the event selection are cuts on $p_T$ of the $b\bar{b}$ pairs
associated with $H_{125}$ and $H_S$, respectively. 

These cuts can be optimised with respect to $M_{4b}$ (defined by the event) and the
tentative value for $M_{H_S}$. To this end we considered many samples of $M_H\sim M_{4b}$
and $M_{H_S}$. In each case we studied the dependence of the sensitivity
$\epsilon_s/\sqrt{\epsilon_B}$ ($\epsilon_{s}$ and $\epsilon_{B}$ denote the efficiencies
of the signal and background, respectively) on the cuts on $p_T$. Maximizing the
sensitivities, we obtain different values for the optimal cuts on $p_T$ for each sample of
$M_{4b}$ and $M_{H_S}$. These different values for the cuts on $p_T$ are well
approximated by the following functions of $M_{4b}$ and $M_{H_S}$:
\bea
\label{eq:pt}
p_T({b\bar{b}}(H_{125})) &>& 1.6\ \text{GeV} + 0.4\, M_{4b}-0.13\, M_{H_S} - \frac{M_{H_S}}{M_{4b}}
\cdot 160\ \text{GeV}\nn \\
p_T({b\bar{b}}(H_S)) &>& 12\ \text{GeV} + 0.4\, M_{4b}-0.15\, M_{H_S} - \frac{M_{H_S}}{M_{4b}}
\cdot 166\ \text{GeV}
\eea

\begin{figure}[!t]
\centering
\includegraphics[scale=0.70]{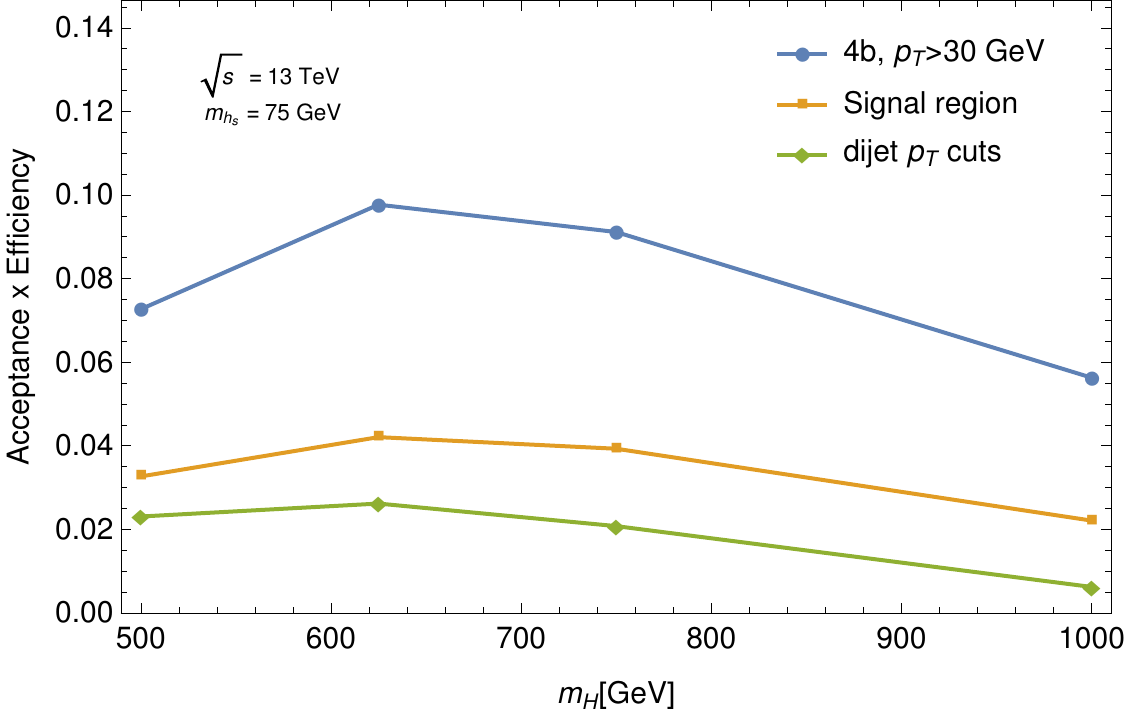}\hfill
\includegraphics[scale=0.70]{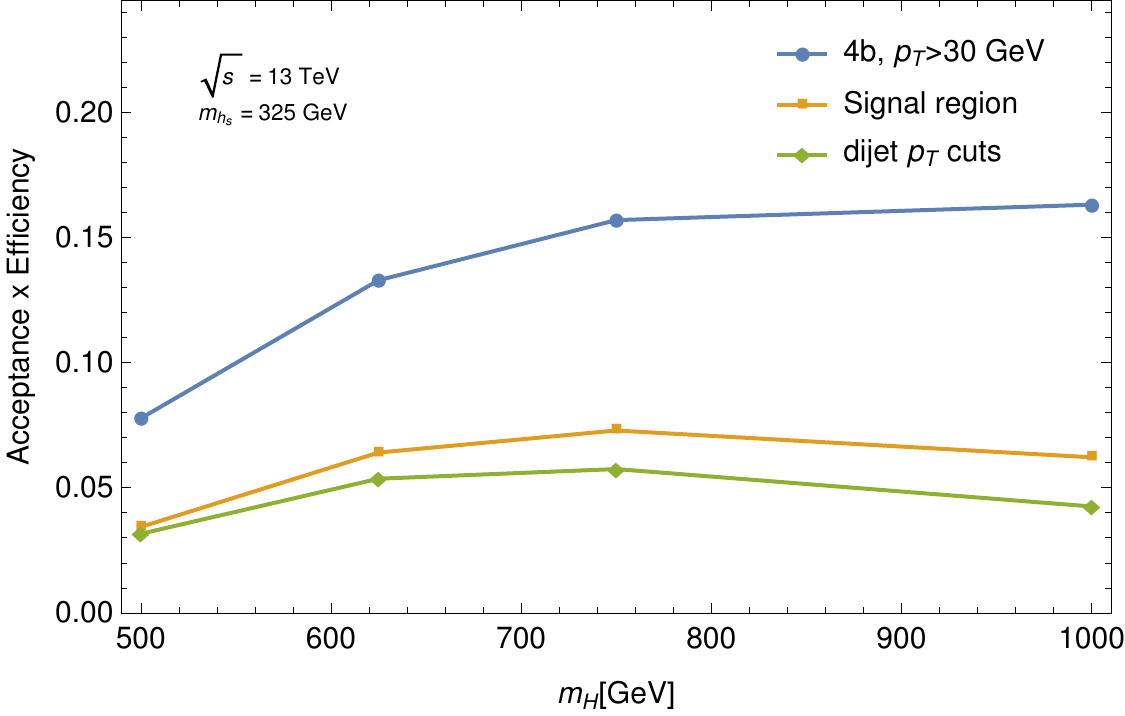}
\caption{Efficiencies of the three cuts ($4b$ with $p_T > 30$~GeV), signal region
(\ref{eq:chi}) and dijet $p_T$ (\ref{eq:pt}) for $M_{H_S}=75$~GeV
and $M_{H_S}=325$~GeV as function of $M_H$.}
\label{fig:efficienH}
\end{figure}

\begin{figure}[!t]
\centering
\includegraphics[scale=0.70]{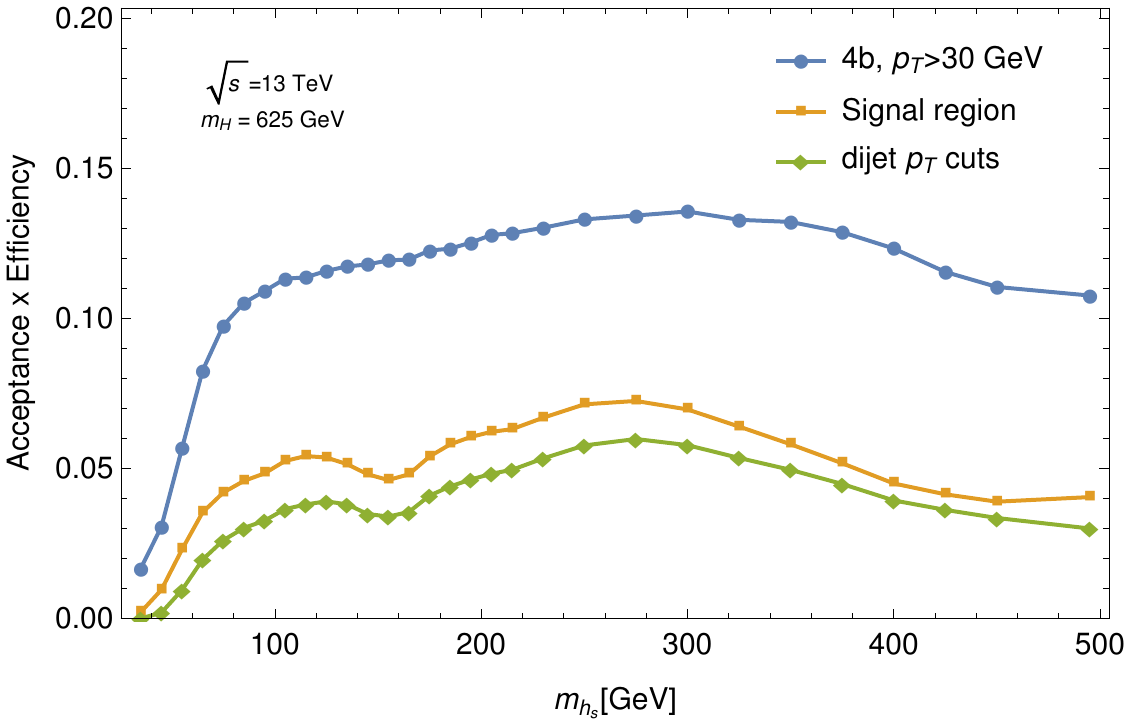}\hfill
\includegraphics[scale=0.70]{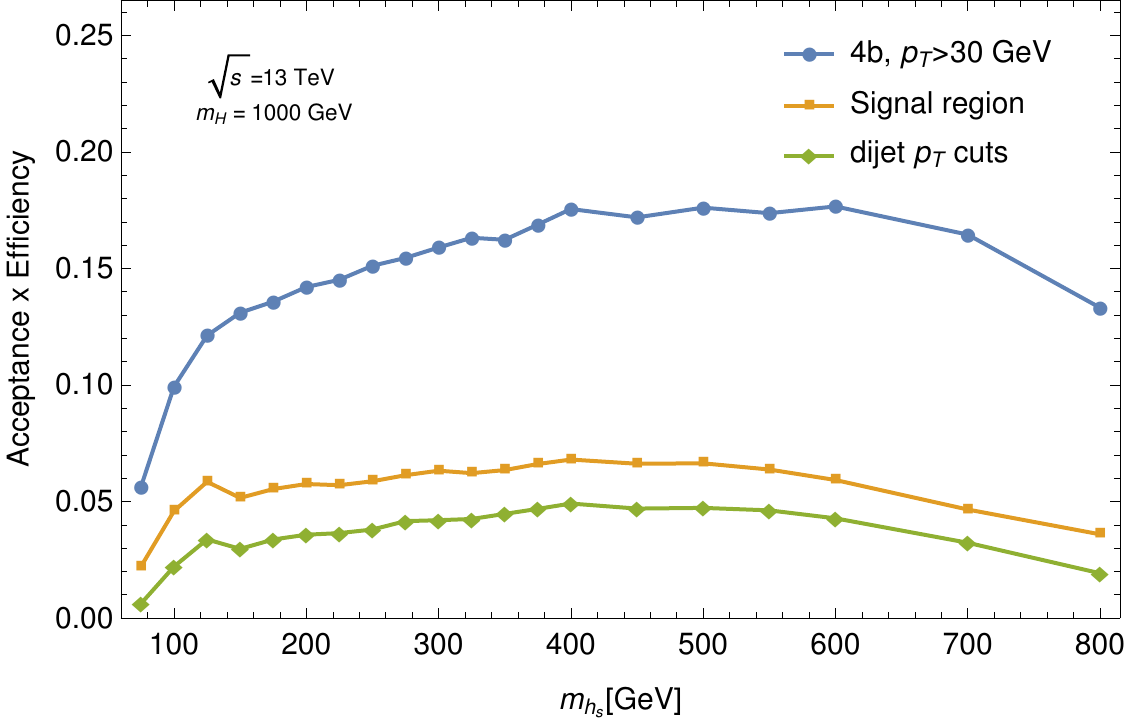}
\caption{Efficiencies of the three cuts ($4b$ with $p_T > 30$~GeV), signal region
(\ref{eq:chi}) and dijet $p_T$ (\ref{eq:pt}) for $M_H=625$~GeV and $M_H=1000$~GeV
as function of $M_{H_S}$.}
\label{fig:efficienHS}
\end{figure}

The efficiencies of the three cuts a)~$4b$ with $p_T > 30$~GeV, b)~signal region
(\ref{eq:chi}) and c)~dijet $p_T$ (\ref{eq:pt}) are shown in Figs.~\ref{fig:efficienH}
for $M_{H_S}=75$~GeV and $M_{H_S}=325$~GeV as function of $M_{H}$, and in
Figs.~\ref{fig:efficienHS} for $M_H=625$~GeV and $M_H=1000$~GeV as function of $M_{H_S}$.
One observes a decrease of the efficiency of the condition of four $b$-tagged jets
for $M_{H_S} \lsim 80$~GeV (for $M_H=625$~GeV) and $M_{H_S} \lsim 100$~GeV (for $M_H=1000$~GeV).
Here the two $b$-jets from $H_S$ are boosted and hardly get resolved by the standard jet clustering
algorithm; dedicated boosted analyses as in \cite{ATL-CONF-2016-049}
could be envoked for this configuration.

For the search for a resonance $H$ we found it useful to replace $M_{4b}$ by
$M_X$, with $M_X$ defined such that uncertainties in the measurements of $M_{b\bar{b}}$ relative
to $M_{H_{125}}$ and $M_{H_S}$ (typically due to radiation out of the jet cones)
are corrected:
\beq
\label{eq:MX}
M_X = M_{4b}+125~\text{GeV}-M_{b\bar{b}}(H_{125}) + M_{H_S}-M_{b\bar{b}}(H_S)\; .
\eeq
The empirical variable $M_X$ was already used by CMS in resonant double Higgs
production search in the $b\bar{b}\gamma\gamma$ channel \cite{CMS:2016tlj}.
We found that replacing $M_X$ by a full Lorentz covariant expression in terms
of $M_{H_{125}}$, $M_{H_S}$ and the two values of $M_{b\bar{b}}$ did not improve
the sensitivities.

\begin{figure}[!t]
\centering
\includegraphics[scale=0.60]{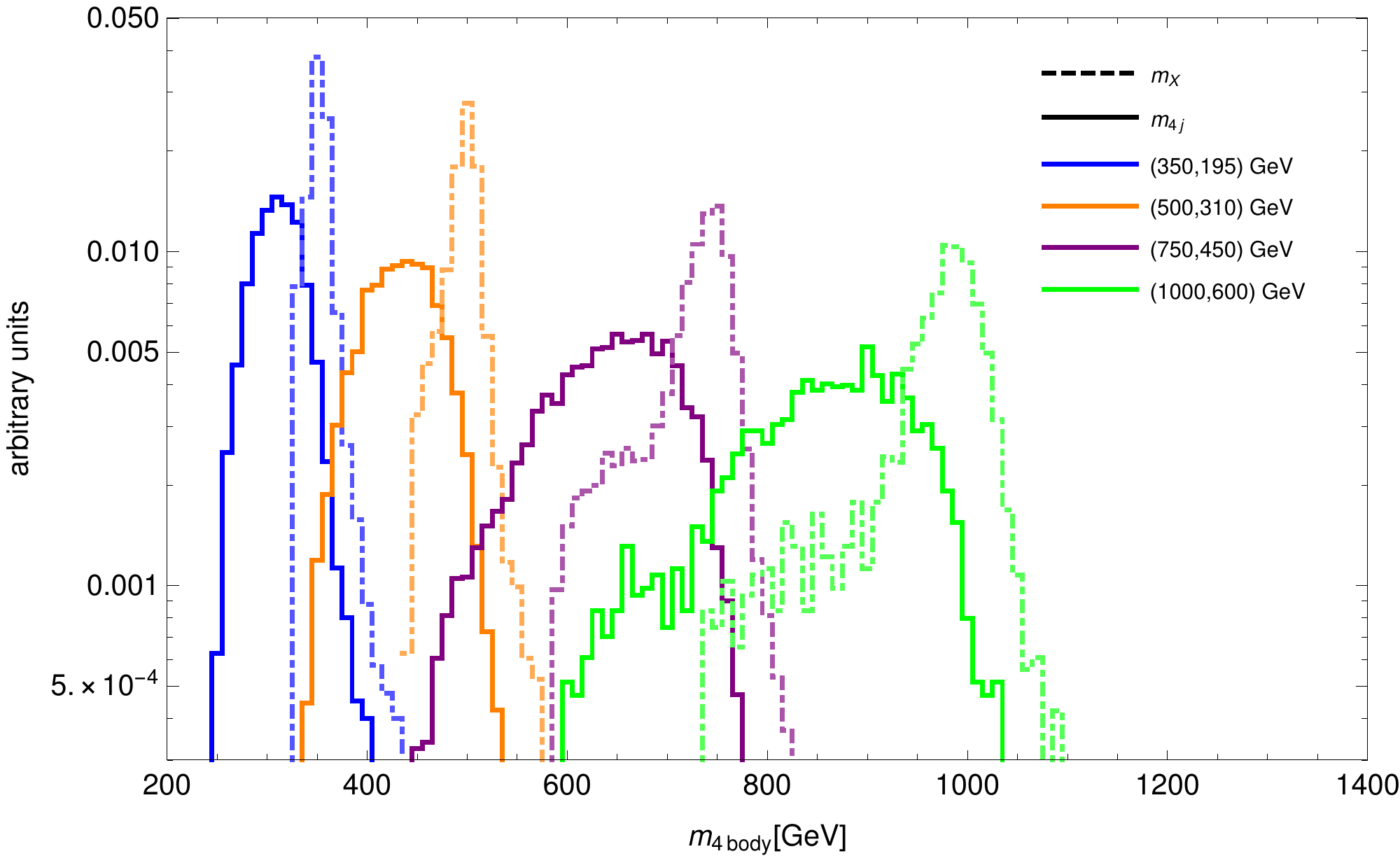}
\caption{Distributions of the reconstructed masses $M_{4b}$ (full lines) and $M_X$
(dashed lines)
for the signal samples $(M_H,M_{H_S})=(350,195),\,(500,310),\,(750,450),\,(1000,600)$~GeV.}
\label{fig:MXvsM4}
\end{figure}

In Fig.~\ref{fig:MXvsM4} we show the distributions of the reconstructed masses $M_{4b}$ and $M_X$
for the signal samples $(M_H,M_{H_S})=(350,195),\,(500,310),\,(750,450),\,(1000,600)$~GeV.
We clearly see a sharpening of the peaks using $M_X$ in each case. 

In order to obtain
the required production cross sections times branching fractions for $5\,\sigma$ discovery or 95\%~CL
exclusion we have to obtain the background distribution of $M_X$.

\subsection{Background Estimation}

Dominant backgrounds for the $b\bar{b}b\bar{b}$ final state are QCD multijet processes
including jet misidentifications, and $t\bar{t}$ \cite{ATL-CONF-2016-049,CMS:2016tlj}.
The QCD multijet background is difficult to obtain from Monte Carlo simulations
alone, and estimated from sidebands in \cite{ATL-CONF-2016-049,CMS:2016tlj}.
Such data is not available for the different values of $M_{H_S}$ studied here,
with one exception: For $M_{H_S}\sim 125$~GeV, the $b\bar{b}b\bar{b}$ final state
coincides with the one searched for in \cite{ATL-CONF-2016-049,CMS:2016tlj}.

We simulated $b\bar{b}b\bar{b}$, $b\bar{b}c\bar{c}$, $b\bar{b}j{j}$ (with
$j\neq c/\bar{c}$) and $t\bar{t}$ processes as we did for the signal samples
in section~\ref{sec:sigsim}.
After applying the cuts of the previous subsection the relative contributions are
$\sim 85-88\%$ from $b\bar{b}b\bar{b}$, $\sim 7-8\%$ from $b\bar{b}c\bar{c}$, 
$\sim 4-8\%$ from $t\bar{t}$ (depending on $M_{H_S}$ and $M_{4b}$).
The $b\bar{b}jj$ contribution is only $0.8\%$
and will subsequently be neglected. We checked that $b\bar{b}b\bar{b}+jets$
processes have little impact on the event shapes. Hence we did not simulate
them separately, but took a NLO K-factor of 1.7 \cite{Alwall:2014hca} into account.

In \cite{ATL-CONF-2016-049} the $M_{4b}$ distribution of the multijet background
has been obtained from a signal free sideband (with two $b$-tags only, and 
$M_{b\bar{b}}$ outside the search window) and appropriate rescaling using data with
four $b$-tags, subtracting the $t\bar{t}$ contribution. The measured $M_{4b}$ distribution
is available in Fig.~5 in \cite{ATL-CONF-2016-049}, where it is compared
to the estimated background.

This allows us to proceed in a similar fashion: 
In order to compare to the data in \cite{ATL-CONF-2016-049}, the previous cuts are slightly
modified: The $b\bar{b}$ pairs are ordered in $p_T$ according to $M_{b\bar{b}}^{lead}$
and $M_{b\bar{b}}^{subl}$ and, as in \cite{ATL-CONF-2016-049}, the signal region
is defined by
\beq
\label{eq:chi_atlas}
\chi = \sqrt{\left(
\frac{M_{b\bar{b}}^{lead}-120\ \text{GeV}}{0.1\; M_{b\bar{b}}^{lead}}\right)^2 +
\left(\frac{M_{b\bar{b}}^{subl}-0.85\; M_{H_S}}{0.1\; M_{b\bar{b}}^{subl}}\right)^2}
< 1.6\; .
\eeq

\begin{figure}[!t]
\centering
\includegraphics[scale=0.53]{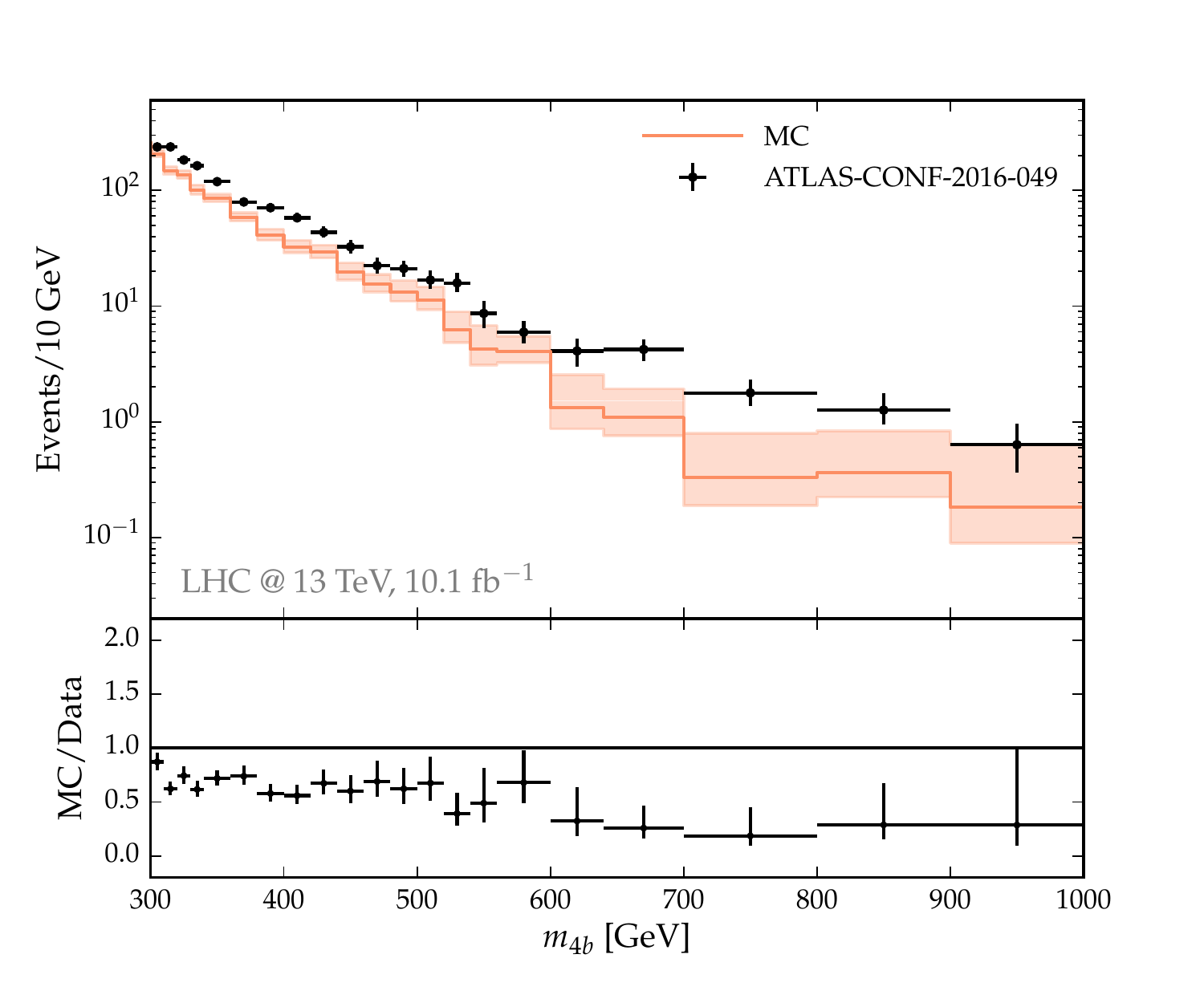}
\hfill
\includegraphics[scale=0.53]{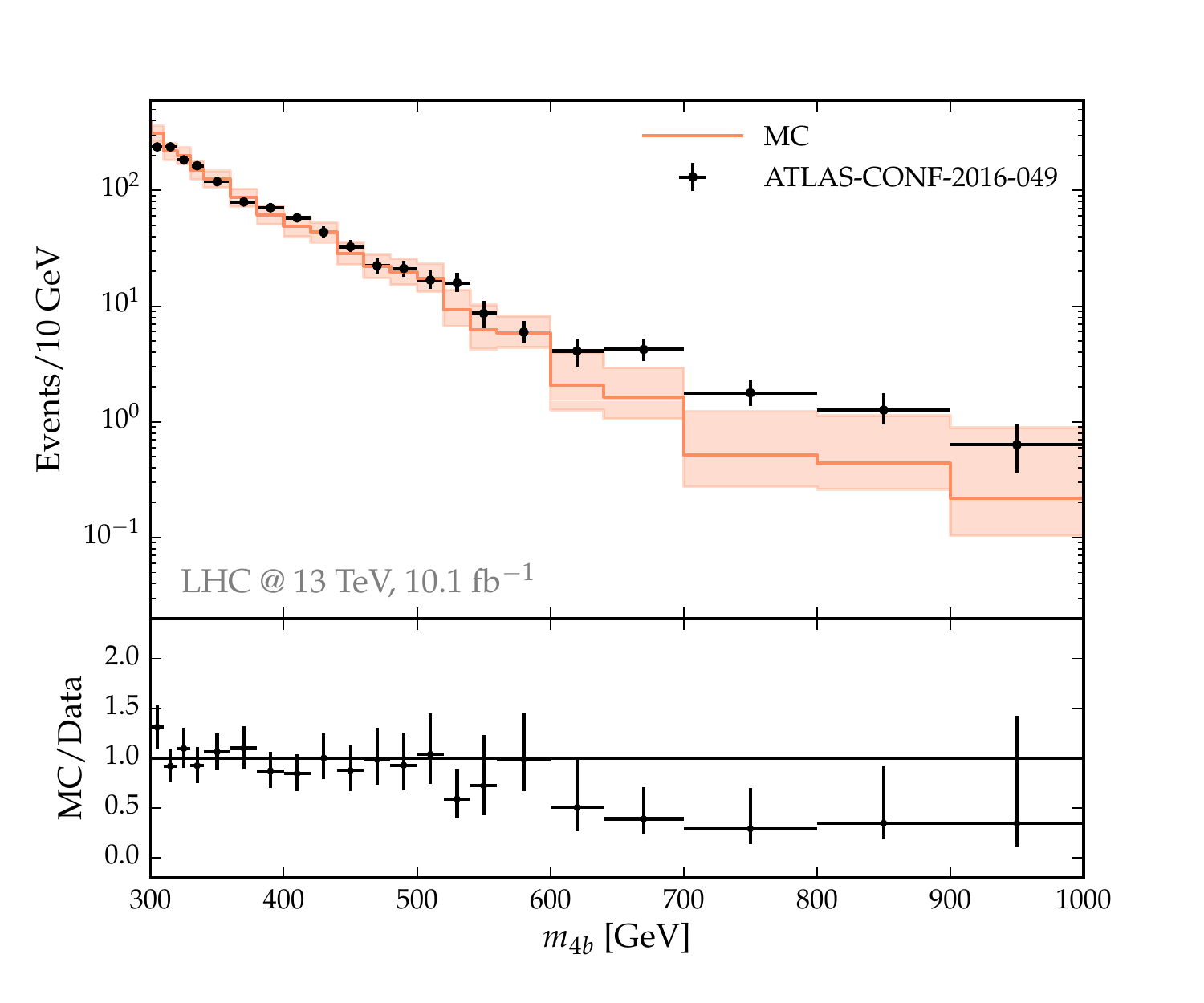}
\caption{Left hand side: measured $M_{4b}$
distribution from ATL-CONF-2016-049 \cite{ATL-CONF-2016-049}, and our MC result
for the background
with statistical errors expected for 10.1~fb$^{-1}$ of integrated luminosity.
The lower part shows the ratio MC/data bin by bin. Right hand side: the $M_{4b}$
distribution of our background after rescaling, compared to the data
in ATL-CONF-2016-049 \cite{ATL-CONF-2016-049}.}
\label{fig:M4b_vs_ATLAS}
\end{figure}

Still (and expectedly) our simulated background falls below the measured data
given in  \cite{ATL-CONF-2016-049}.
On the left hand side of Fig.~\ref{fig:M4b_vs_ATLAS} we show the $M_{4b}$
distribution measured by \cite{ATL-CONF-2016-049} using 10.1~fb$^{-1}$ of integrated luminosity,
and our MC result with statistical errors.
The number of generated MC events corresponds to an equivalent integrated luminosity of
$\sim$13 fb$^{-1}$. The statistical error per bin is thus obtained from the number
of MC events per bin rescaled by ${10}/{13}$.
The lower panels show the ratio MC/data bin by bin, with the uncertainties
from the data and from our MC combined. At least for the interesting
region $M_{4b}\gsim 350$~GeV an overall rescaling of our multijet background, like
it was performed in \cite{ATL-CONF-2016-049}, seems appropriate.
The rescaling factor is obtained taking the average of the data/MC ratio of all bins
weighted by the corresponding uncertainties.
We obtain a rescaling factor of $1.55 \pm 0.27$.
(The $t\bar{t}$ background is left untouched, and remains at $\sim 3-6\%$.)
The comparison of the $M_{4b}$ distribution of our background
after rescaling to the data from \cite{ATL-CONF-2016-049}
is shown on the right hand side of Fig.~\ref{fig:M4b_vs_ATLAS}.

Clearly it is somewhat optimistic to assume that the rescaling of the multijet
background by $1.55 \pm 0.27$ remains valid for $M_{H_S} \neq 125$~GeV. In the
absence of data from sidebands this is, however, the best we can do.
Subsequently $\pm\, 0.27$ will be used as an estimation of the systematic uncertainty of
our background for all $M_{H_S}$, a number to be considered
as indicative.

\begin{figure}[!t]
\centering
\includegraphics[scale=0.50]{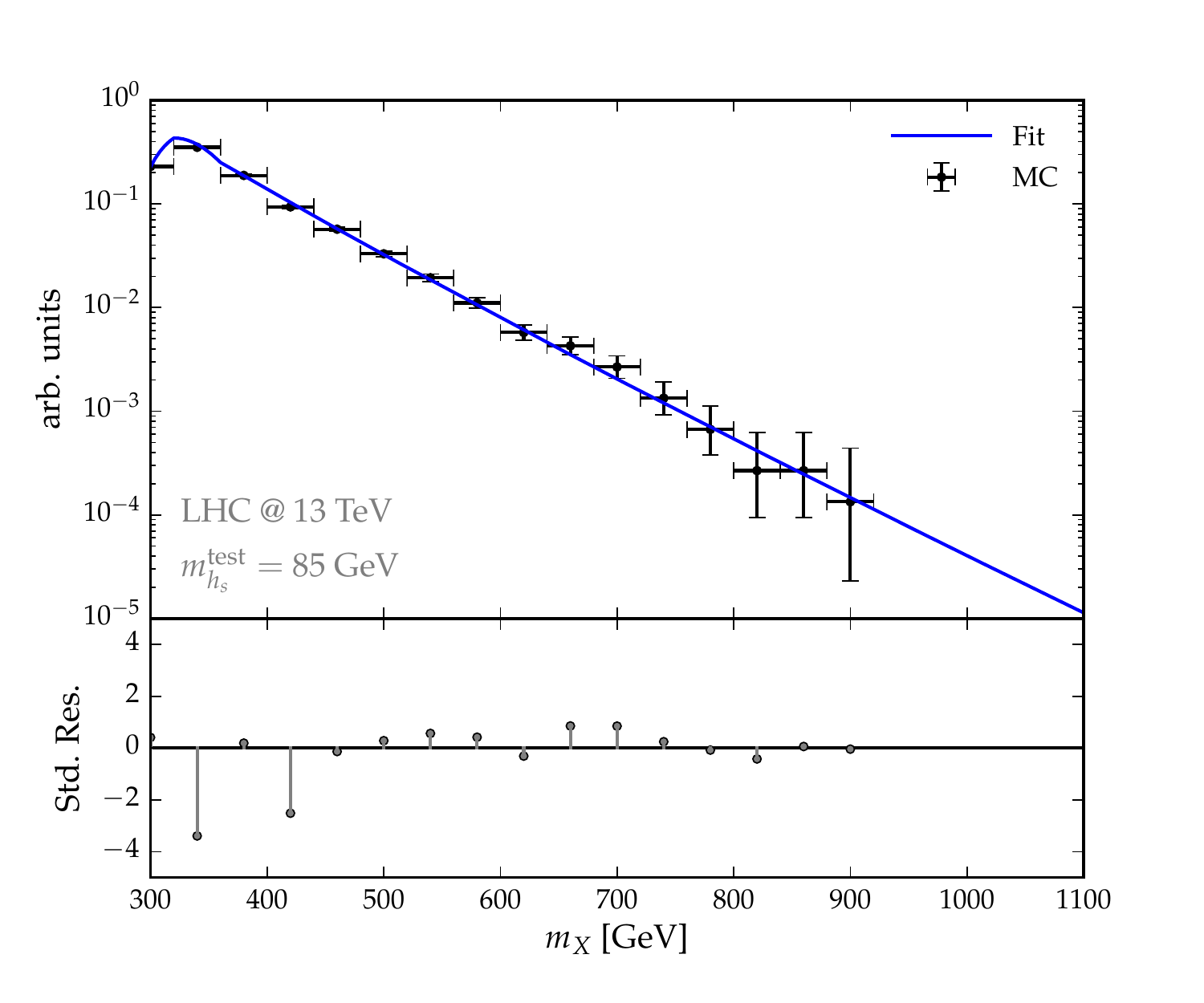}\hfill
\includegraphics[scale=0.50]{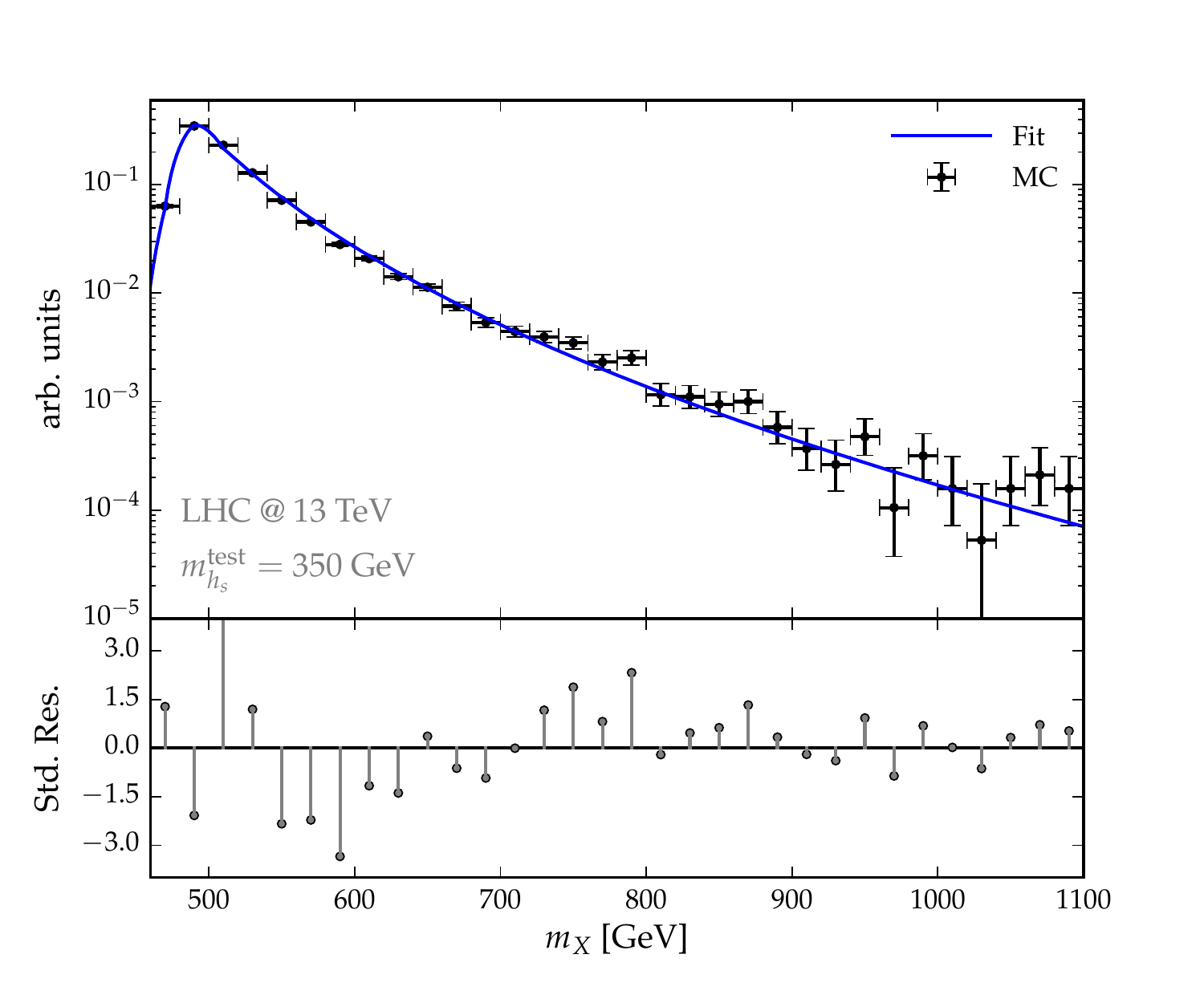}
\caption{Fits by 4~parameter functions (\ref{eq:gammadist}) to distributions of $M_X$
of the background
obtained by Monte Carlo simulations of $b\bar{b}b\bar{b}$, $b\bar{b}c\bar{c}$ and $t\bar{t}$
events after cuts, for $M_{H_S}=85$~GeV (left) and $M_{H_S}=350$~GeV (right). The
indicated uncertainties originate from the Monte Carlo samples.}
\label{fig:backgrdfit}
\end{figure}

For forecasts at 300 or 3000~fb$^{-1}$ integrated luminosity the statistical
uncertainties of the background are much smaller. It is then
convenient to fit the shape of the $M_X$ background distributions (\ref{eq:MX})
after cuts, which will be used in the following, by continuous functions.
We found that the best fits are provided by a four~parameter Gamma distribution
defined in eq.~(\ref{eq:gammadist}) with $M_{H_S}$ dependent fit parameters.
(The $b\bar{b}b\bar{b}$, $b\bar{b}c\bar{c}$, and $t\bar{t}$ background contributions
to the $M_X$ distributions were fitted separately.)
In Figs.~\ref{fig:backgrdfit} we show the sum of these fits for $M_{H_S}=85$~GeV
and $M_{H_S}=350$~GeV.

Of course the remaining statistical fluctuations of the background can still be evaluated and
combined with the systematic uncertainty (not shown in Figs.~\ref{fig:backgrdfit}).
It turns out, however, that for forecasts at 300 or 3000~fb$^{-1}$
the statistical fluctuations are
negligibly small relative to the systematic uncertainty from
the rescaling by~$1.55 \pm 0.27$.

\subsection{Future 95\% CL Exclusion Limits and $5\,\sigma$ Discovery Cross Sections}

Given the $M_X$ distribution of the background for various hypothetical values
of $M_{H_S}$ and the $M_X$ distributions of signals
one can, following the statistical methods from \cite{Cowan:2010js}
and described in the appendix B,
obtain values for 95\%~CL exclusion limits and $5\,\sigma$ discovery limits for cross sections
times branching fractions into the $b\bar{b}b\bar{b}$ final state as function
of the integrated luminosity, $M_H$ and $M_{H_S}$.

\begin{figure}[!t]
\centering
\includegraphics[scale=0.53]{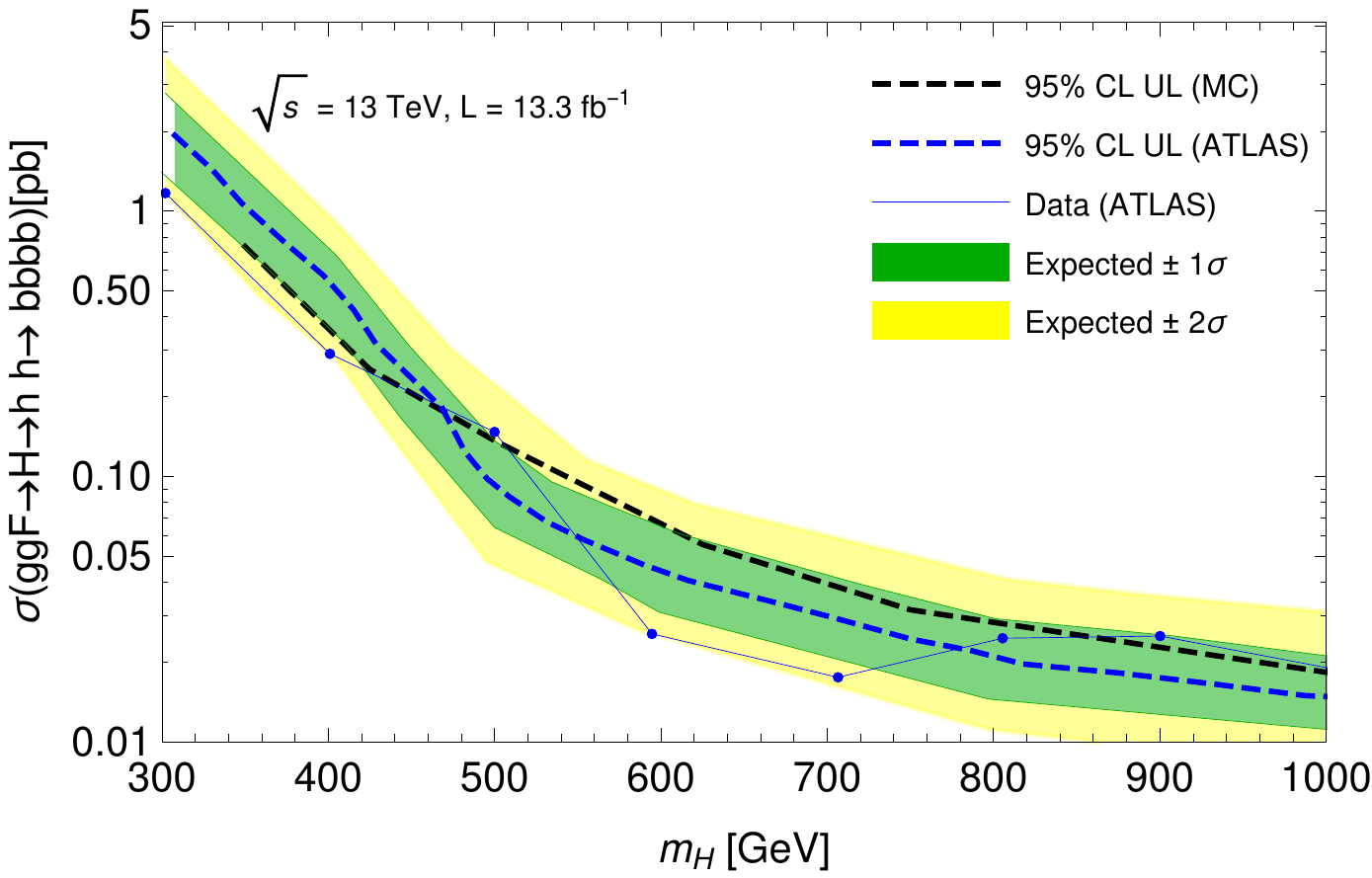}
\caption{Expected 95\%~CL upper limits from ATLAS \cite{ATL-CONF-2016-049} in blue,
their $\pm 2\,\sigma$ incertainty bands, the expected
95\%~CL upper limits from our Monte Carlo in black and, for completeness, the 95\%~CL upper
limits obtained from the data.}
\label{fig:compATLASlimitsbbbb}
\end{figure}

In the case of an integrated luminosity of 13.3~fb$^{-1}$ at 13~TeV we can compare
the expected 95\%~CL exclusion limits on cross sections times branching fractions
to the ones given by ATLAS in Fig.~11 in \cite{ATL-CONF-2016-049}, for
$M_X = 300\dots 1000$~GeV and $M_{H_S}\sim 125$~GeV.
(This ATLAS search was actually dedicated to spin~2 resonances decaying to SM Higgs
pairs, but the differences to spin-0 resonances are expected to be small.)
In Fig.~\ref{fig:compATLASlimitsbbbb} we show the expected 95\%~CL upper
limits from ATLAS, their $\pm 2\,\sigma$ incertainty bands, the expected
95\%~CL upper limits from our Monte Carlo and, for completeness, the 95\%~CL upper
limits obtained from the data. We see that our expected 95\%~CL upper limits
coincide well with the ones expected by ATLAS.

\begin{figure}[!t]
\centering
\includegraphics[scale=0.50]{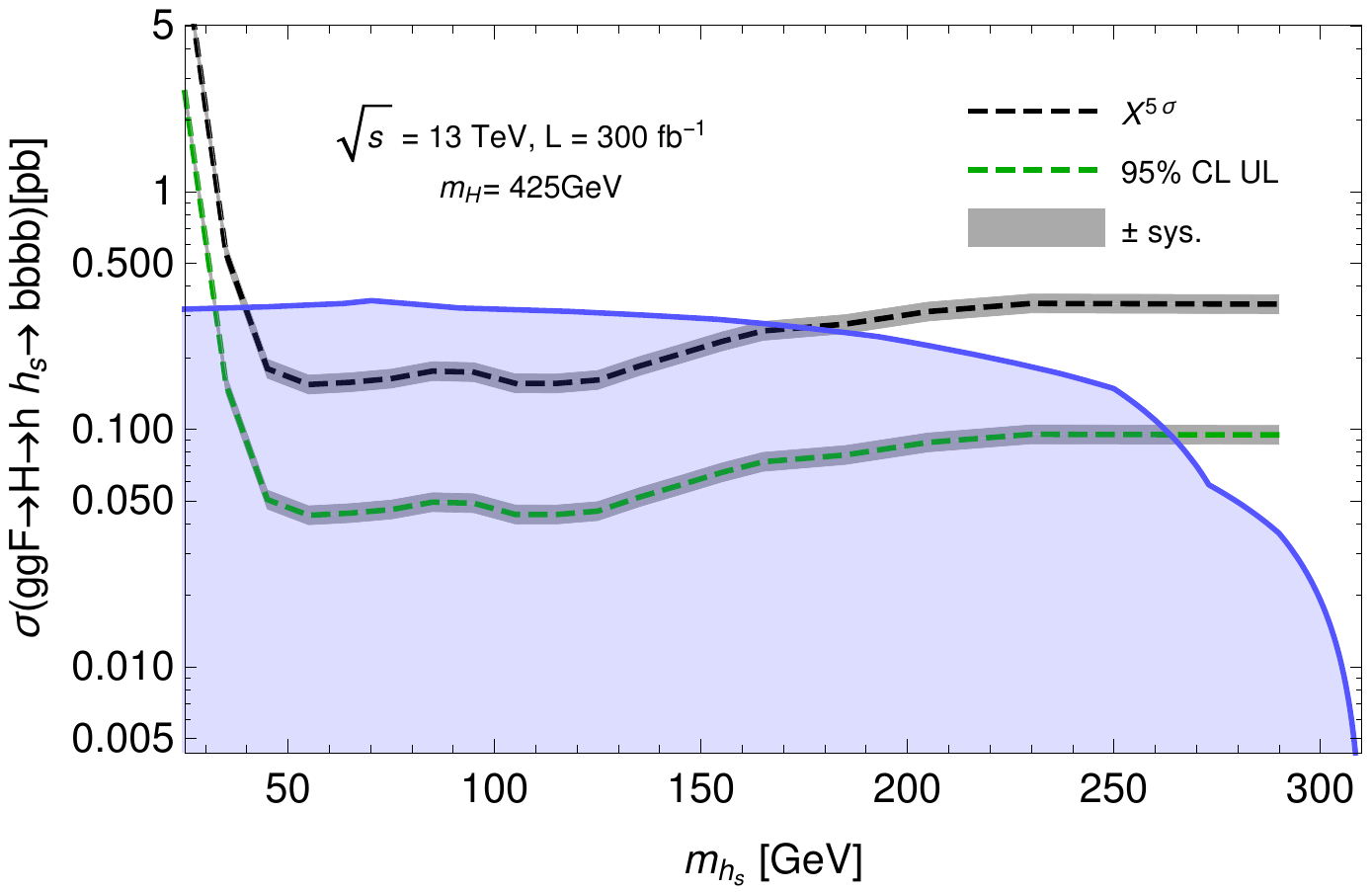}\hfill
\includegraphics[scale=0.50]{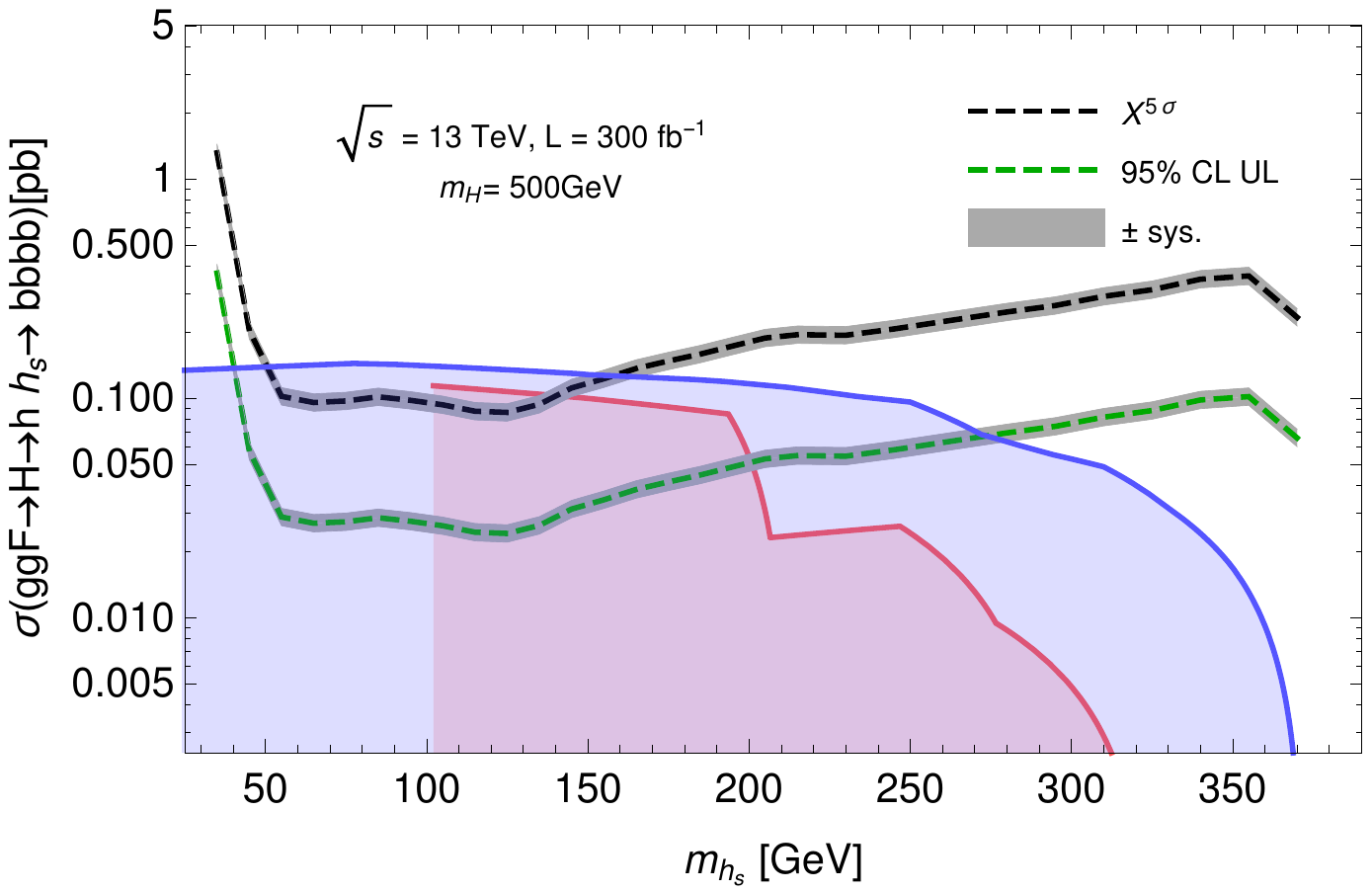}\\
\includegraphics[scale=0.50]{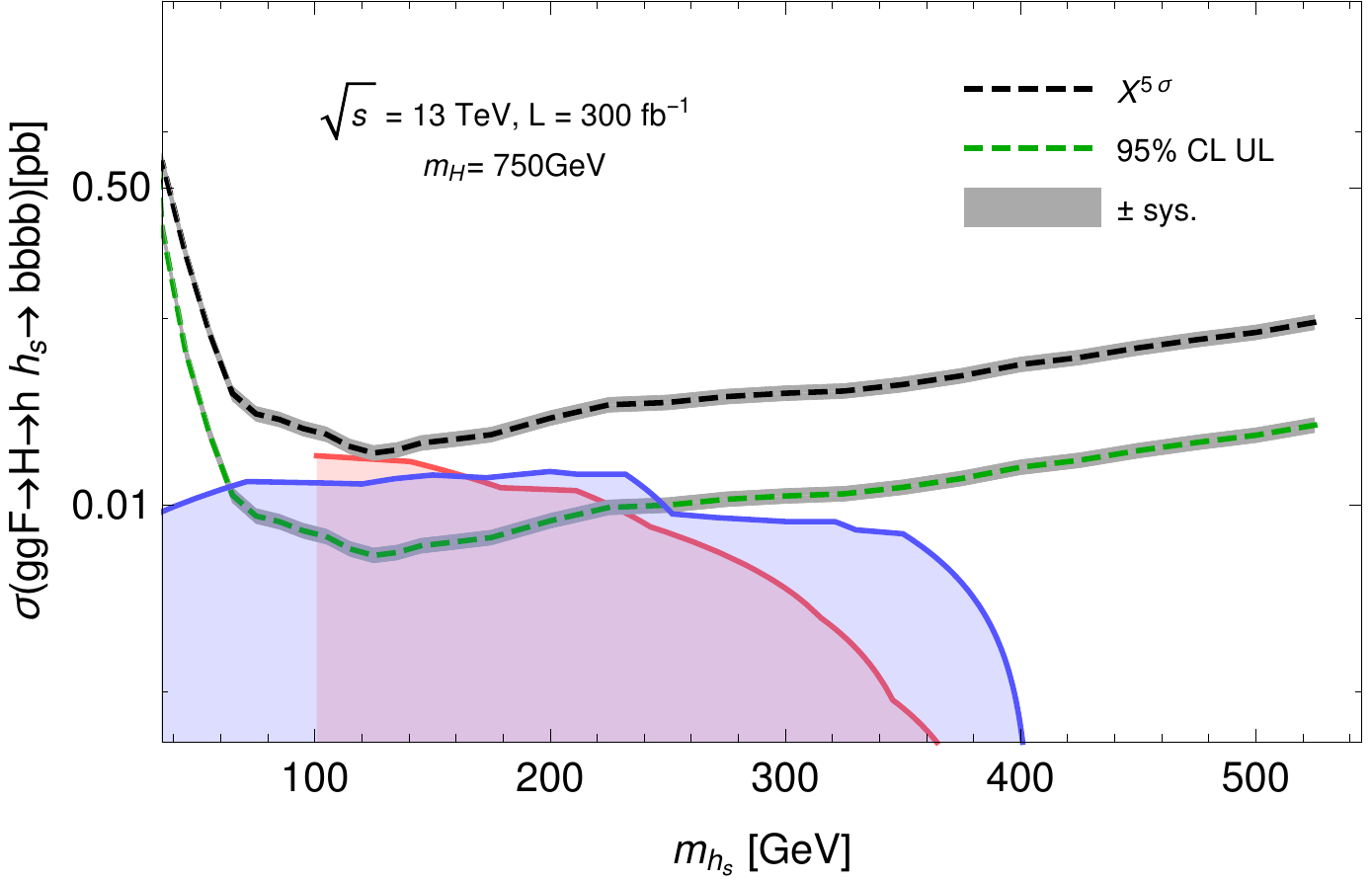}\hfill
\includegraphics[scale=0.50]{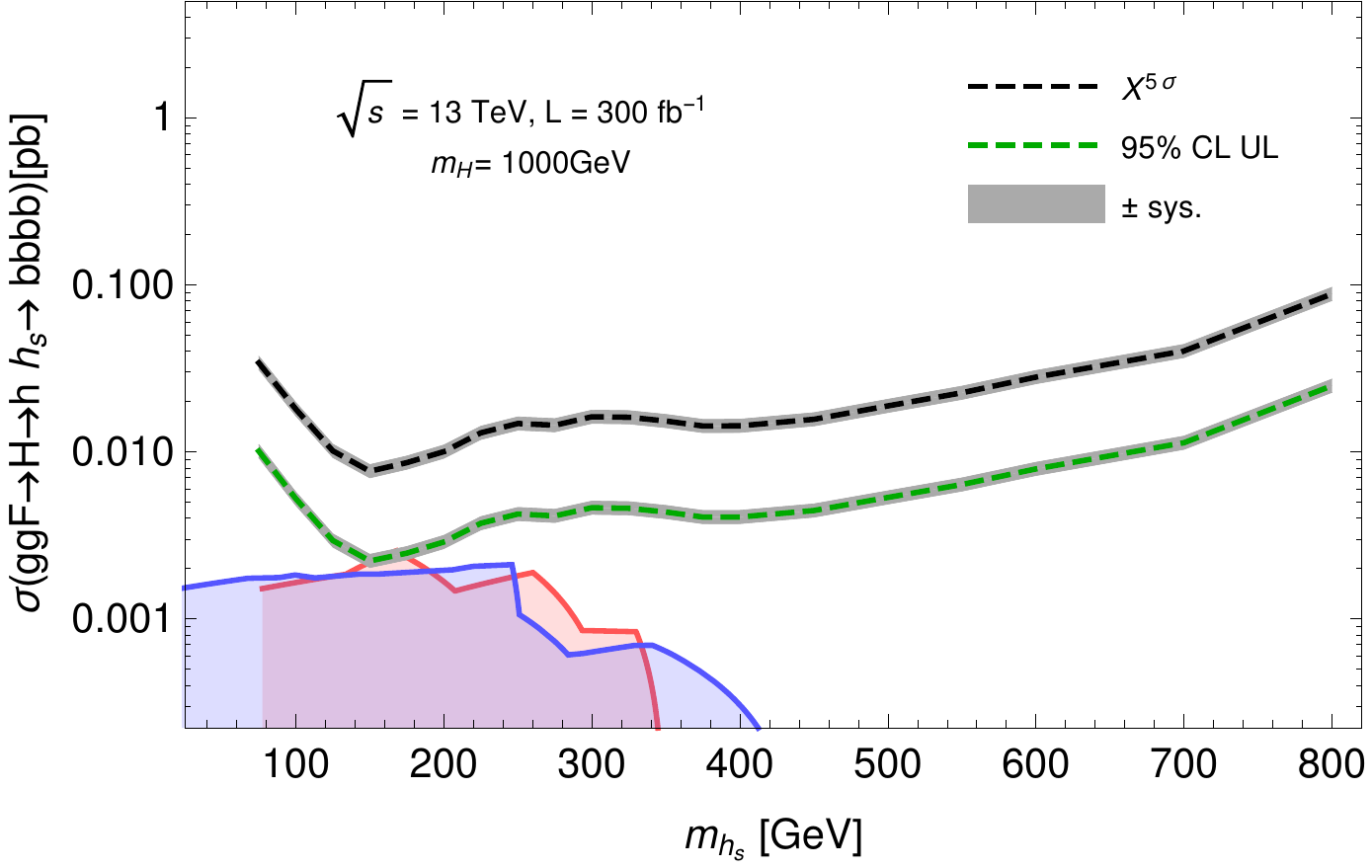}
\caption{95\% CL exclusion limits and $5\,\sigma$ discovery cross sections
in the $b\bar{b}b\bar{b}$ final state
as function of $M_{H_S}$ for 300~fb$^{-1}$ integrated luminosity and
$M_H=425$~GeV (upper left), $M_H=500$~GeV (upper right), $M_H=750$~GeV (lower left),
$M_H=1000$~GeV (lower right). Shaded blue regions: viable values for
$\sigma(ggF\to H\to H_{125}+H_S\to b\bar{b}b\bar{b})$ in the parameter space of the NMSSM.
Shaded red regions: viable values for
$\sigma(ggF\to A\to H_{125}+A_S\to b\bar{b}b\bar{b})$ in case they exceed
potentially the ones for
$\sigma(ggF\to H\to H_{125}+H_S\to b\bar{b}b\bar{b})$.}
\label{fig:sensbbbb_300}
\end{figure}

\begin{figure}[!h]
\centering
\includegraphics[scale=0.50]{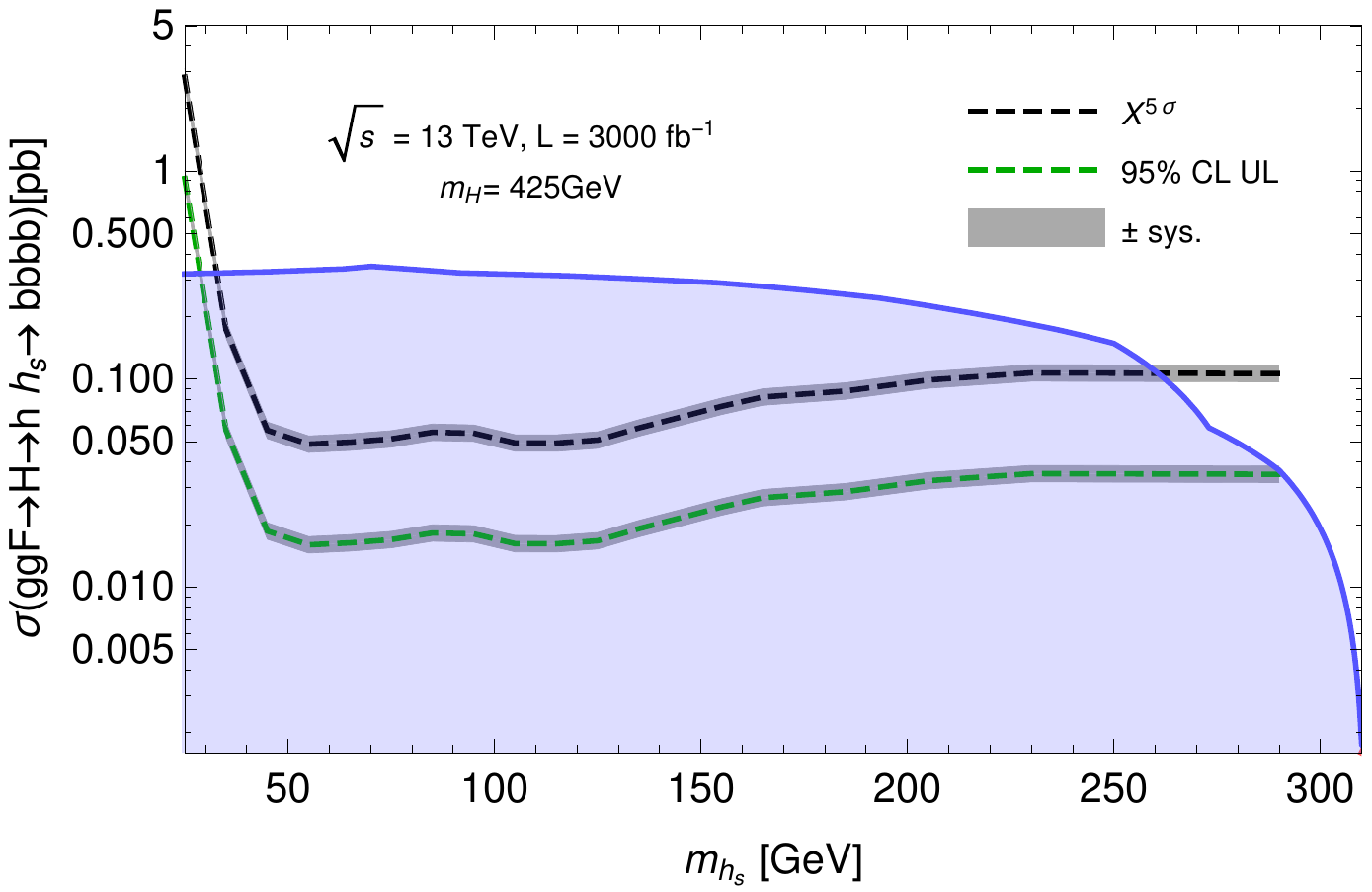}\hfill
\includegraphics[scale=0.50]{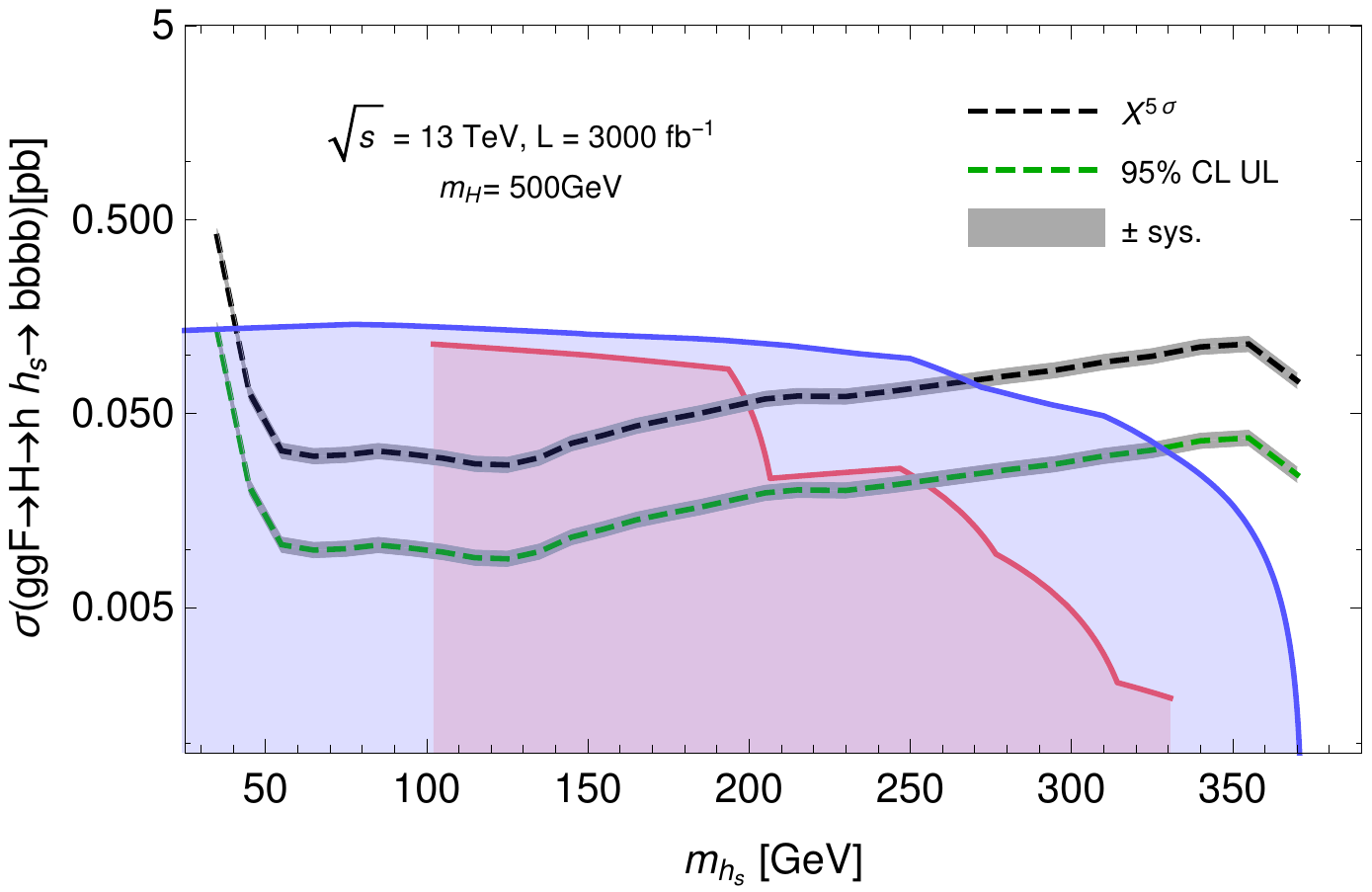}\\
\includegraphics[scale=0.50]{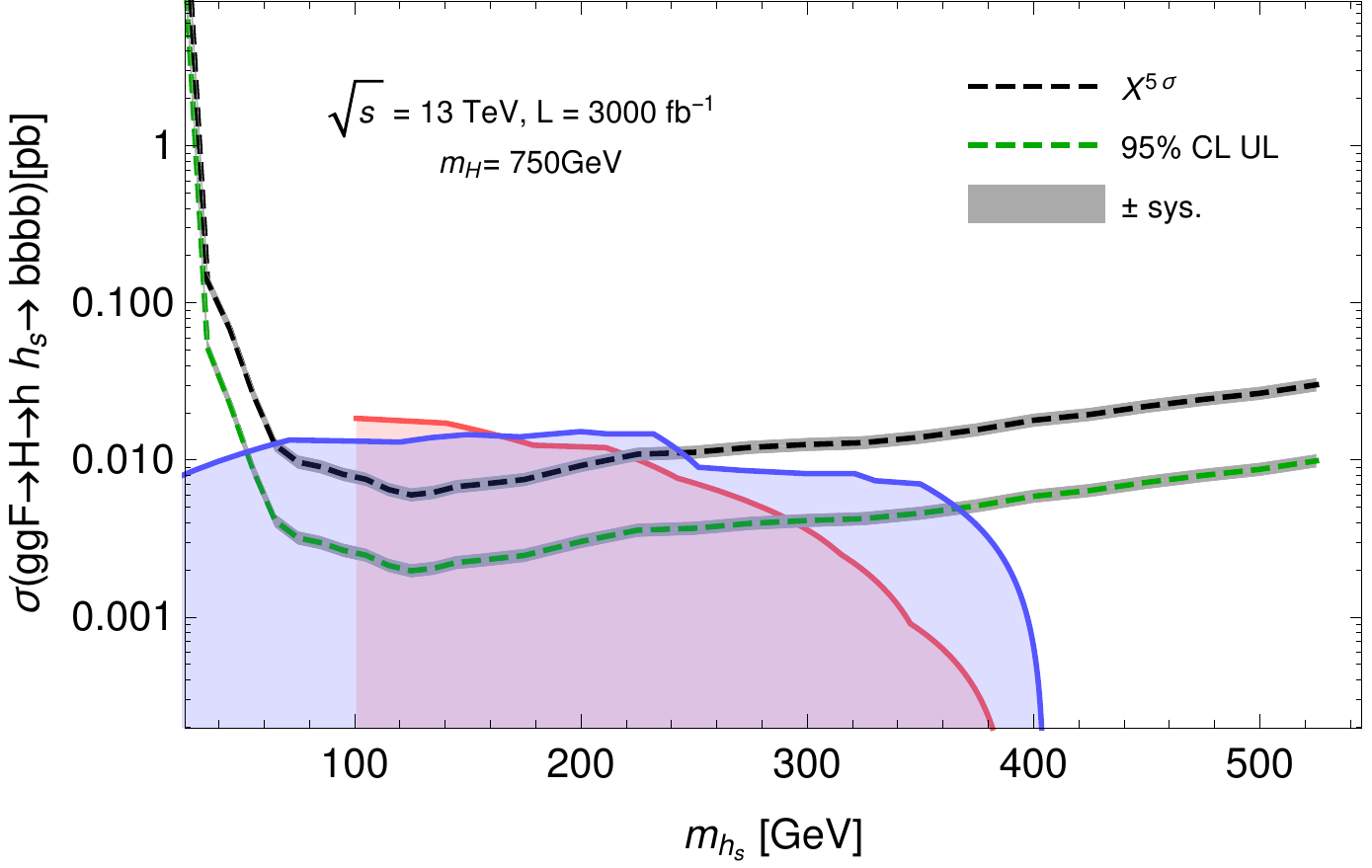}\hfill
\includegraphics[scale=0.50]{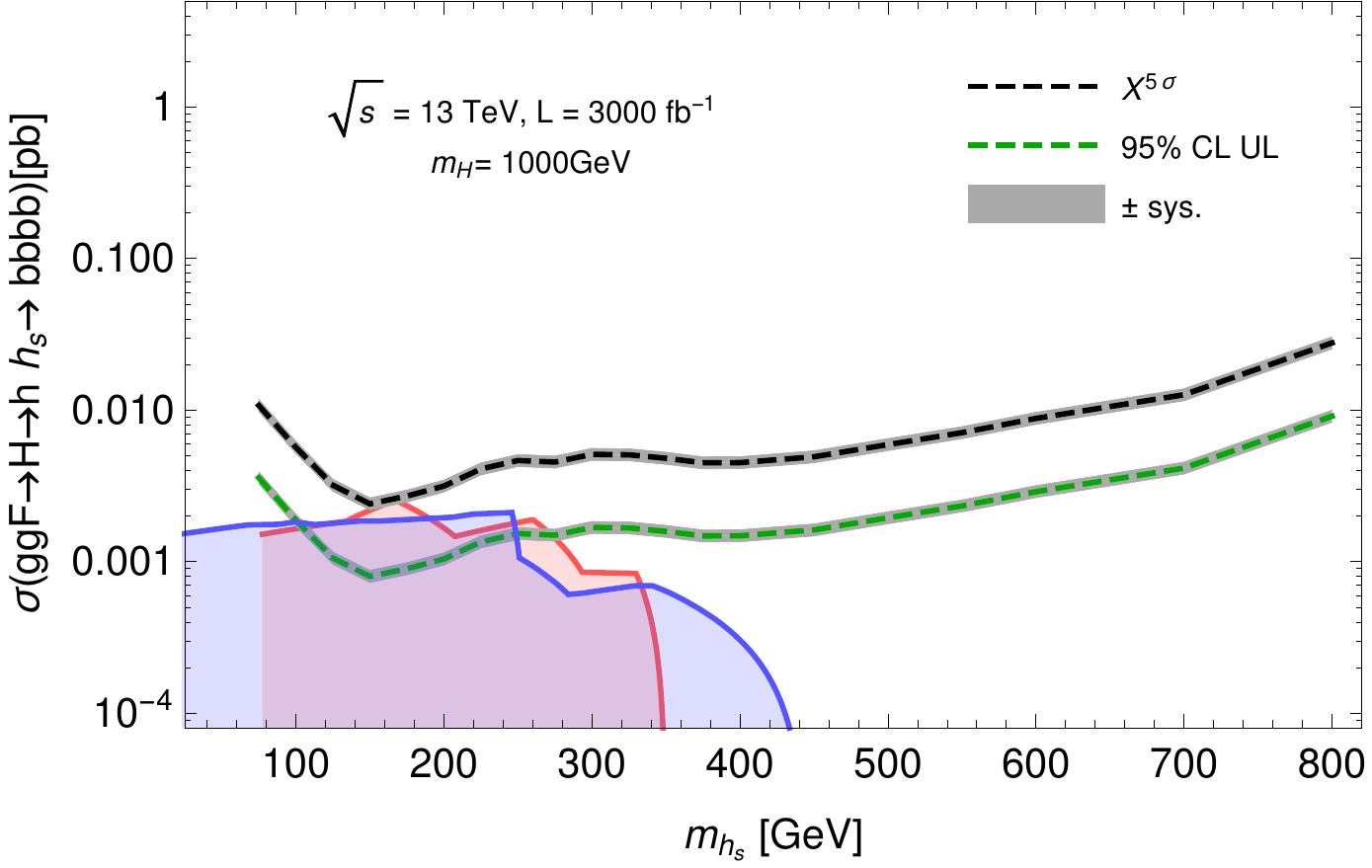}
\caption{95\% CL exclusion limits and $5\,\sigma$ discovery cross sections
in the $b\bar{b}b\bar{b}$ final state
as function of $M_{H_S}$ for 3000~fb$^{-1}$ integrated luminosity and
$M_H=425$~GeV (upper left), $M_H=500$~GeV (upper right), $M_H=750$~GeV (lower left),
$M_H=1000$~GeV (lower right). The colored regions are explained in the
caption of Fig.~\ref{fig:sensbbbb_300}.}
\label{fig:sensbbbb_3000}
\end{figure}

Since the background was fitted to data at 13~TeV c.m. energy we will show our
results also for 13~TeV, for 300 and 3000~fb$^{-1}$ integrated luminosity.
We choose four representative values for $M_H=425,\,500,\,750\ \text{and}\ 1000$~GeV,
and show the 95\%~CL exclusion limits and $5\,\sigma$ discovery cross sections
as function of $M_{H_S}$ in each case. For 300~fb$^{-1}$ integrated luminosity
these are shown in Figs.~\ref{fig:sensbbbb_300}, for 3000~fb$^{-1}$ integrated
luminosity in Figs.~\ref{fig:sensbbbb_3000}.

The expected limits become weaker for $M_{H_S}\lsim 50$~GeV (for $M_X=425-500$~GeV)
and $M_{H_S}\lsim 100$~GeV (for $M_H=1000$~GeV). As stated in subsection~\ref{sec:bbbbsignals}
here the $b\bar{b}$ pair from $H_S$ becomes too boosted and is no longer resolved
by the standard jet clustering algorithm.

The shaded blue regions in Figs.~\ref{fig:sensbbbb_300} and \ref{fig:sensbbbb_3000}
indicate viable values for the cross sections times branching fractions
for $\sigma(ggF\to H\to H_{125}+H_S\to b\bar{b}b\bar{b})$ in
the parameter space of the NMSSM, see section~\ref{sec:thehiggssector}.
Typically the viable values for $\sigma(ggF\to A\to H_{125}+A_S\to b\bar{b}b\bar{b})$
are smaller; if not we show them as shaded red regions.
In the region of the NMSSM parameter space corresponding to $M_H \gsim 500$~GeV,
the partial width for $H_S\to H_{125}+H_{125}$ becomes relatively large ($\approx 10$~MeV)
if kinematically allowed. As a consequence the branching fractions of $H_S$ into
$b\bar{b}$ (and the other channels considered in this paper) decrease, leading
to a decrease of the possible production cross sections times branching fractions
for $M_H \gsim 500$~GeV, $M_{H_S} \gsim 250$~GeV.


\begin{figure}
\centering
\includegraphics[scale=0.7]{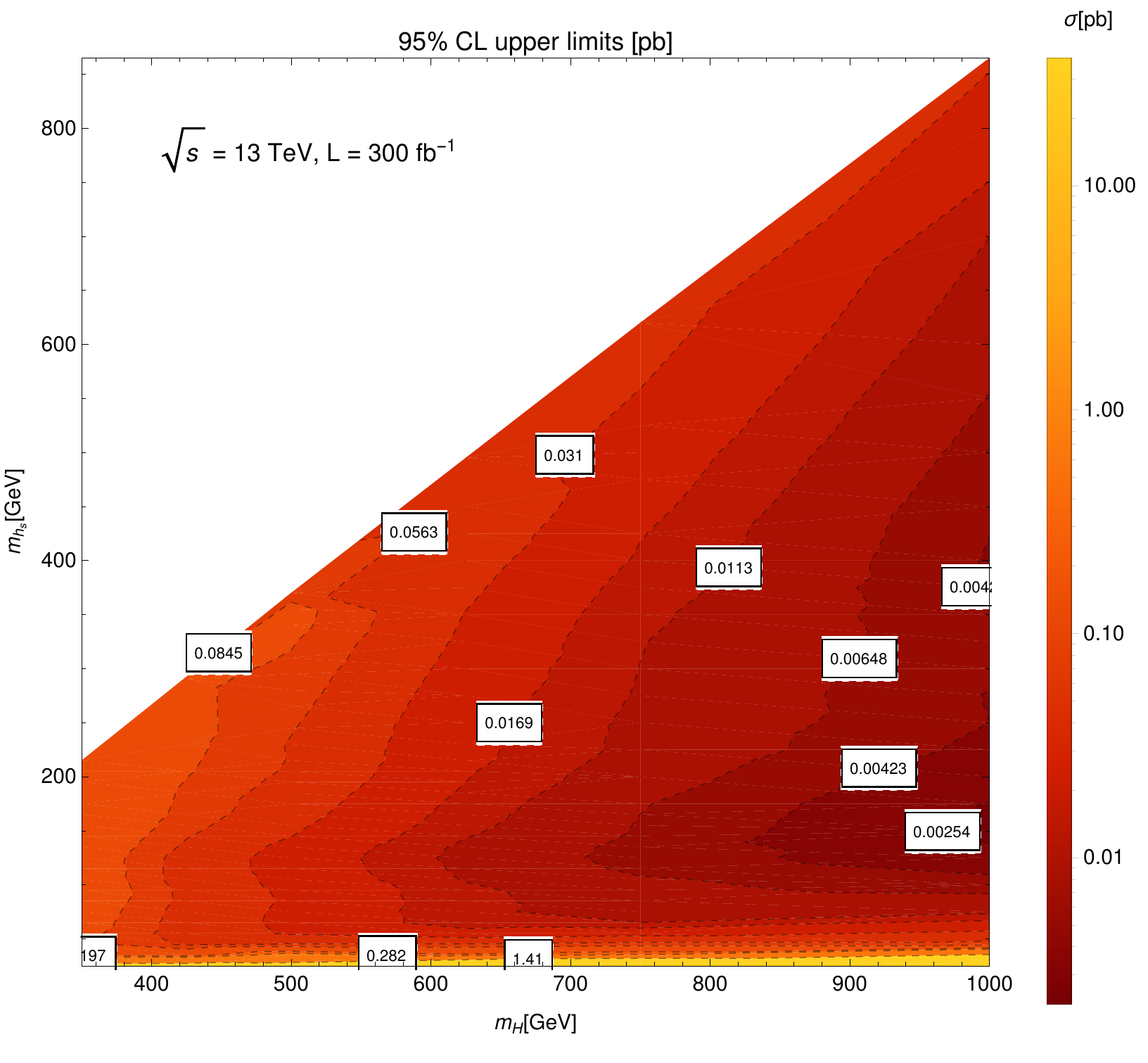}
\includegraphics[scale=0.7]{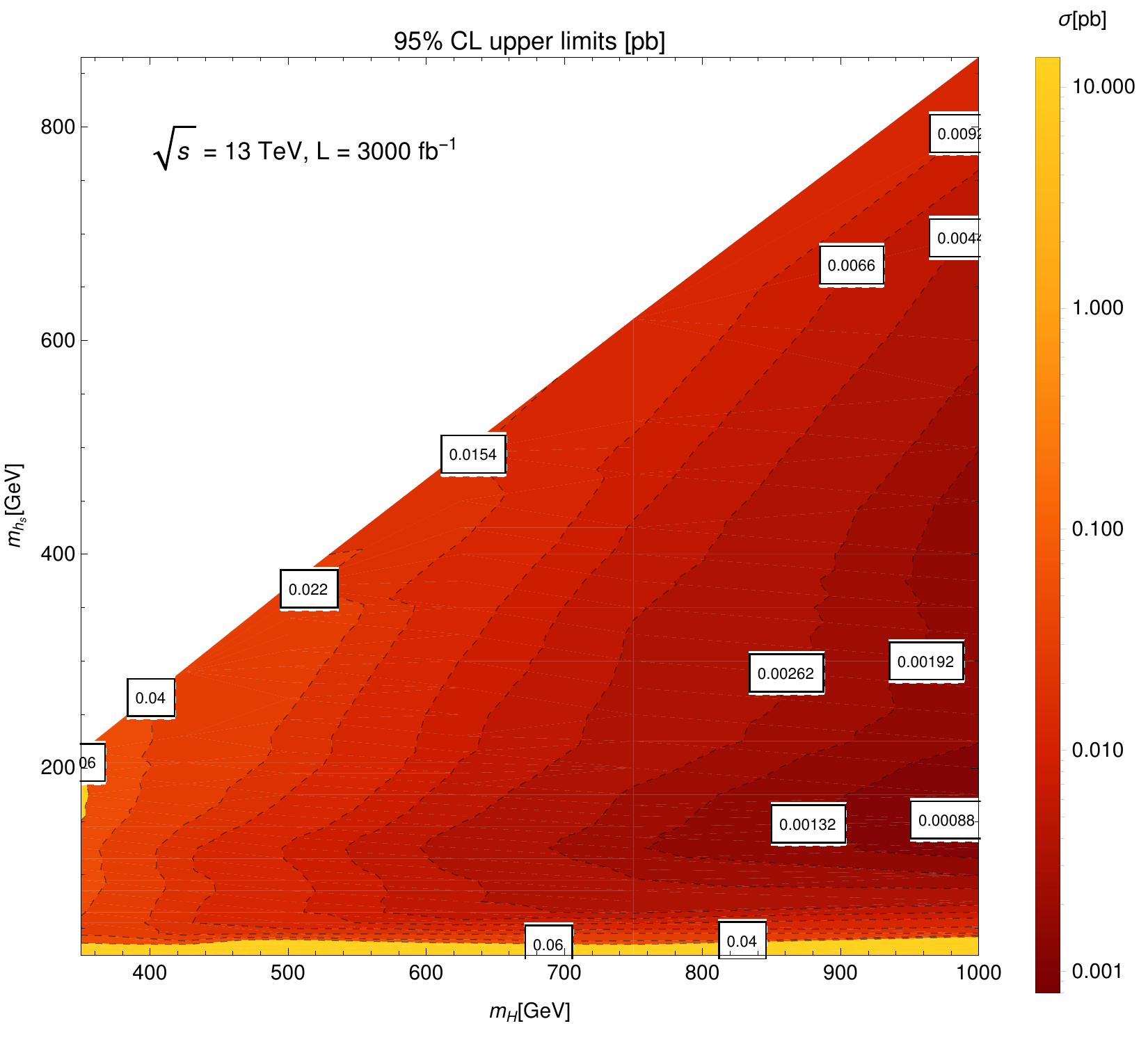}
\caption{95\% CL expected upper limits for the process $H\rightarrow H_{125} H_S
\rightarrow b \bar{b} b \bar{b}$ for L=300 \fb (end of Run III) (up) and L=3000 \fb (HL-LHC).}
\label{fig:ExpectedSensitivityContoursSophMX}
\end{figure}

The following conclusions can be drawn from Figs.~\ref{fig:sensbbbb_300}
and \ref{fig:sensbbbb_3000}:
For $M_{H} \lsim 500$~GeV wide ranges of $M_{H_S}$
in the NMSSM parameter space can be discovered
or, at least, excluded. For larger $M_{H}$ testable regions
in the NMSSM parameter space exist, but for $M_{H}\sim 1$~TeV
only for 3000~fb$^{-1}$ integrated luminosity.
In fig. \ref{fig:ExpectedSensitivityContoursSophMX} 
we summarize these results showing the 95\% C.L. expected upper limits in the $M_H$ vs $
M_{H_S}$ plane.

We recall, however, that the sensitivities to cross sections in
Figs.~\ref{fig:sensbbbb_300} and \ref{fig:sensbbbb_3000} are model independent
and valid for arbitrary (e.g. non-supersymmetric) extensions of the
Higgs sector.


\section{Search strategies for the $b\bar{b}\tau\tau$ final state}

Searches for resonant $H_{125}$ pair production in the $b\bar{b}\tau\tau$
final state have been performed by ATLAS at 8~TeV \cite{Aad:2015xja}, and
by CMS at 13~TeV in \cite{CMS-PAS-HIG-16-013,CMS-PAS-HIG-16-029,Sirunyan:2017djm}.
Following these searches we concentrate on the $\tau_h\tau_h$, $\tau_h\tau_e$
and $\tau_h\tau_\mu$ modes. As in the case of the $b\bar{b}b\bar{b}$ final state
we optimise the cuts as function of a tentative value for $M_{H_S}$.

A priori the $\tau\tau$ pair can originate from $H_S$ or $H_{125}$;
both cases will be studied below. For the analysis we will make no
assumptions on the relative branching ratios $BR(H_S\to b\bar{b})$ and
$BR(H_S\to \tau\tau)$. The aim is to obtain separate 95\%~CL
exclusion limits and $5\,\sigma$ discovery cross sections for the processes
$ggF\to H\to H_S(\to b\bar{b})+H_{125}(\to \tau\tau)$, and
$ggF\to H\to H_S(\to \tau\tau)+H_{125}(\to b\bar{b})$.

\subsection{Analyses of Signal Samples}

For the simulation of signal samples the same series of codes as for the
$b\bar{b}b\bar{b}$ final state was used, see section~\ref{sec:sigsim}.
Events are required to have exactly two $b$-tagged jets with $p_T(b) > 30$~GeV
and $|\eta|<2.5$. For the $b$-tagging efficiency a working point with
$\varepsilon_b = 70\%$ is chosen.
If the event has exactly two hadronic taus $\tau_h$, both are required to have
$p_T(\tau_h) > 45$~GeV. Events with one hadronic tau are required to have
exactly one additional isolated lepton $\ell = e,\mu$ of opposite charge and with 
$p_T(\ell) > 20$~GeV. For the transverse mass $m_T^\ell$ of leptons we
require
\beq
\label{eq:mt}
m_T^\ell \equiv \sqrt{2p_T(\ell) E_T^{miss}\left(1-\cos\left(\phi(E_T^{miss})
-\phi(\ell)\right)\right)} < 40\ \text{GeV}\; .
\eeq
All objects are required to have $|\eta|<2.47$.
In the case of $\tau_h \tau_\ell$ final states the invariant mass $M_{\tau\tau}$
was reconstructed using the collinear mass, i.e. assuming that the neutrino
from the $\tau_\ell$ decay is emitted collinear to the lepton and responsible
for all $E_T^{miss}$. 

Considering first the case where the $\tau\tau$ pair originates from $H_S$
with an assumed mass $M_{H_S}$ (and hence that $M_{b\bar{b}}$ should be close to
125~GeV), a $M_{H_S}$ dependent signal region is defined
by
\beq
\label{eq:chitau}
\chi = \sqrt{\left(
\frac{M_{b\bar{b}}-110\ \text{GeV}}{0.35\cdot M_{b\bar{b}}}\right)^2 +
\left(\frac{M_{\tau\tau}-0.92\cdot M_{H_S}}
{\text{Max}(0.35\cdot M_{\tau\tau},\ 30\ \text{GeV})}\right)^2}
< 1\; .
\eeq

\begin{figure}[!t]
\centering
\includegraphics[scale=0.55]{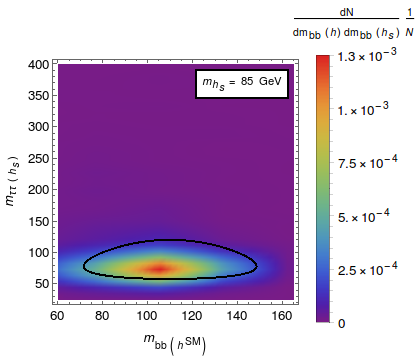}\hfill
\includegraphics[scale=0.55]{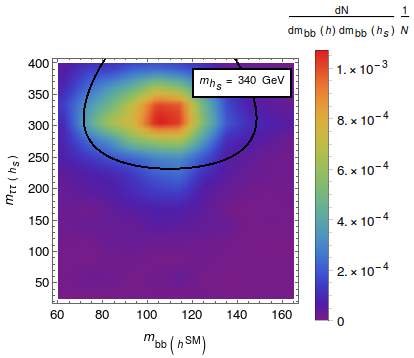}
\caption{Dijet mass distributions $M_{b\bar{b}}(H_{125})$ and $M_{\tau\tau}(H_S)$
for two different benchmark points $M_{H_S}=85$~GeV and $M_{H_S}=340$~GeV. The black
contours indicate the signal regions defined by $\chi < 1$.}
\label{fig:bbttmasses}
\end{figure}

If different pairings within a given event satisfy (\ref{eq:chitau}), the combination
that minimizes $\chi$ is chosen.
In Figs.~\ref{fig:bbttmasses} we show, for $M_H=500$~GeV, the distributions of the dijet mass
$M_{b\bar{b}}(H_{125})$ and $M_{\tau\tau}(H_S)$ for the pairing minimizing $\chi$,
for two different benchmark points $M_{H_S}=85$~GeV and $M_{H_S}=340$~GeV in the case where
$M_{H_S}$ for the analysis was chosen correctly. The black
contours indicate the signal regions defined by $\chi < 1$.

Cuts on the transverse momenta of Higgs candidates are chosen as
\bea
\label{eq:pttau}
p_T({b\bar{b}}) &>& 52\ \text{GeV} + 0.14\, M_{X}-0.2\, M_{H_S} - \frac{M_{H_S}}{M_{X}}\cdot
202\ \text{GeV}\nn \\
p_T({\tau\tau}) &>& 24\ \text{GeV} + 0.19\, M_{X}-0.02\, M_{H_S} - \frac{M_{H_S}}{M_{X}}\cdot
128\ \text{GeV}
\eea
with
\beq
\label{eq:mxtau}
M_X = M_{b\bar{b}\tau\tau} + 125\ \text{GeV}\ - M_{b\bar{b}}\; .
\eeq
The numerical coefficients in eqs.~(\ref{eq:chitau})--(\ref{eq:mxtau}) were
obtained by optimizing the relative signal to background efficiency.

In the case where the $b\bar{b}$ pair originates from $H_S$
with an assumed mass $M_{H_S}$ (and hence that $M_{\tau\tau}$ should be close to
125~GeV), the signal region~(\ref{eq:chitau}) is replaced by
\beq
\chi = \sqrt{\left(
\frac{M_{\tau\tau}-120\ \text{GeV}}{0.35\cdot 120\ \text{GeV}}\right)^2 +
\left(\frac{M_{b\bar{b}}-0.85\cdot M_{H_S}}
{\text{Max}(0.35\cdot M_{H_S},\ 35\ \text{GeV})}\right)^2}
< 1
\eeq
and the cuts on the transverse momenta of Higgs candidates are
\bea
p_T({\tau\tau}) &>& 118\ \text{GeV} + 0.02\, M_{X}-0.55\, M_{H_S} - \frac{M_{H_S}}{M_{X}}\cdot
380\ \text{GeV}
\nn \\
p_T({b\bar{b}}) &>& 16\ \text{GeV} + 0.19\, M_{X}-0.02\, M_{H_S} - \frac{M_{H_S}}{M_{X}}\cdot
137\ \text{GeV}
\eea
with
\beq
M_X = M_{b\bar{b}\tau\tau} + M_{H_S} - M_{b\bar{b}}\; .
\eeq
Hence, for each tentative value
of $M_{H_S}$ two different analyses using different cuts are to be performed, resulting
in two (slightly) different distributions of $M_X$.

\subsection{Background Estimation}

Backgrounds originate from $t\bar{t}$ (and single top) and QCD+electroweak
$b\bar{b}\tau\tau$ production. Contributions from one or more jets misidentified as
$\tau_h$ are seen to become small after the cuts on $p_T({\tau\tau})$.
We have generated $1.5\cdot 10^7$ $t\bar{t}$ events using  \texttt{MadGraph5\_aMC@NLO}
\cite{Alwall:2014hca}; the LO cross section was rescaled by a (NNLO+NNLL) K-factor 1.7
obtained from \texttt{top++2.0} 
\cite{{Cacciari:2011hy,Czakon:2011xx,Baernreuther:2012ws,Czakon:2012zr,Czakon:2012pz,Czakon:2013goa}}.
\texttt{MadGraph5\_aMC@NLO} was also used to generate QCD+electroweak $b\bar{b}\tau\tau$
events; the LO cross section was rescaled by a NLO K-factor 2.9.

After applying the cuts of the previous subsection, the relative contributions of
the SM backgrounds depend on $M_{H_S}$ and on whether the $\tau\tau$ pair originates from $H_S$ or
$H_{125}$, although $t\bar{t}$ is always dominant:
For $H_S\to \tau\tau$ the $t\bar{t}$ contribution increases from $\sim 60\%$
for $M_{H_S}\sim 50$~GeV to $\sim 100\%$ for $M_{H_S}\gsim 350$~GeV, the remaining
background stems from QCD+electroweak $b\bar{b}\tau\tau$ production.
For $H_{125}\to \tau\tau$ the $t\bar{t}$ contribution is always $\sim 90\%$.

In order to validate the background contribution to the $M_X$ distribution after
cuts we use again a search for
$H_{125}$ pair production at 13~TeV, now in the $b\bar{b}\tau\tau$ channel.
Measurements of distributions of the (slightly corrected) total invariant mass
by CMS, separately
in the $b\bar{b}e\tau_h$, $b\bar{b}\mu\tau_h$ and $b\bar{b}\tau_h\tau_h$
channels, can be found in Fig.~1 in \cite{CMS-PAS-HIG-16-029}.

\begin{figure}[!h]
\centering
\includegraphics[scale=0.35]{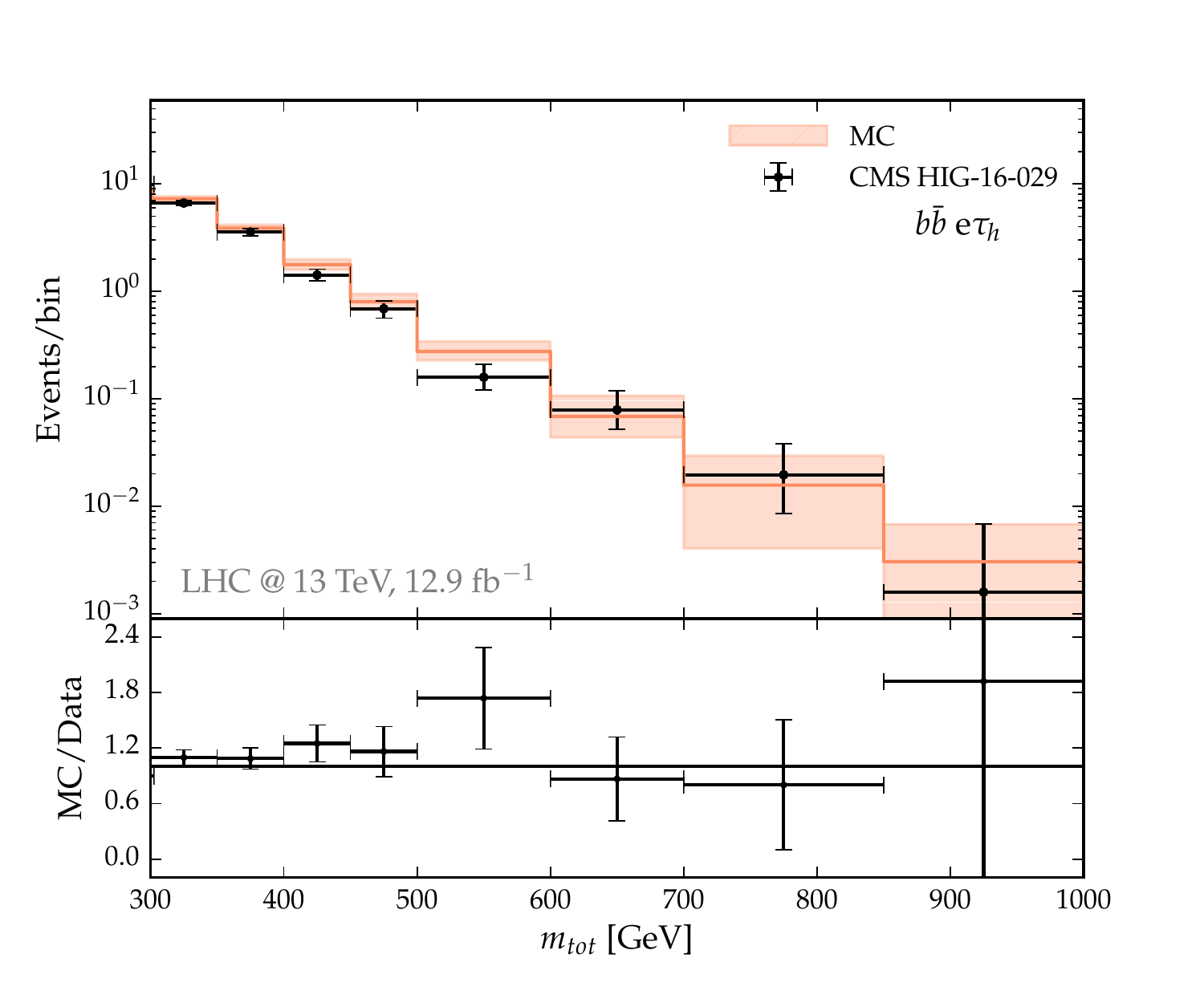}
\includegraphics[scale=0.35]{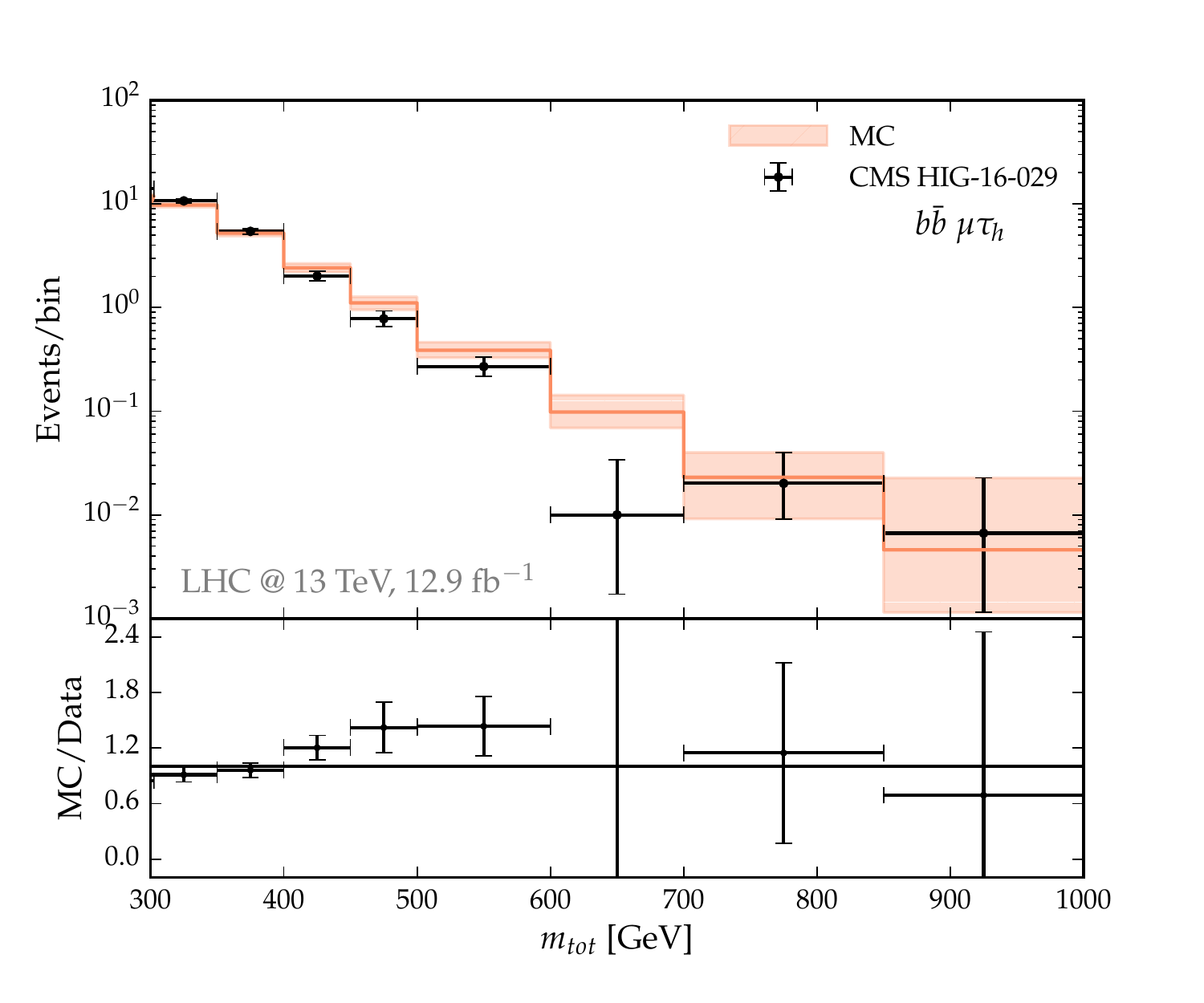}
\includegraphics[scale=0.35]{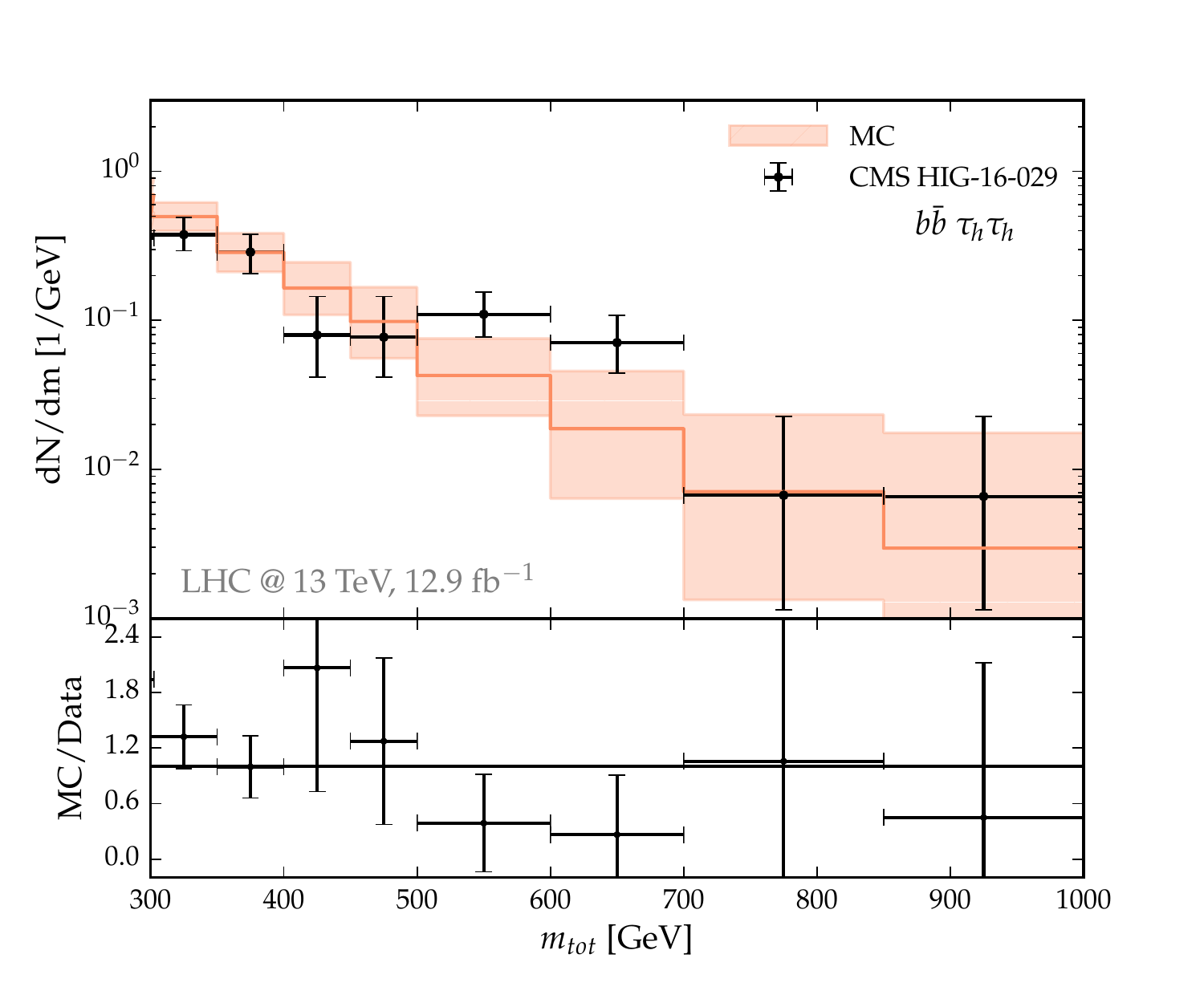}
\caption{Measured total invariant
mass distribution from \cite{CMS-PAS-HIG-16-029} in black, and our MC results
in orange
including the statistical uncertainties corresponding to 12.9~fb$^{-1}$ of
integrated luminosity in the three channels
$b\bar{b}e\tau_h$, $b\bar{b}\mu\tau_h$ and $b\bar{b}\tau_h\tau_h$.}
\label{fig:Mbbtt_vs_CMS}
\end{figure}

We have reproduced the cuts in \cite{CMS-PAS-HIG-16-029} using our background
samples. In Figs.~\ref{fig:Mbbtt_vs_CMS} we show the measured total invariant
mass distribution from Fig.~1 in \cite{CMS-PAS-HIG-16-029} in black, and our MC results
including the statistical uncertainties corresponding to 12.9~fb$^{-1}$ of
integrated luminosity, in the three channels, in orange. Due to the smaller number of events
the statistical uncertainties are now larger than in the $4b$ case.

Still we can ask which rescaling of our simulated background, independent of
the total invariant mass and common to all three channels (to improve the
statistics), provides a best
fit to the data. We find a factor $1.01 \pm 0.24$, and will subsequently use
$\pm 0.24$ as an estimate of the systematic uncertainty of the background normalisation.

For forecasts at 300 or 3000~fb$^{-1}$ integrated luminosity
the shape of the $M_X$ background distributions (\ref{eq:mxtau})
will again be parametrized by continuous functions with $M_{H_S}$ dependent
parameters: For the $t\bar{t}$ background the Frechet distribution, and for
the $b\bar{b}\tau\tau$ background (all channels combined) the GaussExp function
already used in \cite{CMS:2016tlj}. Both functions are defined in the appendix~A.
In Figs.~\ref{fig:backgrdfitbbtt} we show these fits for $M_{H_S}=85$~GeV and
$M_{H_S}=350$~GeV where the $\tau\tau$ pair originates from $H_S$.

\begin{figure}[!t]
\centering
\includegraphics[scale=0.45]{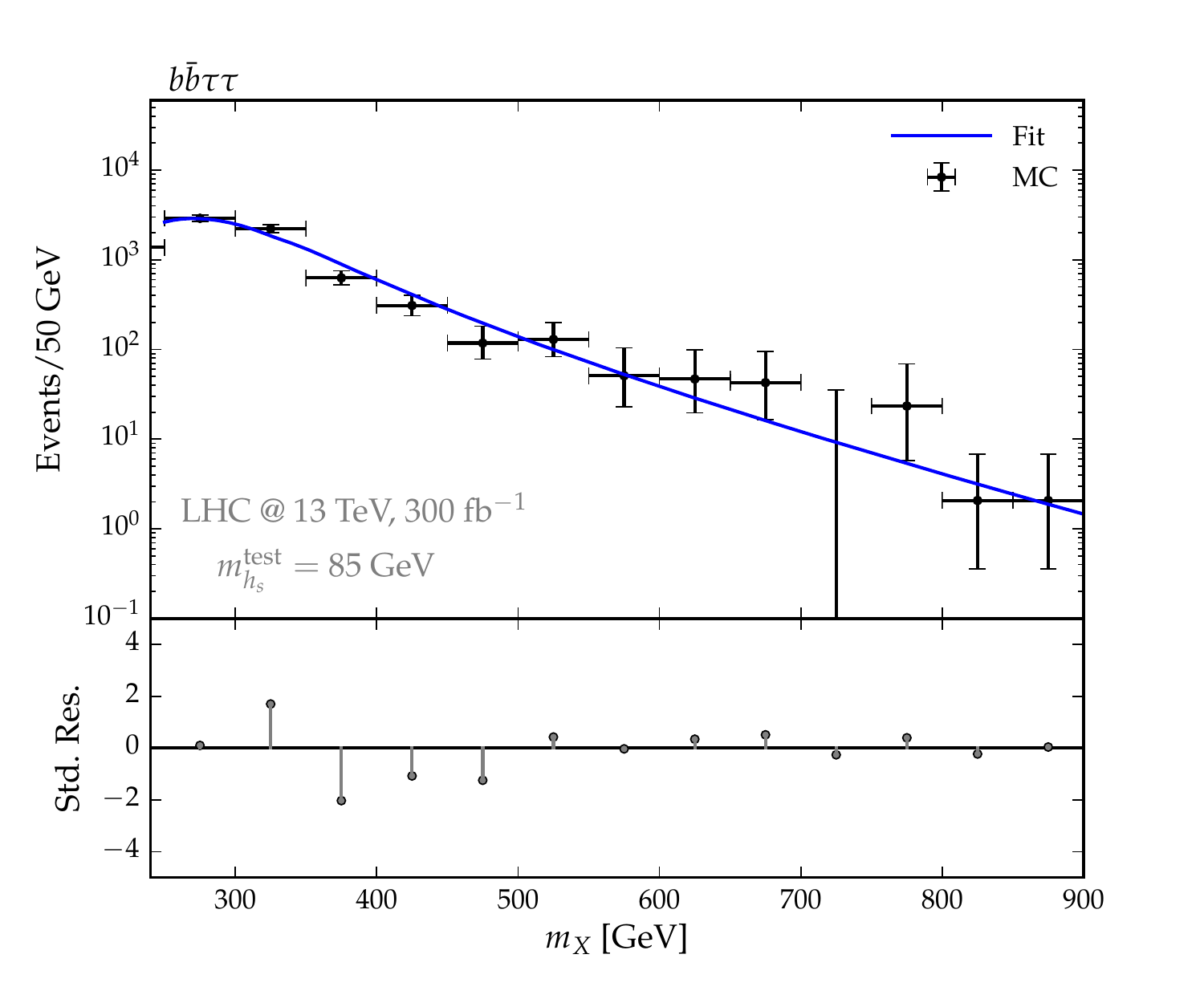}\hfill
\includegraphics[scale=0.45]{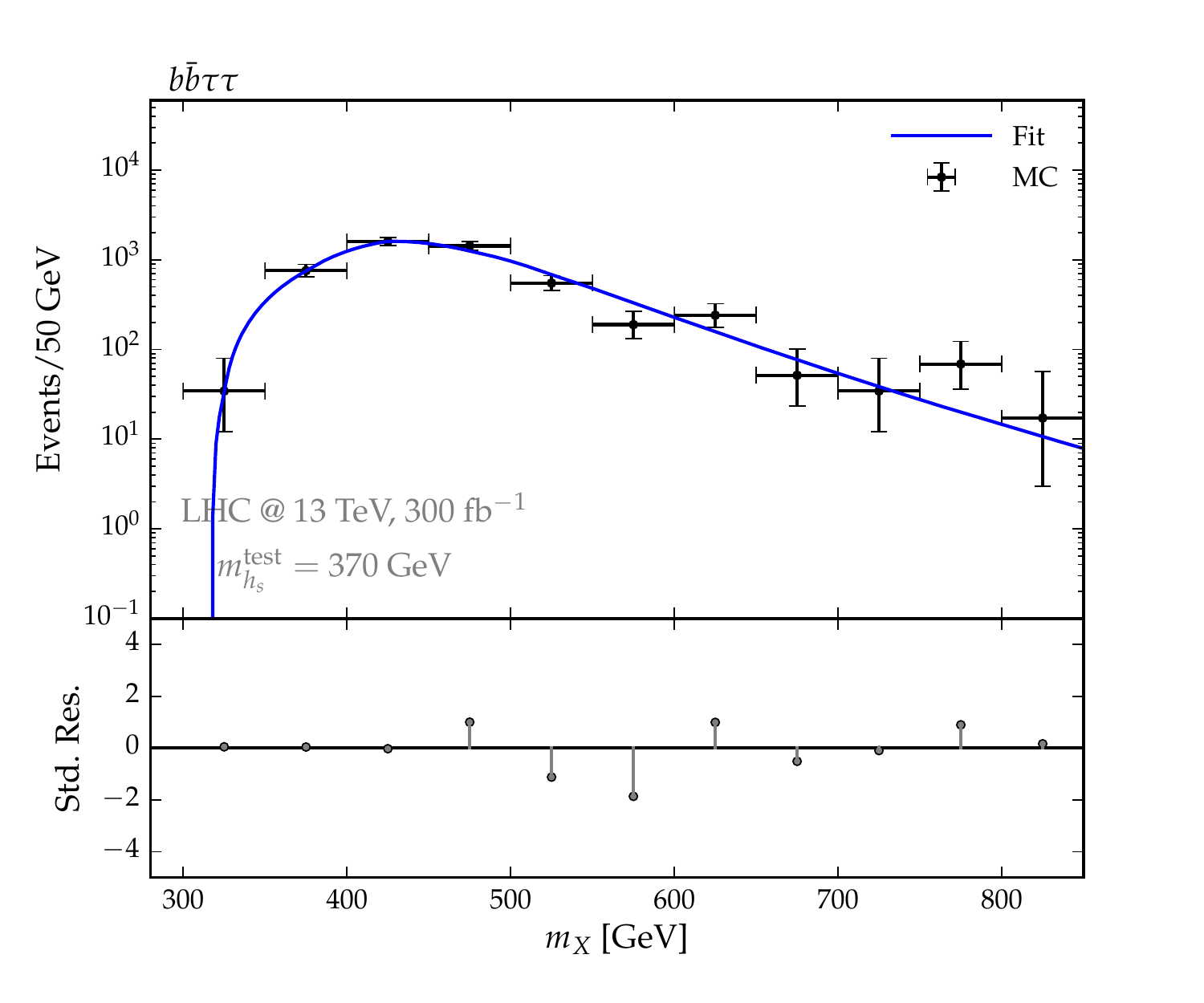}
\caption{Fits to distributions of $M_X$
of the background
obtained by Monte Carlo simulations of $t\bar{t}$ and $b\bar{b}\tau\tau$
events after cuts eq.~\ref{eq:mt}--\ref{eq:pttau} corresponding to
$M_{H_S}=85$~GeV (left) and $M_{H_S}=370$~GeV (right). The
indicated uncertainties originate from the Monte Carlo samples.}
\label{fig:backgrdfitbbtt}
\end{figure}

\subsection{Future 95\% CL Exclusion Limits and $5\,\sigma$ Discovery Cross Sections}

\begin{figure}[!h]
\centering
\includegraphics[scale=0.50]{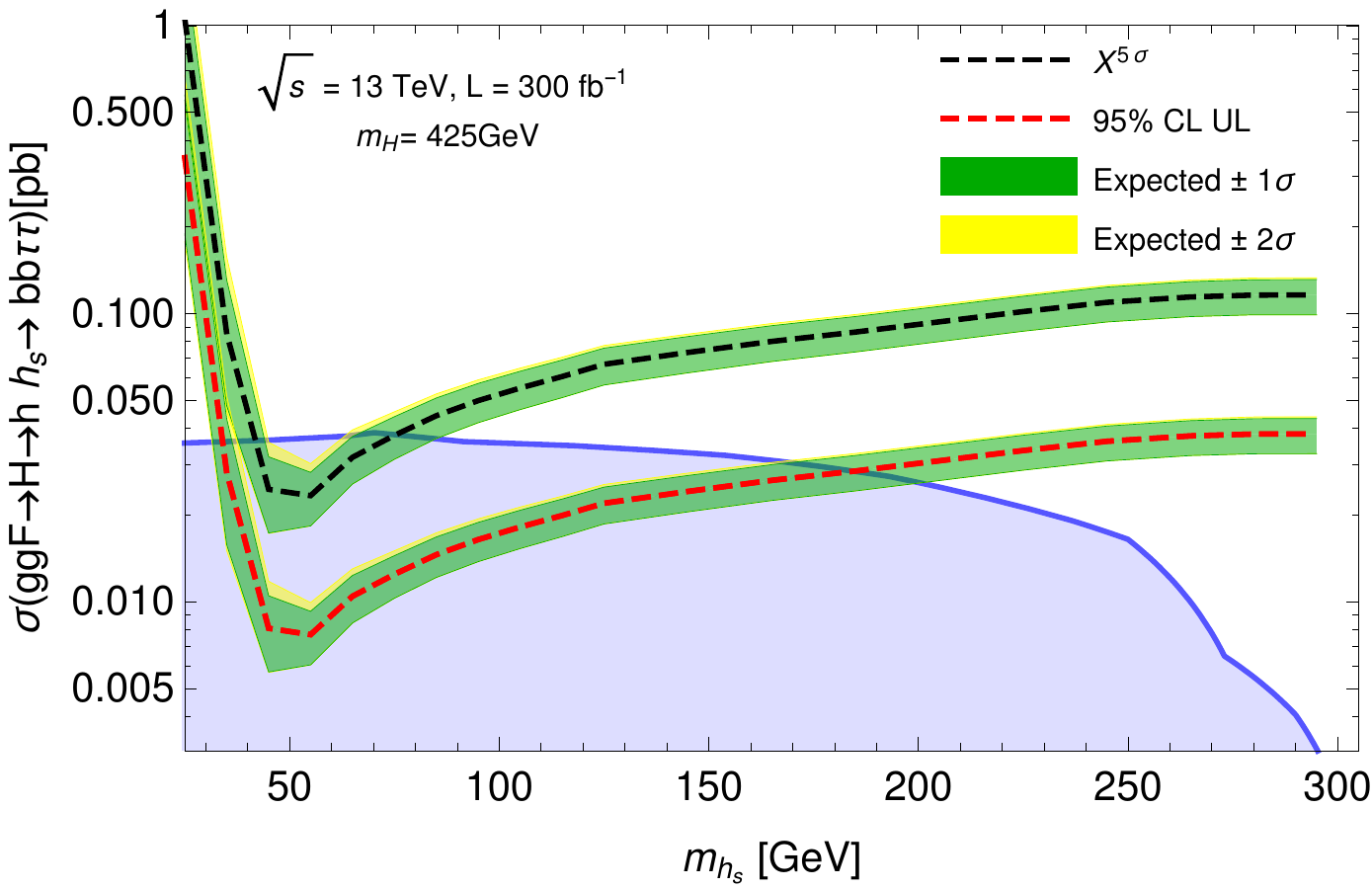}\hfill
\includegraphics[scale=0.50]{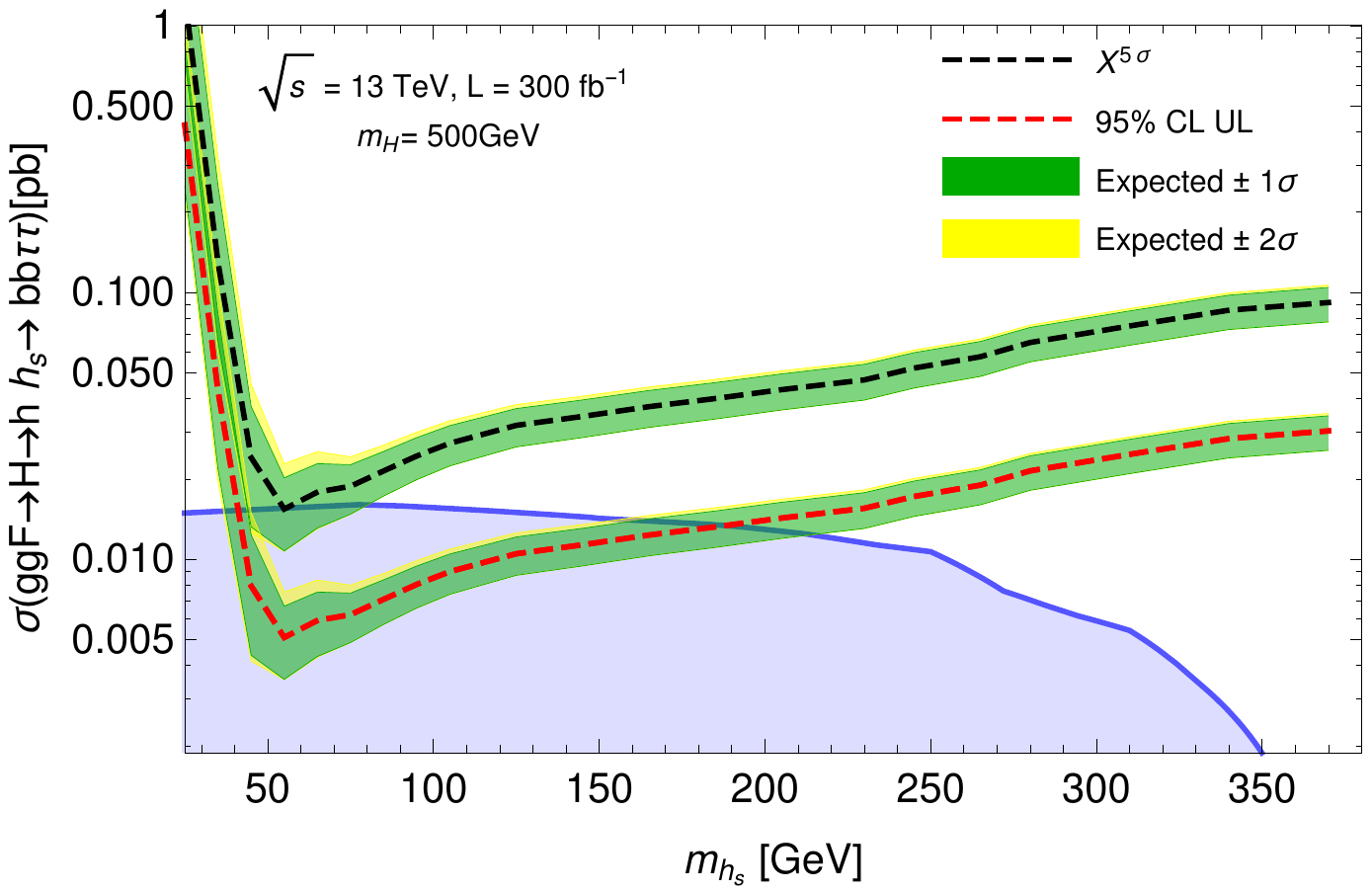}\\
\includegraphics[scale=0.50]{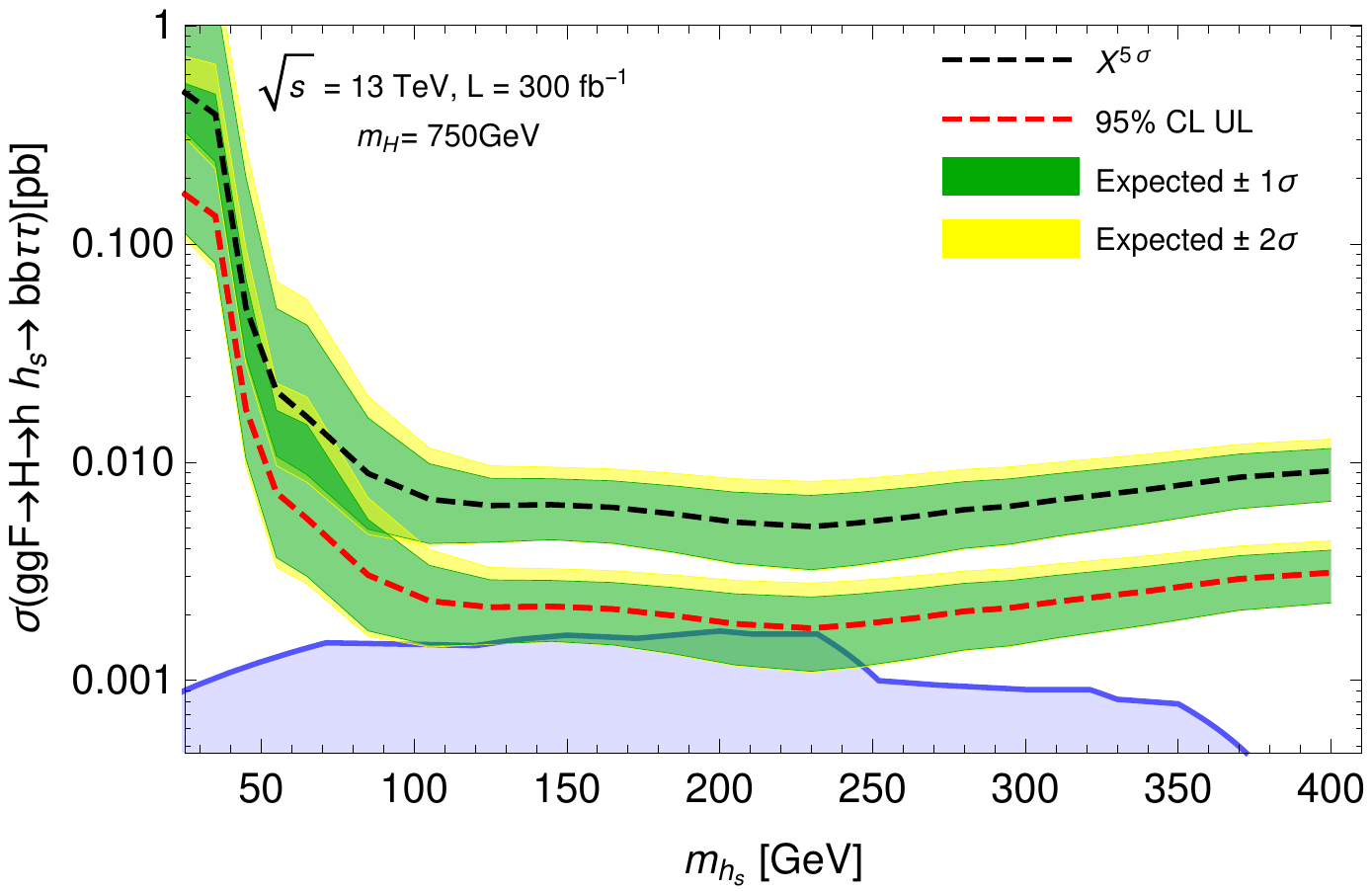}\hfill
\includegraphics[scale=0.50]{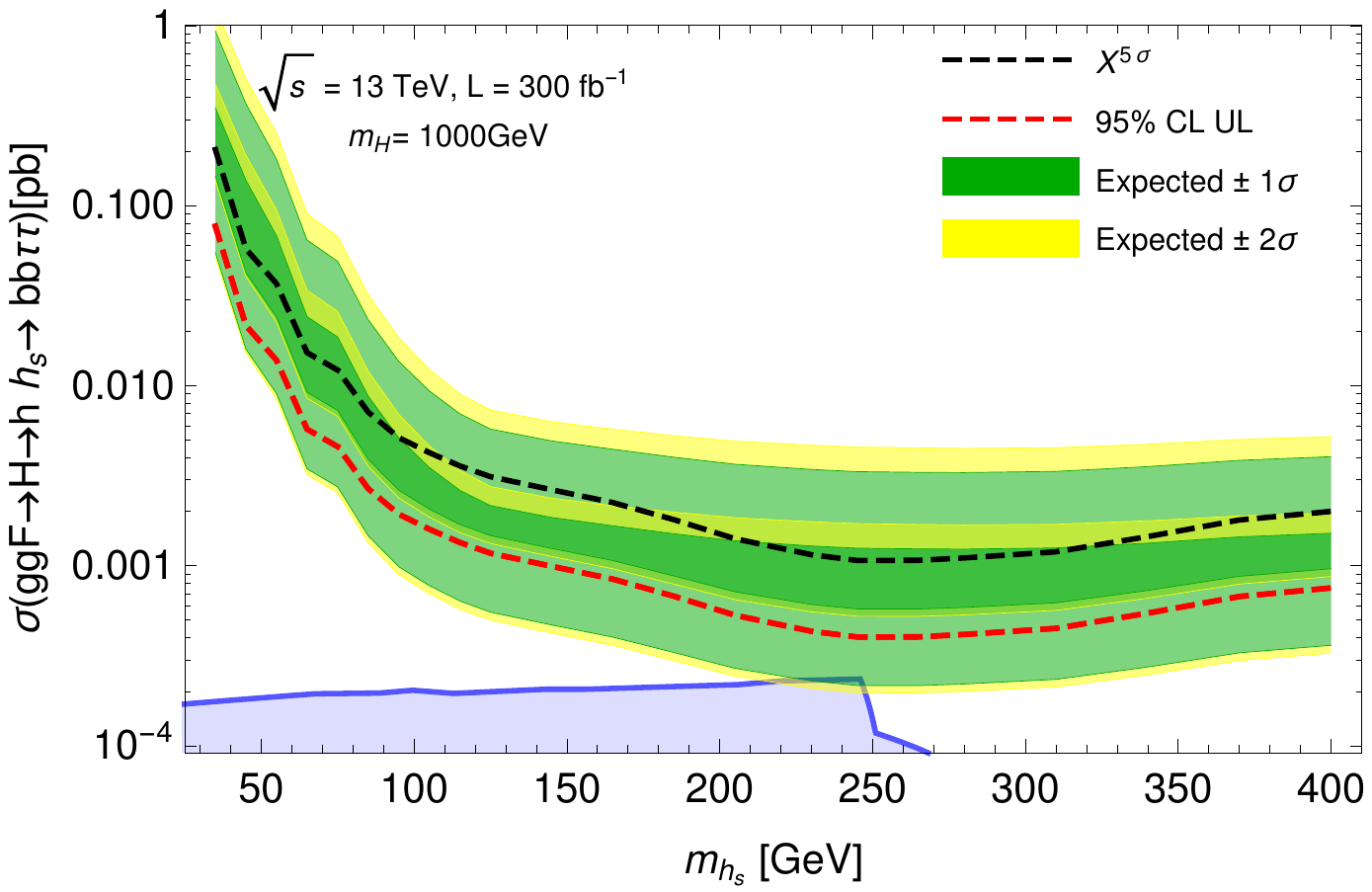}
\caption{95\% CL exclusion limits and $5\,\sigma$ discovery cross sections
for $H_{125}\to b\bar{b}$ and $H_S\to \tau\tau$
as function of $M_{H_S}$ for 300~fb$^{-1}$ integrated luminosity and
$M_H=425$~GeV (upper left), $M_H=500$~GeV (upper right), $M_H=750$~GeV (lower left),
$M_H=1000$~GeV (lower right).}
\label{fig:plotSensULbbtt300}
\end{figure}

\begin{figure}[!h]
\centering
\includegraphics[scale=0.50]{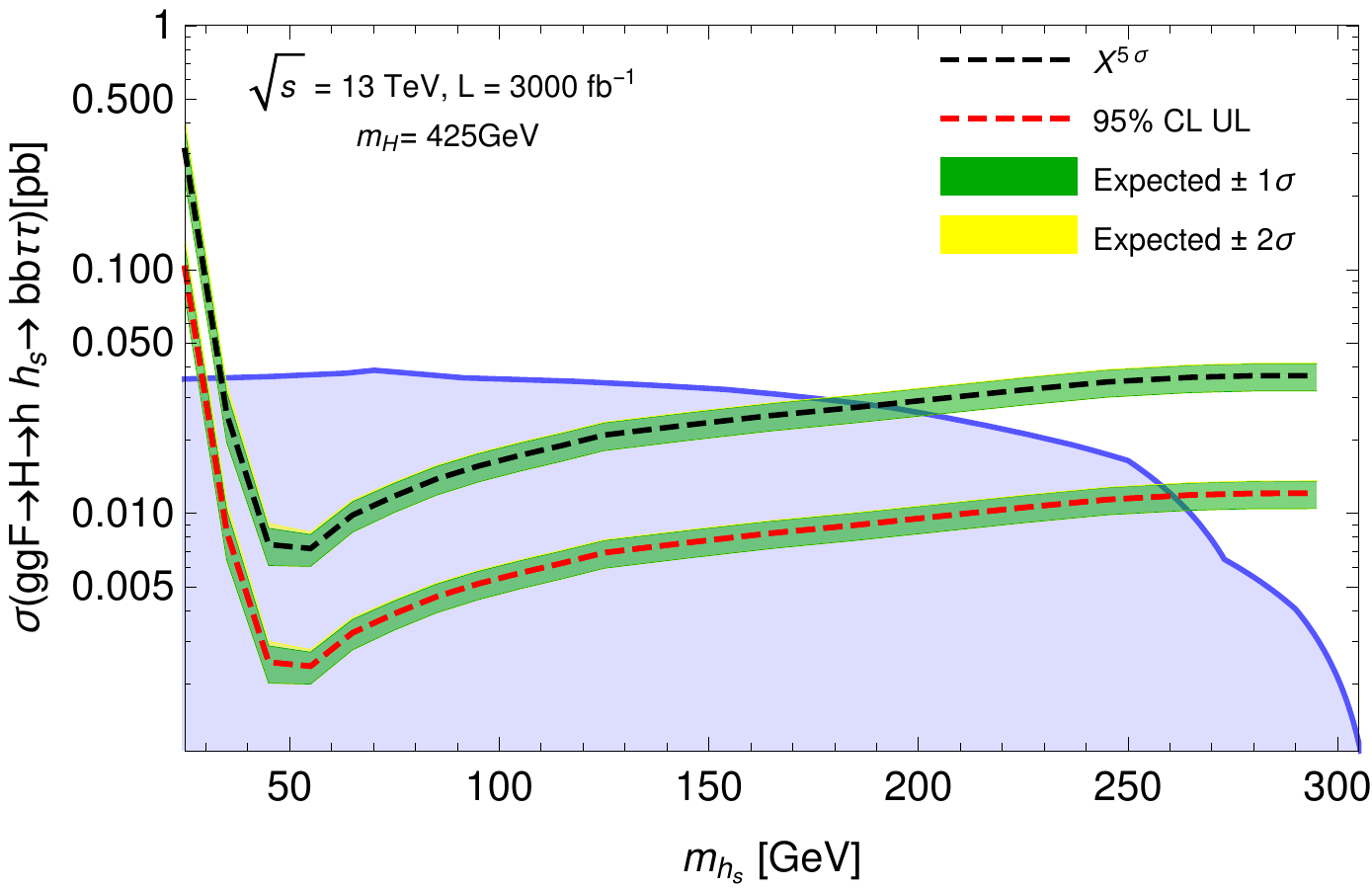}\hfill
\includegraphics[scale=0.50]{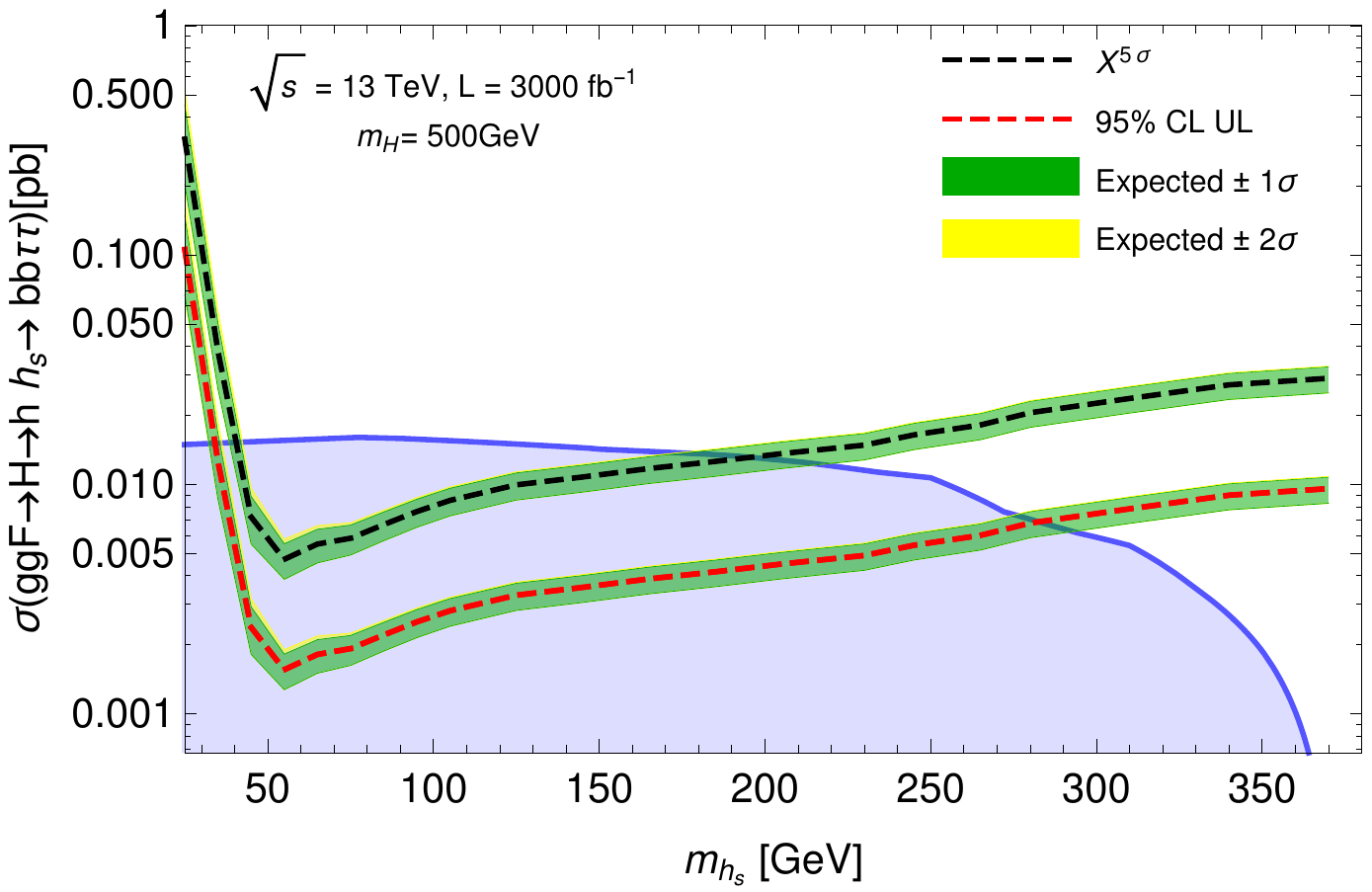}\\
\includegraphics[scale=0.50]{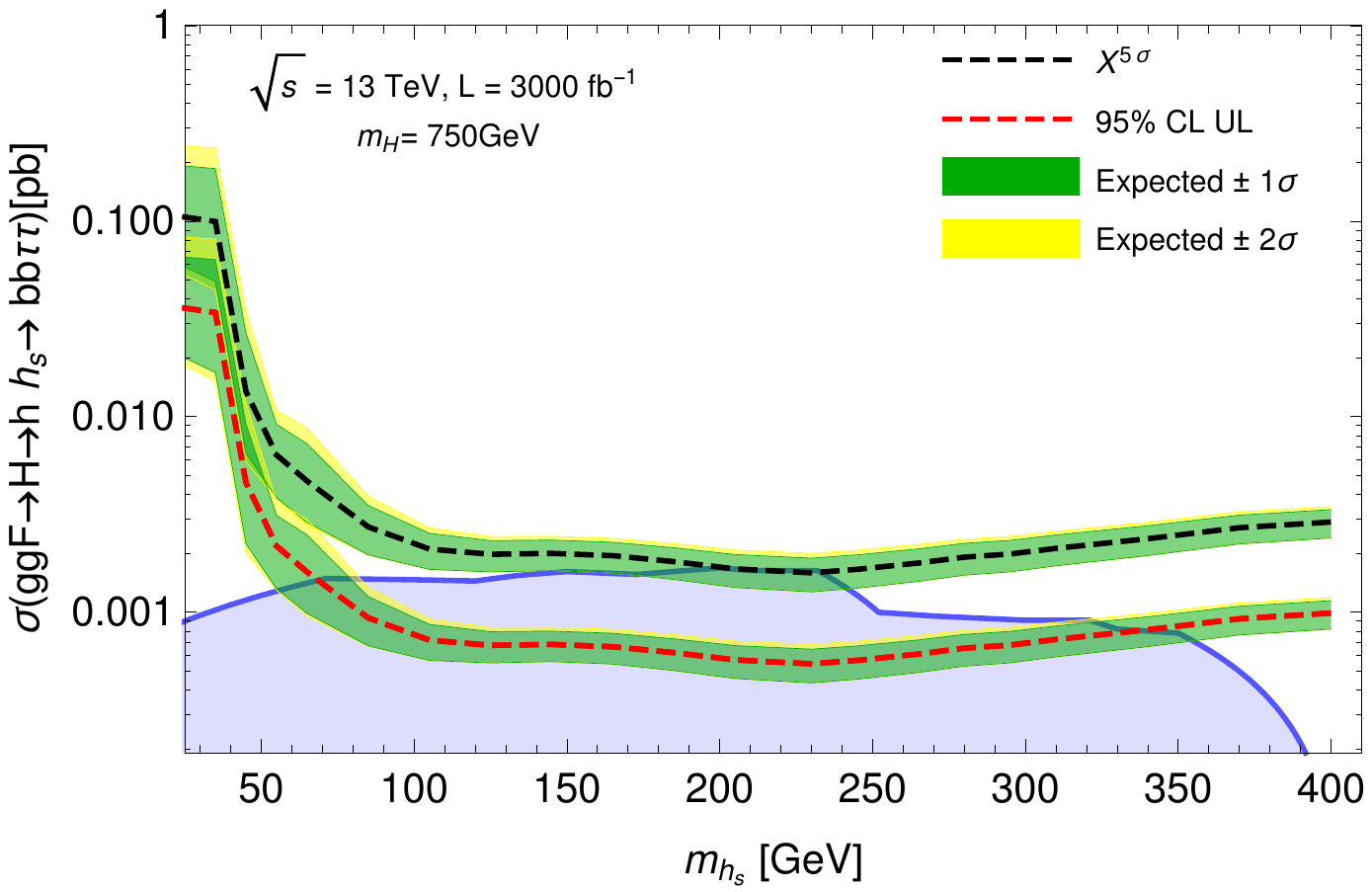}\hfill
\includegraphics[scale=0.50]{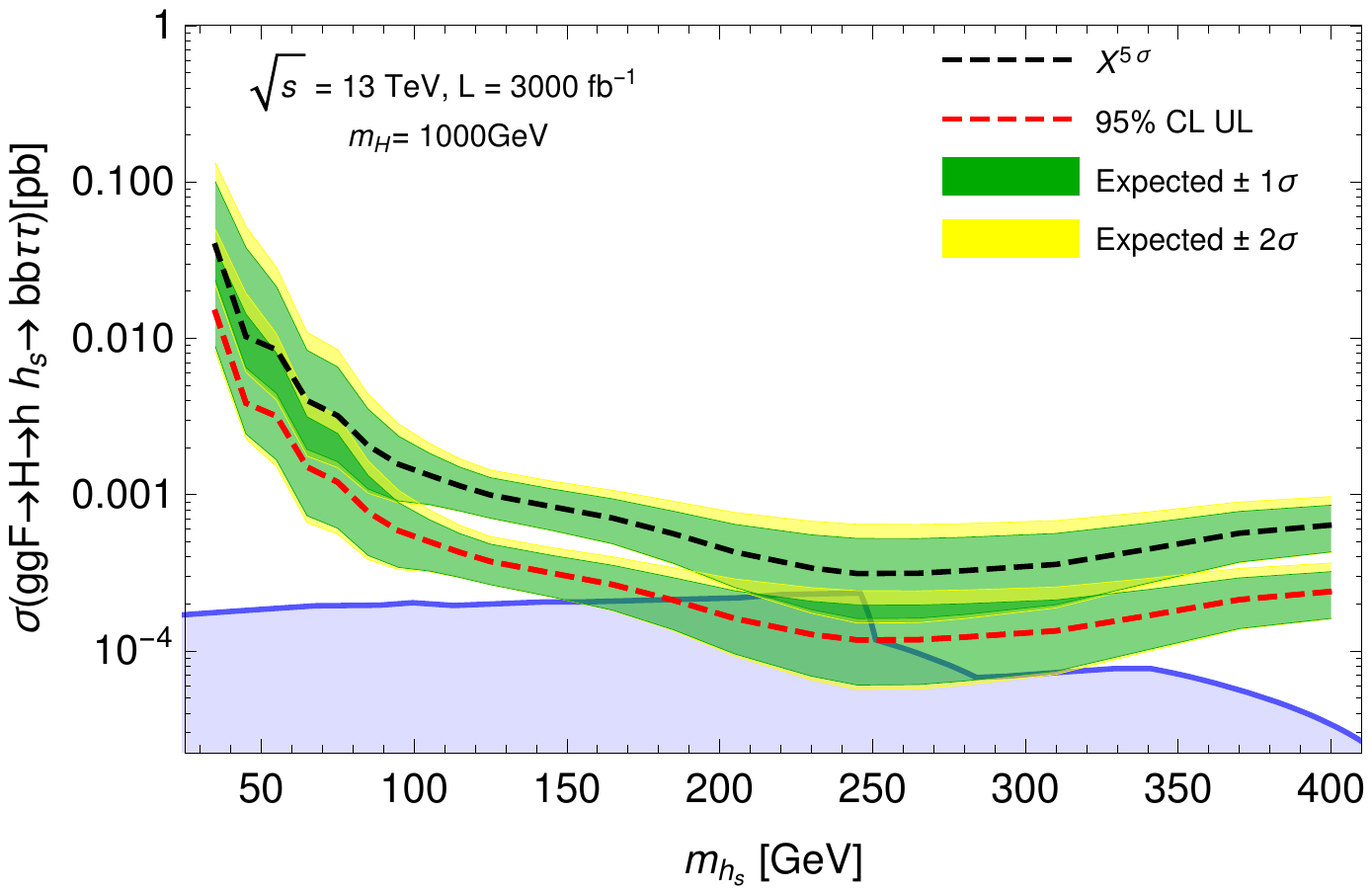}
\caption{95\% CL exclusion limits and $5\,\sigma$ discovery cross sections
for $H_{125}\to b\bar{b}$ and $H_S\to \tau\tau$
as function of $M_{H_S}$ for 3000~fb$^{-1}$ integrated luminosity and
$M_H=425$~GeV (upper left), $M_H=500$~GeV (upper right), $M_H=750$~GeV (lower left),
$M_H=1000$~GeV (lower right).}
\label{fig:plotSensULbbtt3000}
\end{figure}

\begin{figure}[!h]
\centering
\includegraphics[scale=0.50]{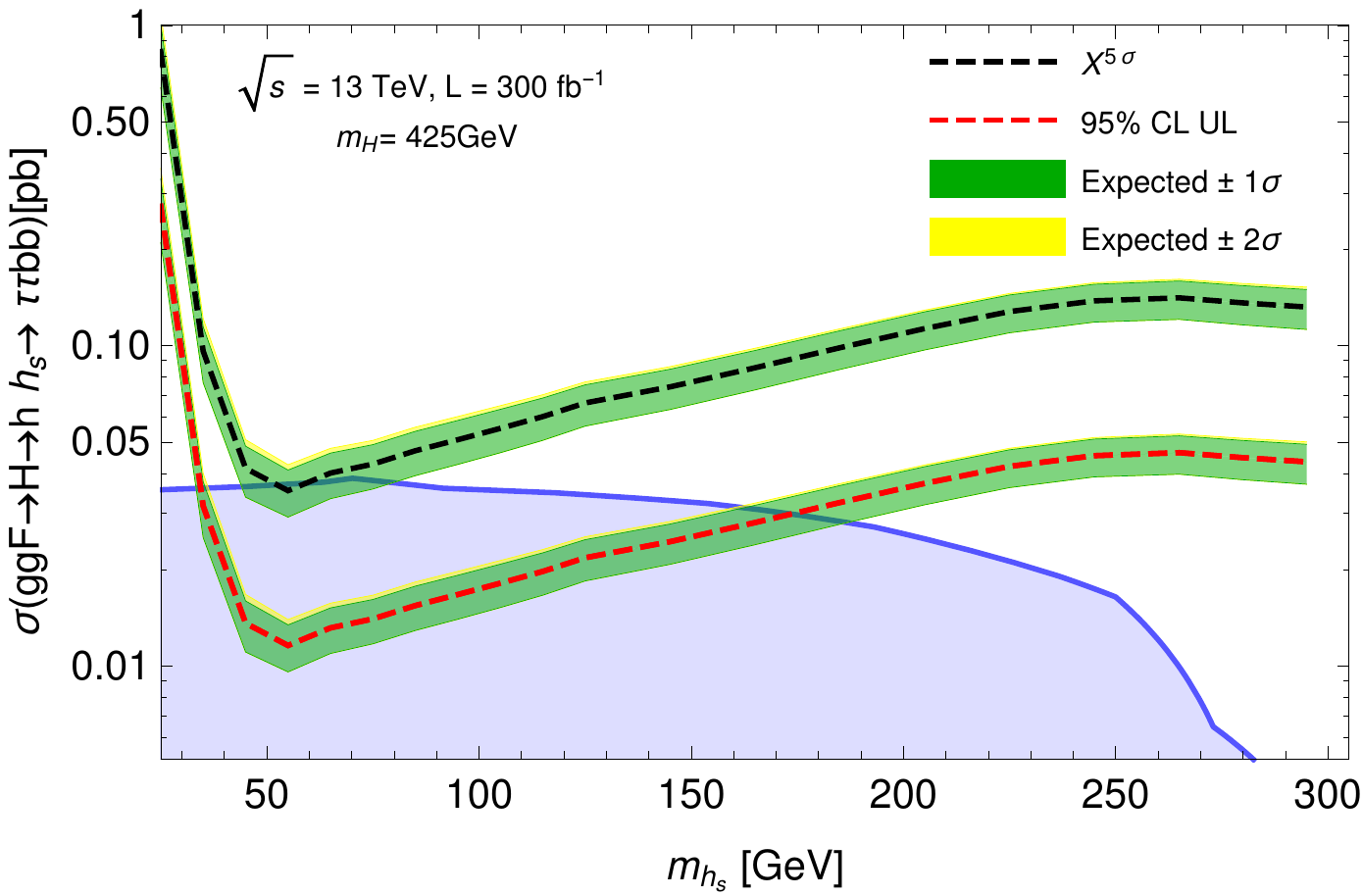}\hfill
\includegraphics[scale=0.50]{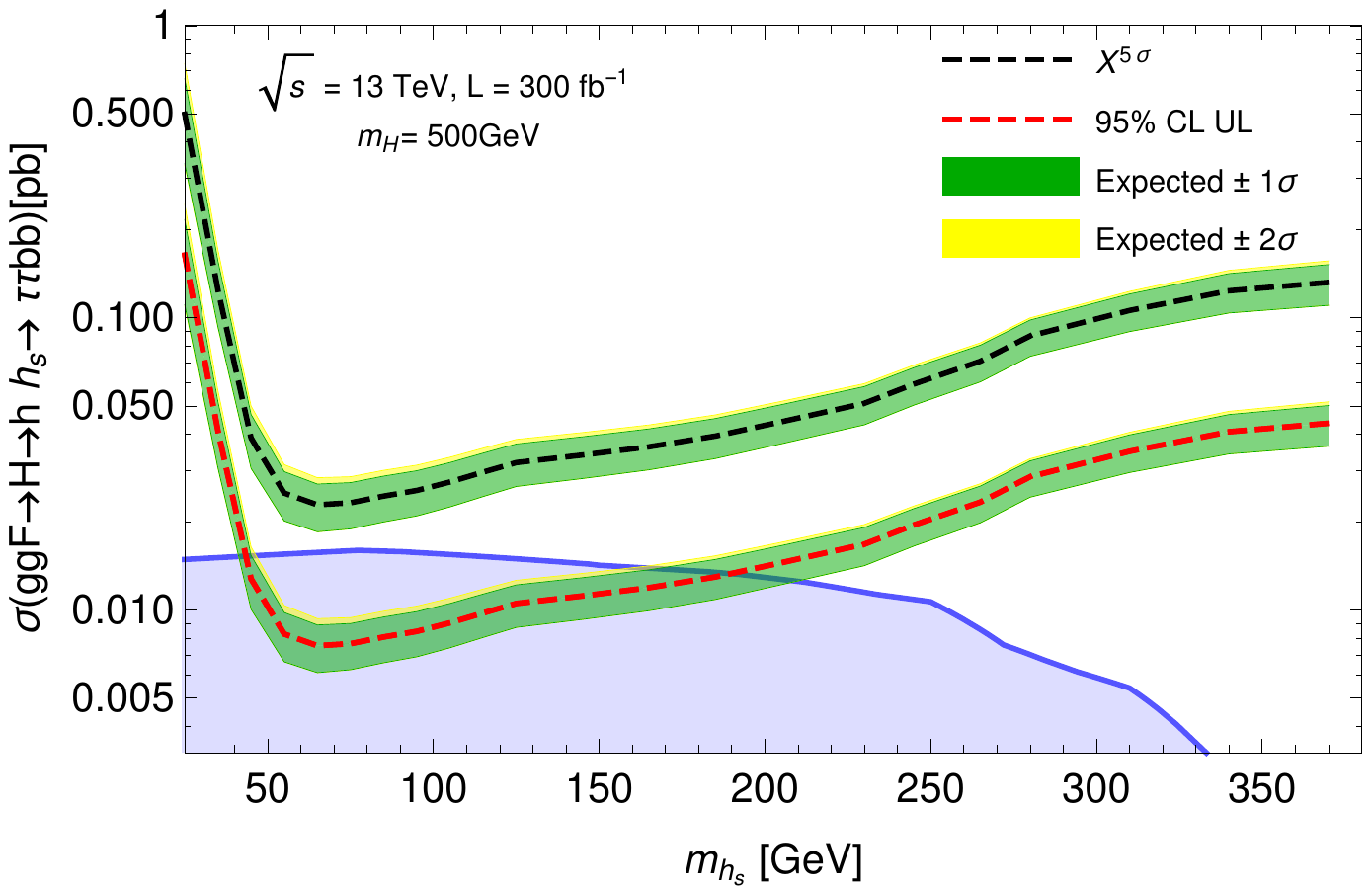}\\
\includegraphics[scale=0.50]{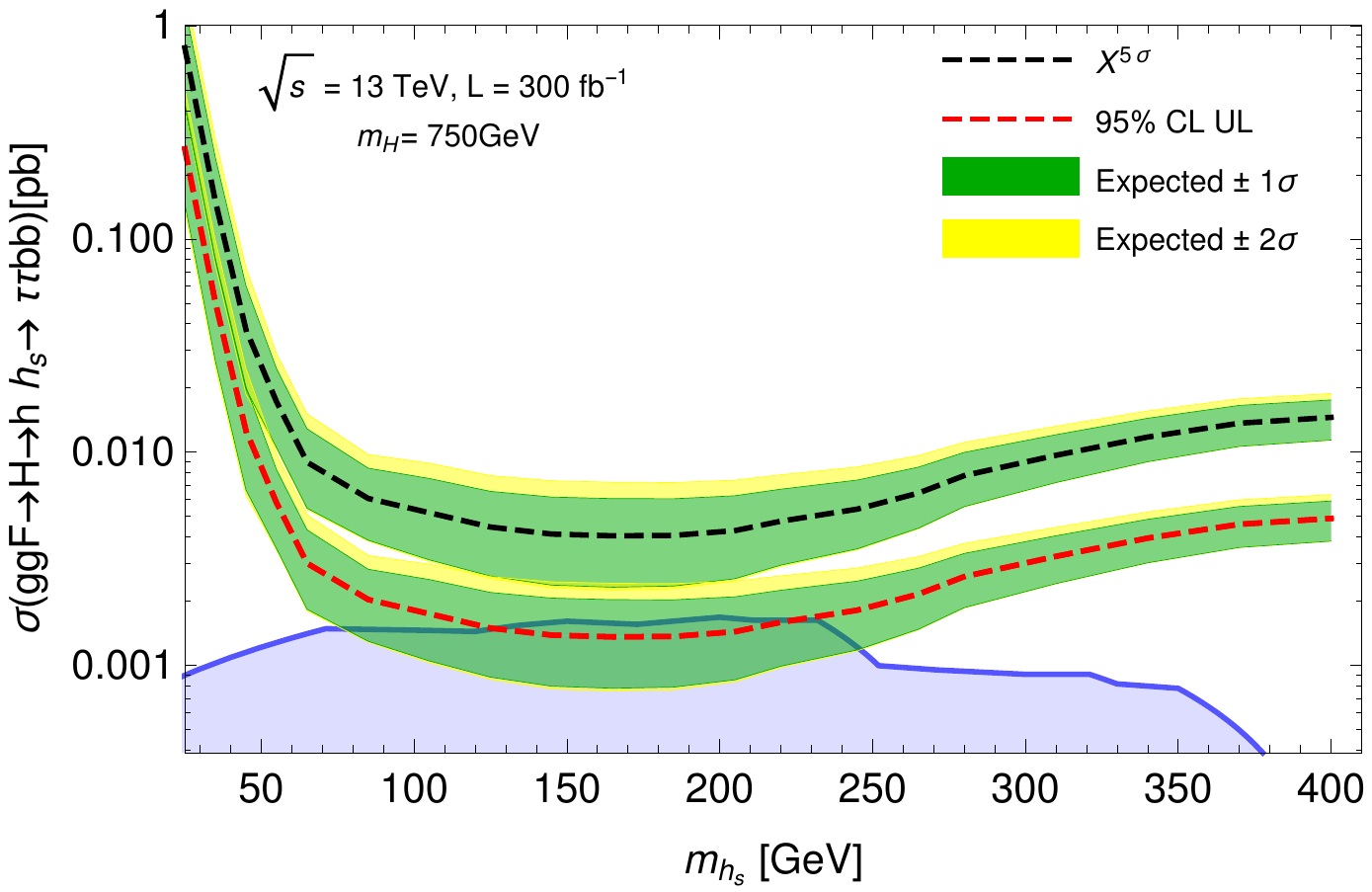}\hfill
\includegraphics[scale=0.50]{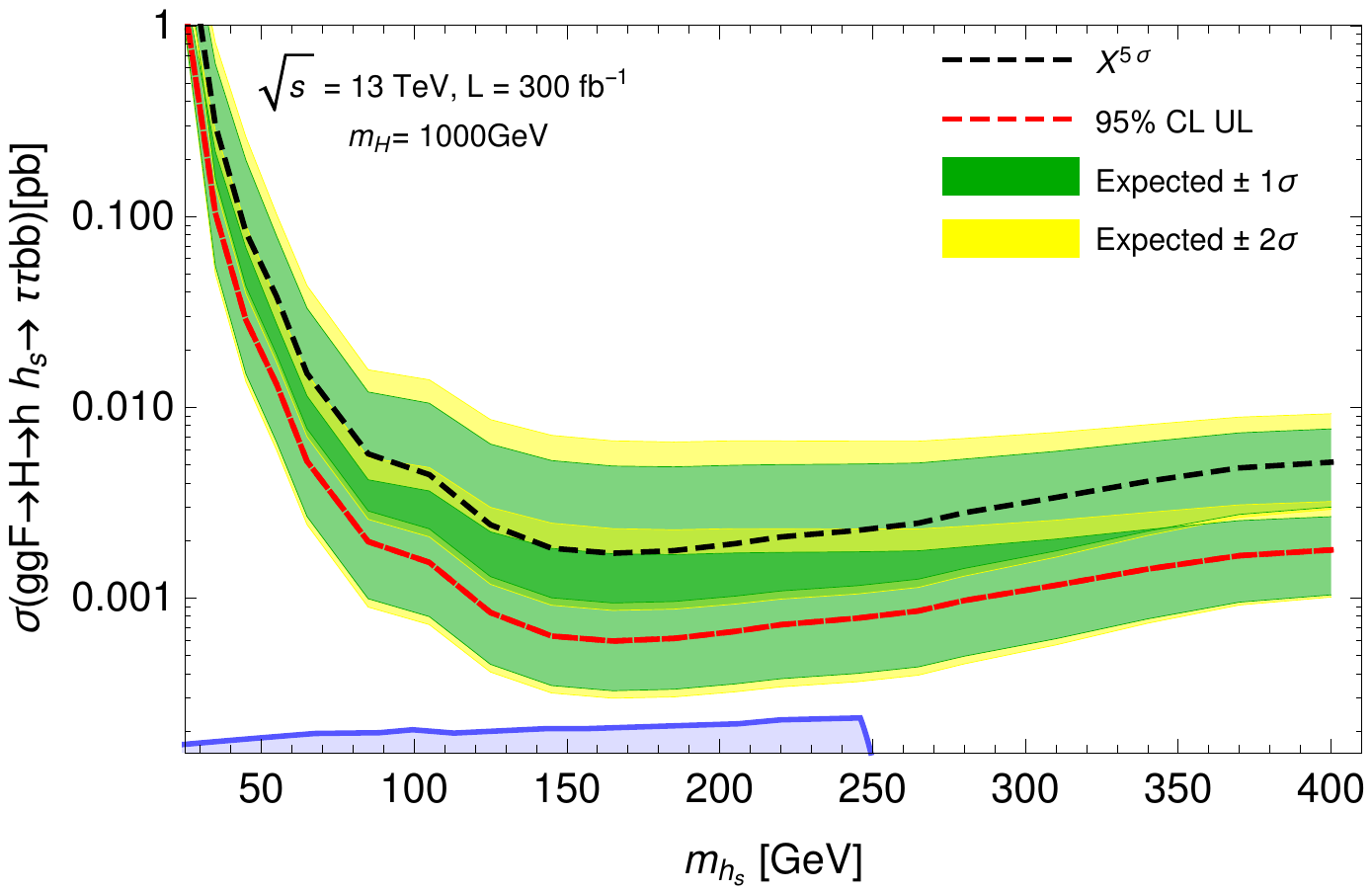}
\caption{95\% CL exclusion limits and $5\,\sigma$ discovery cross sections
for $H_{125}\to \tau\tau$ and $H_S\to b\bar{b}$
as function of $M_{H_S}$ for 300~fb$^{-1}$ integrated luminosity and
$M_H=425$~GeV (upper left), $M_H=500$~GeV (upper right), $M_H=750$~GeV (lower left),
$M_H=1000$~GeV (lower right).}
\label{fig:plotSensULttbb300}
\end{figure}

\begin{figure}[!h]
\centering
\includegraphics[scale=0.50]{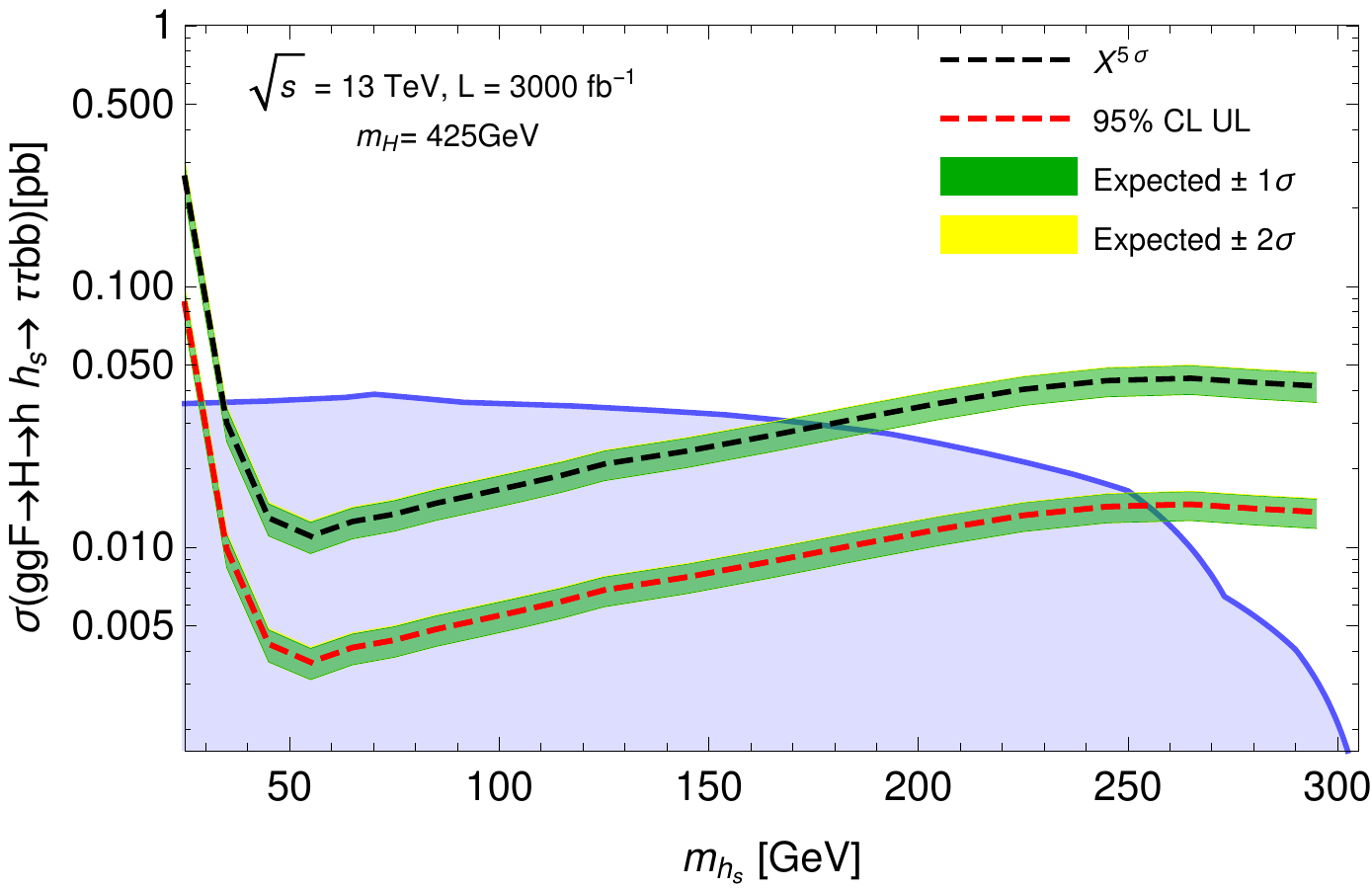}\hfill
\includegraphics[scale=0.50]{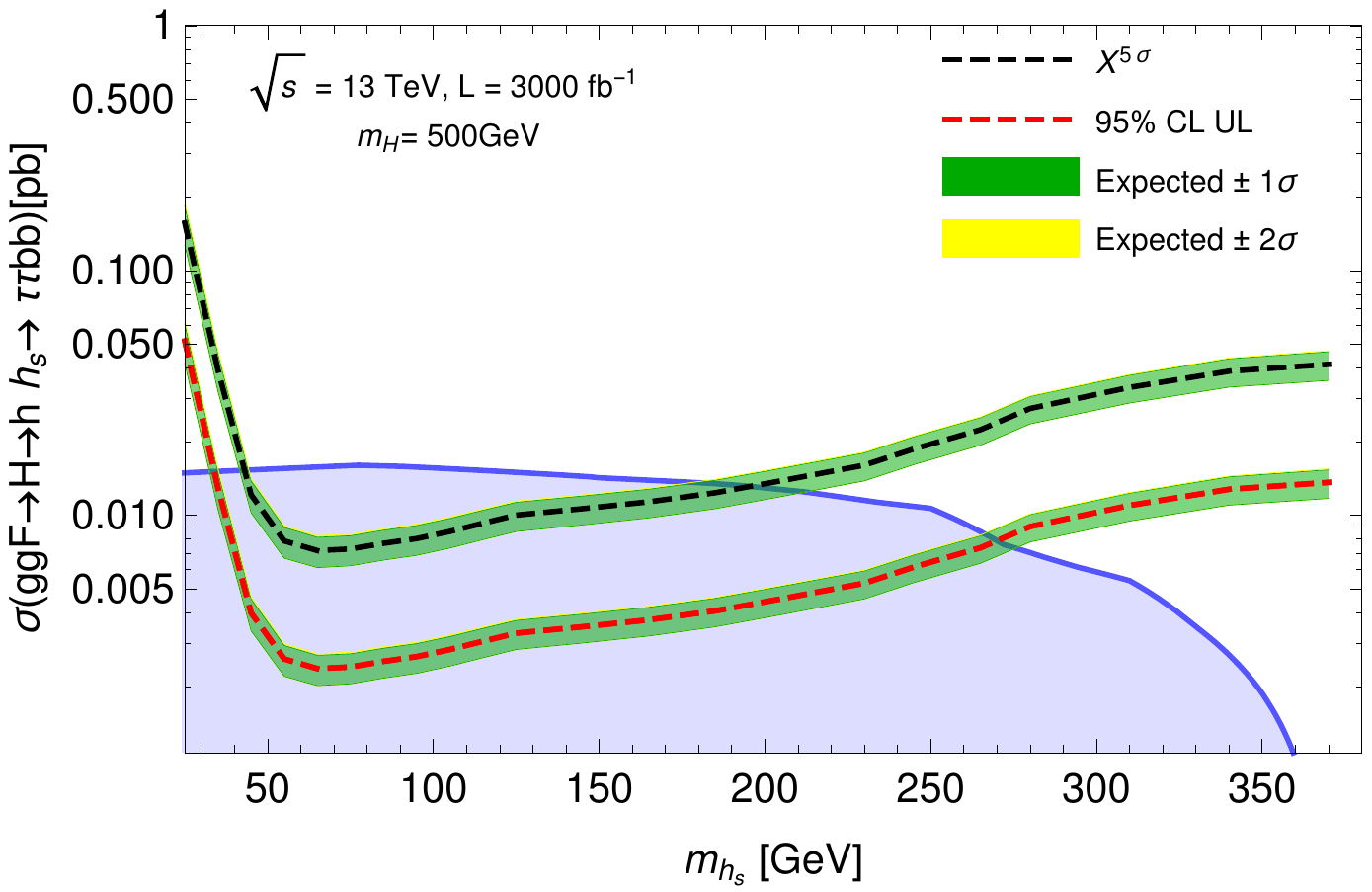}\\
\includegraphics[scale=0.50]{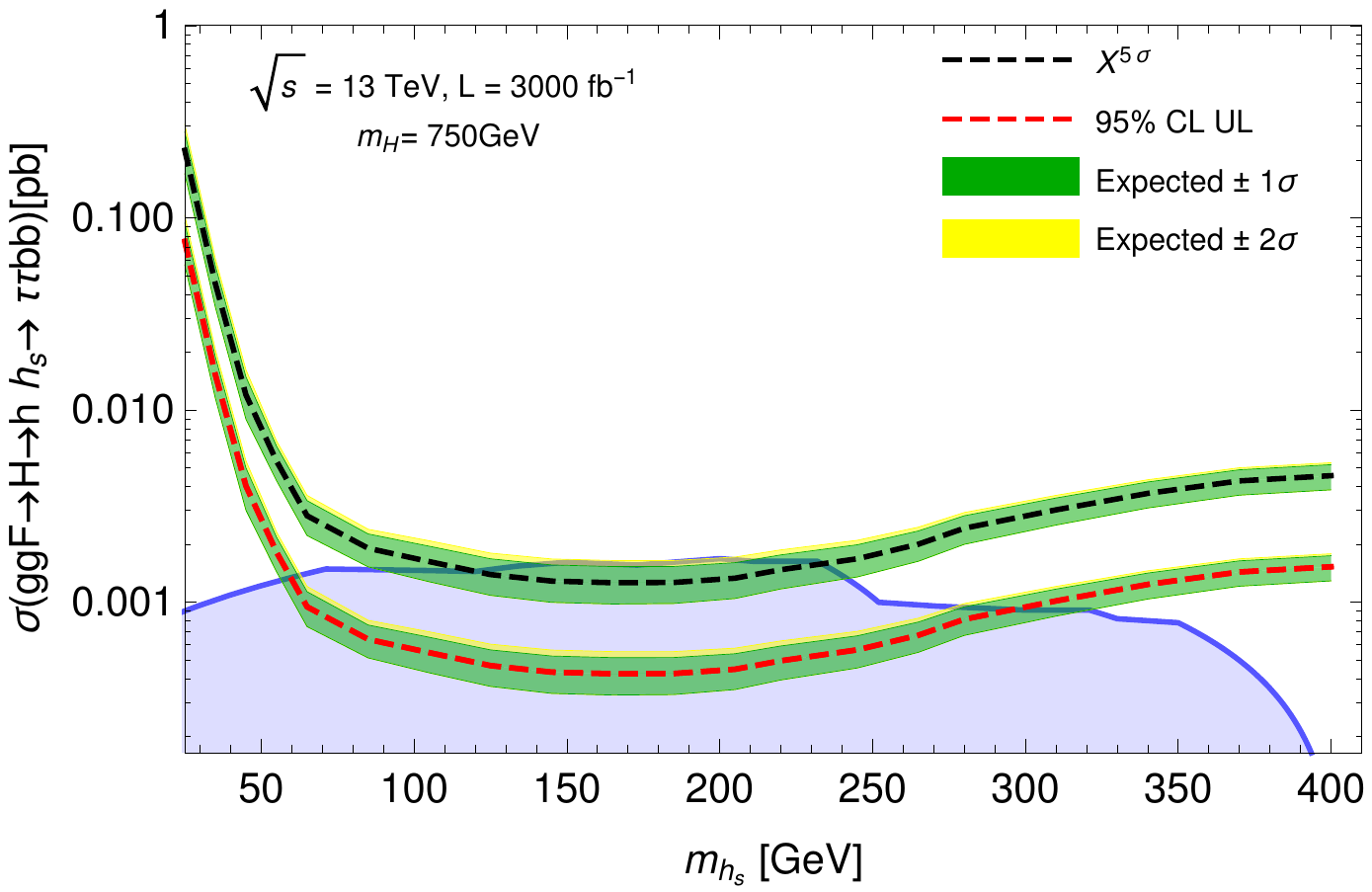}\hfill
\includegraphics[scale=0.50]{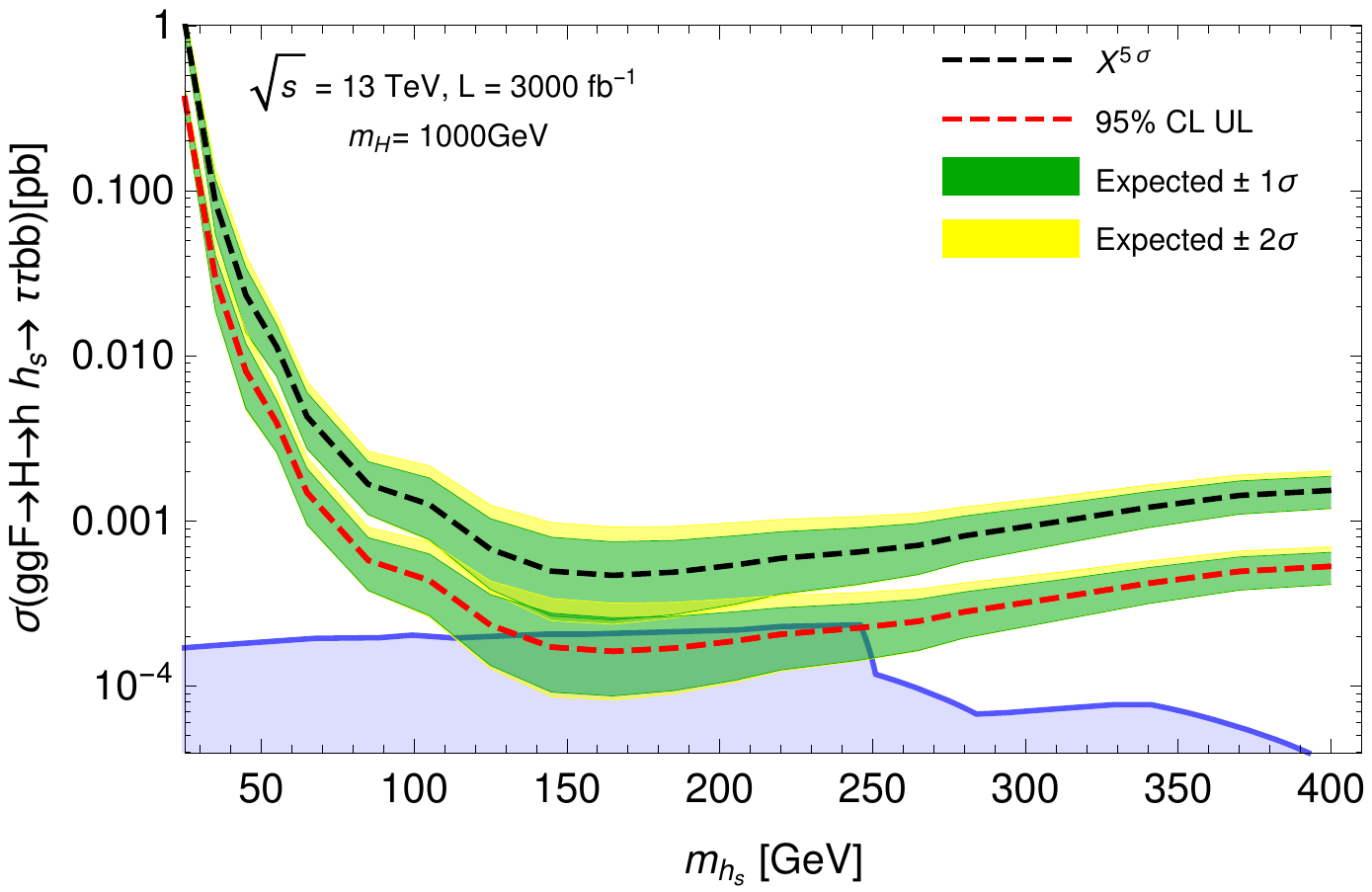}
\caption{95\% CL exclusion limits and $5\,\sigma$ discovery cross sections
for $H_{125}\to \tau\tau$ and $H_S\to b\bar{b}$
as function of $M_{H_S}$ for 3000~fb$^{-1}$ integrated luminosity and
$M_H=425$~GeV (upper left), $M_H=500$~GeV (upper right), $M_H=750$~GeV (lower left),
$M_H=1000$~GeV (lower right).}
\label{fig:plotSensULttbb3000}
\end{figure}

Given the $M_X$ distribution of the background for various hypothetical values
of $M_{H_S}$ and the $M_X$ distributions of signals we can, as before,
obtain values for 95\%~CL exclusion and $5\,\sigma$ discovery for cross sections
times branching fractions into the $H_{125}\to b\bar{b}$, $H_S\to \tau\tau$ and
$H_{125}\to \tau\tau$, $H_S\to b\bar{b}$ final states as function
of the integrated luminosity, $M_H$ and $M_{H_S}$.
We choose four representative values for $M_H=425,\,500,\,750\ \text{and}\ 1000$~GeV,
and show the 95\%~CL exclusion limits and $5\,\sigma$ discovery cross sections
as function of $M_{H_S}$ in each case. For $H_{125}\to b\bar{b}$, $H_S\to \tau\tau$
at 300~fb$^{-1}$ integrated luminosity
these are shown in Figs.~\ref{fig:plotSensULbbtt300}, for 3000~fb$^{-1}$ integrated
luminosity in Figs.~\ref{fig:plotSensULbbtt3000}. For $H_{125}\to \tau\tau$,
$H_S\to b\bar{b}$ at 300~fb$^{-1}$ integrated luminosity
these are shown in Figs.~\ref{fig:plotSensULttbb300}, for 3000~fb$^{-1}$ integrated
luminosity in Figs.~\ref{fig:plotSensULttbb3000}.
The uncertainties include
statistical uncertainties and, added linearly, $\pm 0.24$ considered as an estimate of the
systematic uncertainty originating from the normalisation of the background.

The following observations can be made: First, the expected sensitivities on
cross sections times branching ratios differ hardly among the
cases $H_{125}\to b\bar{b}$ and $H_S\to \tau\tau$ versus
$H_{125}\to \tau\tau$ and $H_S\to b\bar{b}$; if at all, the analyses aiming
at $H_{125}\to b\bar{b}$ and $H_S\to \tau\tau$ are typically somewhat
more sensitive.

Second, in Two-Higgs-Doublet models of type~II as well as in the NMSSM the branching
fractions into $b\bar{b}$ and $\tau\tau$ of both $H_{125}$ and $H_S$ are
always related by a factor $\sim 9:1$. Accordingly the possible 
cross sections times branching fractions in
the NMSSM parameter space for both
$\sigma(ggF\to H\to H_{125}+H_S\to b\bar{b}\tau\tau)$ and
$\sigma(ggF\to H\to H_{125}+H_S\to \tau\tau b\bar{b})$, indicated in blue in 
Figs.~\ref{fig:plotSensULbbtt300} -- \ref{fig:plotSensULttbb3000},
are $\sim 1/9$ of the ones in
Figs.~\ref{fig:sensbbbb_300} -- \ref{fig:sensbbbb_3000} for the
$b\bar{b}b\bar{b}$ final state. (The same reasoning applies to
$\sigma(ggF\to A\to H_{125}+A_S\to b\bar{b}\tau\tau)$ and
$\sigma(ggF\to A\to H_{125}+A_S\to \tau\tau b\bar{b})$; the viable NMSSM
points correspond to the ones in Figs.~\ref{fig:sensbbbb_300} and
\ref{fig:sensbbbb_3000}.)

Then one can ask, for a given point in parameter space, which of the analyses
considered up to now is the most sensitive. According to our results this is
the search in the $b\bar{b}b\bar{b}$ final state which allows to test
somewhat larger regions in parameter space.


\section{Search strategies for the $b\bar{b}\gamma\gamma$ final state}

Searches for resonant $H_{125}$ pair production in the $b\bar{b}\gamma\gamma$
final state have been performed by ATLAS at 8~TeV \cite{Aad:2014yja,Aad:2015xja}
and at 13~TeV \cite{TheATLAScollaboration:2016ibb},
by CMS at 8~TeV in \cite{Khachatryan:2016sey} and at 13~TeV in \cite{CMS-PAS-HIG-16-032,
CMS-PAS-HIG-17-008}.
A priori the diphotons can originate from $H_S$ or $H_{125}$;
both cases will be studied below. As in case of the previous final states
we optimise the cuts as function of a tentative value for $M_{H_S}$.

\subsection{Analyses of Signal Samples}

For the simulation of signal samples we used again \texttt{MadGraph5\_aMC@NLO} 
\cite{Alwall:2014hca}.
Events are required to have exactly two $b$-tagged jets with $p_T(b) > 40$~GeV
and $|\eta|<2.5$. Following ATLAS \cite{TheATLAScollaboration:2016ibb}
a working point with $\epsilon_b=0.85$ was chosen for the $b$-tagging efficiency
in order to increase the statistics.

At least two photons are required in each event which have to satisfy the isolation
criteria
\beq
\frac{\sum_i p_{T,i}}{p_{T,\gamma}} < 0.1
\eeq
where the sum over $i$ includes all tracks within a cone $\Delta R = 0.4$ around the
photon. The two leading photons are required to satisfy
\beq
E_T^{lead}/M_{\gamma\gamma}>0.35, \qquad E_T^{subl}/M_{\gamma\gamma}>0.25, \qquad
|\eta|<2.37\; .
\eeq

Additional cuts depend on whether the diphoton pair is assumed to originate from
$H_{125}$ or $H_S$, and the assumed value of $M_{H_S}$. First we consider the case
$H_S\to \gamma\gamma$. Then  the $b\bar{b}$ pair is assumed to originate from $H_{125}$,
and we require
\beq
100\ \text{GeV} < M_{b\bar{b}} < 150\ \text{GeV}\; .
\eeq
As in the previous cases it is useful to define a corrected invariant
mass $M_X$ for the $b\bar{b}\gamma\gamma$ system:
\beq
M_X=M_{b\bar{b}\gamma\gamma}+125\ \text{GeV} - M_{b\bar{b}}\; .
\eeq

Given an assumed value for $M_{H_S}$ it turned out to optimize the signal
to background
efficiency applying a $M_X$ and $M_{H_S}$ dependent cut on the diphoton
invariant mass $M_{\gamma\gamma}$, since the measured distribution
$|M_{\gamma\gamma} - M_{H_S}|$ broadens somewhat with $M_X$:
\beq
|M_{\gamma\gamma} - M_{H_S}| < 4.3\ \text{GeV} + 0.016 M_X\; .
\eeq
Also the cuts on the $H_{125}$ and $H_S$ candidates depend on $M_X$:
\bea
p_T(b\bar{b}) &>& 17\ \text{GeV} + 0.18 M_X\; ,\\
E_T(\gamma\gamma) &=& 68.3\ \text{GeV} + 0.25 M_X\; .
\eea
($E_T(\gamma\gamma)$ instead of $p_T(\gamma\gamma)$ allows for $M_{H_S}$
independent cuts.)

Assuming $H_{125}\to \gamma\gamma$ and $H_S\to b\bar{b}$ the previous cuts
are modified as follows: First, for $M_X$ we take
\beq
\label{eq:MX_bbgg}
M_X=M_{b\bar{b}\gamma\gamma}+ M_{H_S} - M_{b\bar{b}}\; ,
\eeq
and for the diphoton invariant mass we require
\beq
|M_{\gamma\gamma} - 125\ \text{GeV}| < 2\ \text{GeV} + 0.02 M_X\; .
\eeq
The mass window for $M_{b\bar{b}}$ is now
\beq
0.9 M_{H_S} - 30\ \text{GeV} < M_{b\bar{b}} < 0.9 M_{H_S} + 20\ \text{GeV}\; .
\eeq
The cuts on the $H_{125}$ and $H_S$ candidates are:
\bea
p_T(b\bar{b}) &>& -5.7\ \text{GeV} + 0.29 M_X\; ,\\
E_T(\gamma\gamma) &=& 7.45\ \text{GeV} + 0.33 M_X\; .
\eea
All the numerical values above have been obtained by optimising signal-to-background
ratios using the MC events.

\subsection{Background Estimation}

SM backgrounds originate from $b\bar{b}\gamma\gamma$, $c\bar{c}\gamma\gamma$,
$jj\gamma\gamma$ ($j\neq c/\bar{c}$), $b\bar{b}j\gamma$ and $t\bar{t}H$. We
simulated these backgrounds again using \texttt{MadGraph5\_aMC@NLO}.
Due to the relatively large $b$-tagging efficiency $\epsilon=0.85$ mistagging
rates are relatively large. $b\bar{b}j\gamma$ contribute if another fake photon
appears.

After applying the cuts of the previous subsection the relative contributions of
the SM backgrounds depend strongly on $M_{H_S}$ and $M_H$, and
on whether the $\gamma\gamma$ pair originates from $H_S$ or $H_{125}$.
For a light $H_S\sim 85$~GeV the background from $b\bar{b}j\gamma$ is always important,
$\sim 60\%$ for $H_S\to\gamma\gamma$ and $\sim 30\%$ for $H_S\to b\bar{b}$.
However, all other $\gamma\gamma + X$ SM backgrounds can also contribute
several 10\% individually. The contribution from $t\bar{t}H$ is always
$\lsim 3\%$.

For $b\bar{b}\gamma\gamma$, $c\bar{c}\gamma\gamma$ and $jj\gamma\gamma$
an extra jet was allowed in the final state, $b\bar{b}j\gamma$ was multiplied
by a NLO K~factor 1.58 from \cite{Alwall:2014hca}, and $t\bar{t}H$ (with SM
couplings) was simulated at NLO.

Still we cannot expect that the total SM background cross section is
completely captured by the MC simulation; in searches by ATLAS and CMS
in \cite{TheATLAScollaboration:2016ibb,CMS-PAS-HIG-16-032,CMS-PAS-HIG-17-008}
sidebands are
used for this purpose. Thus we proceed as before and correct the total
background cross section using data driven methods in the particular case
$M_{H_S}\sim 125$~GeV equivalent to (resonant) SM Higgs pair production.
To this end we modify slightly the cuts on $p_T(b)$ ($> 55$~GeV and $> 35$~GeV
for the leading and next-to-leading $b$-jets), $M_{b\bar{b}}$ and
$M_{\gamma\gamma}$ such that they coincide with the ones in the ATLAS search
 \cite{TheATLAScollaboration:2016ibb}.
(In the case of $M_{\gamma\gamma}$ we could check that the efficiency coincides.)
Extrapolating from sidebands with less $b$-jets, ATLAS \cite{TheATLAScollaboration:2016ibb}
obtained $1.63\pm 0.3$ expected background events in the signal region, whereas we found
$1.06\pm 0.14$ events from summing all MC events. Since all our backgrounds were
simulated to similar order (NLO) in the QCD coupling and separate higher order
K~factors are not available, we multiply their sum by $1.54\pm 0.35$ where the
latter uncertainty will again be treated as an estimate of the systematic uncertainty
contributing to our final results.

The dependence of the background on the total invariant mass $M_{b\bar{b}\gamma\gamma}$
was parametrized in \cite{TheATLAScollaboration:2016ibb} by a two parameter Landau
distribution given in appendix~A. We found that the Landau distribution fits the
$M_X$ distribution from eq.~(\ref{eq:MX_bbgg}) as well (with $M_{H_S}$ dependent
parameters), and used it for the expected $M_X$ distributions of the various background
contributions for our forecasts at 300~fb$^{-1}$ and 3000~fb$^{-1}$.

\subsection{Future 95\% CL Exclusion Limits and $5\,\sigma$ Discovery Cross Sections}

Given the $M_X$ distribution of the background for various hypothetical values
of $M_{H_S}$ and the $M_X$ distributions of signals we can, as before,
obtain values for 95\%~CL exclusion and $5\,\sigma$ discovery for cross sections
times branching fractions into the $H_{125}\to b\bar{b}$, $H_S\to \gamma\gamma$ and
$H_{125}\to \gamma\gamma$, $H_S\to b\bar{b}$ final states as function
of the integrated luminosity, $M_H$ and $M_{H_S}$.

\begin{figure}[!h]
\centering
\includegraphics[scale=0.65]{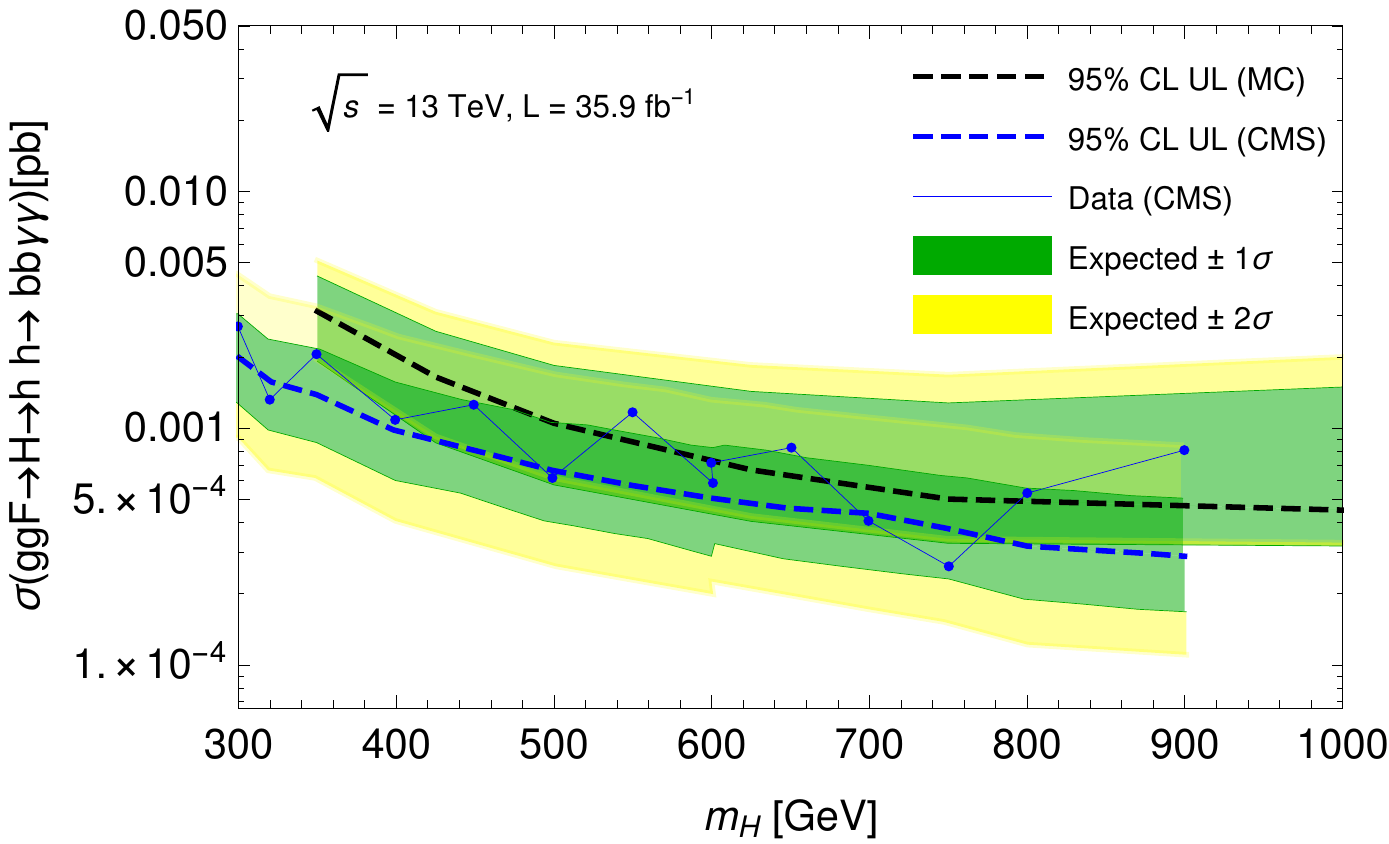}
\caption{Expected 95\% CL exclusion limits
for $ggF\to H_{125}H_{125}\to b\bar{b}\gamma\gamma$ for 35.9~fb$^{-1}$
from CMS \cite{CMS-PAS-HIG-17-008} and from our MC simulation as function
of $M_H$. For completeness the observed limits are shown.}
\label{fig:compLimits_CMS_bbgg_log}
\end{figure}

After completing our analysis the CMS search \cite{CMS-PAS-HIG-17-008} for resonant
Higgs pair production in the channel
$H_{125}+H_{125}\to b\bar{b}\gamma\gamma$ based on 35.9~fb$^{-1}$ appeared. The expected
95\% CL exclusion limits given in \cite{CMS-PAS-HIG-17-008}
can be compared to ours for $M_{H_S}=125$~GeV
 for the same integrated luminosity; this comparison as function of $M_H$ is shown in
Fig.~\ref{fig:compLimits_CMS_bbgg_log}. The expected limits coincide within $1\,\sigma$
for $M_H\gsim 500$~GeV, and within $2\,\sigma$ everywhere. Our expected limits are
systematically more conservative; we note that the CMS analysis employs a trained
boosted decision tree in order to separate the signal from the backgrounds which is
not available here.

\begin{figure}[!t]
\centering
\includegraphics[scale=0.50]{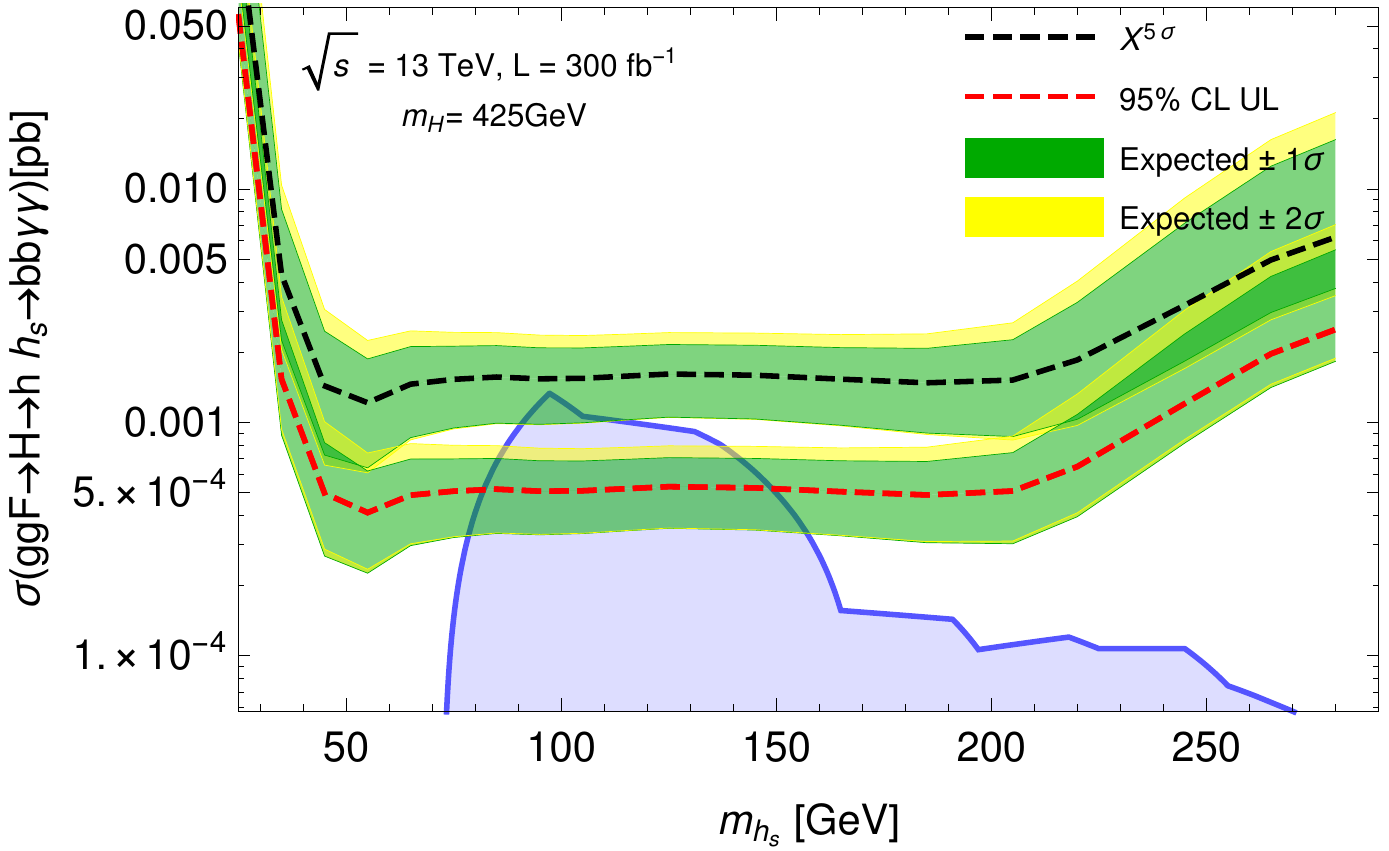}\hfill
\includegraphics[scale=0.50]{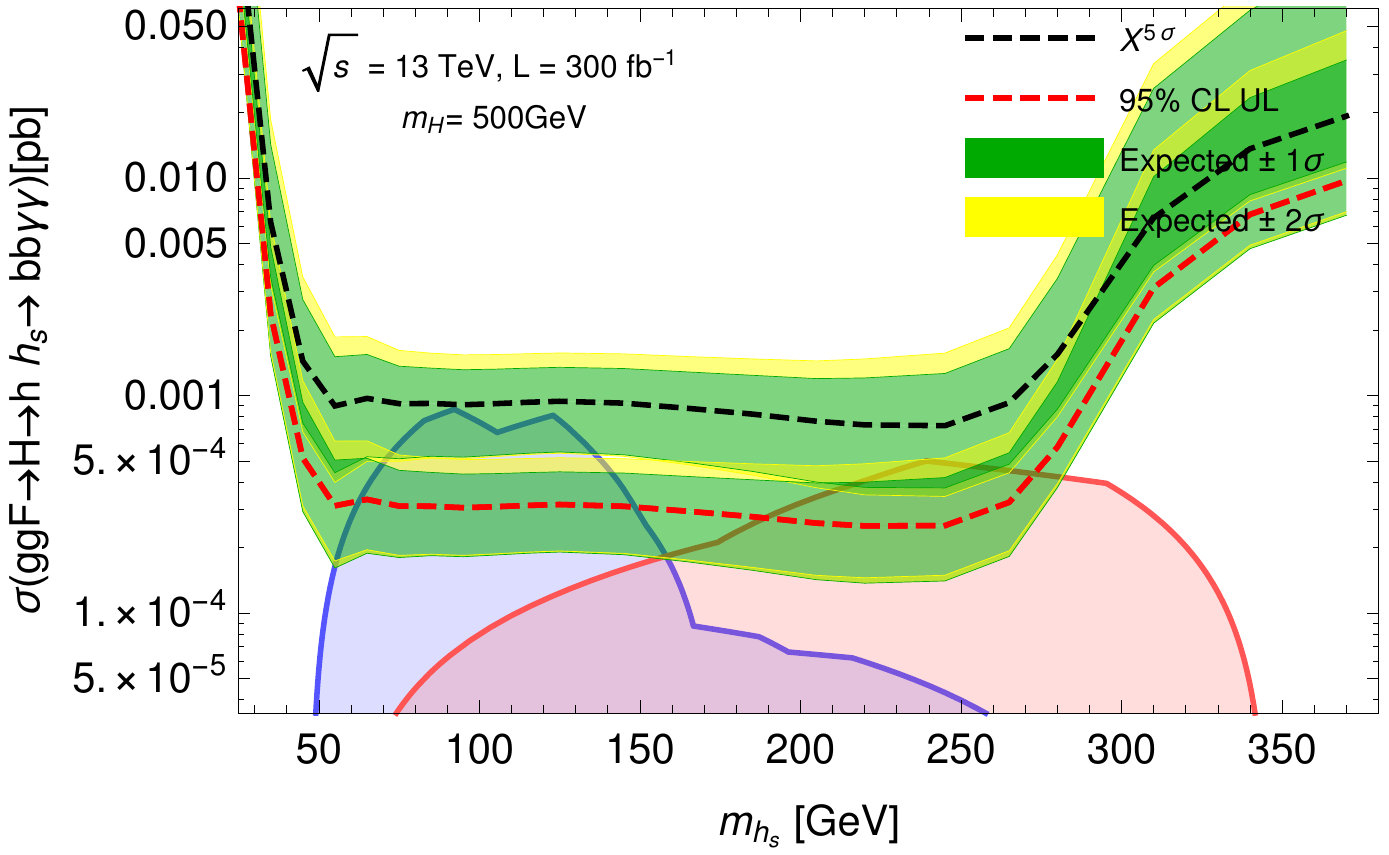}\\
\includegraphics[scale=0.50]{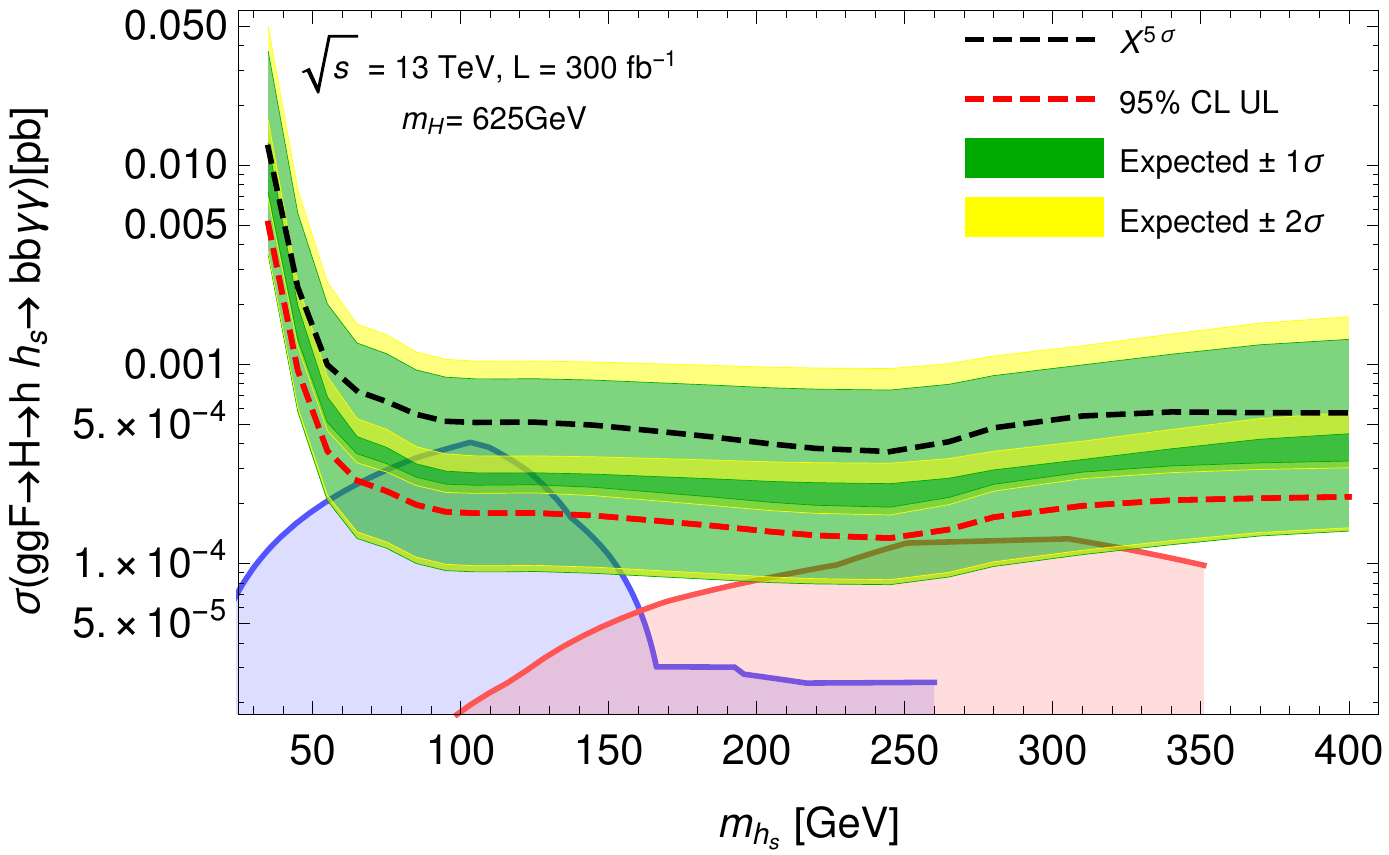}\hfill
\includegraphics[scale=0.50]{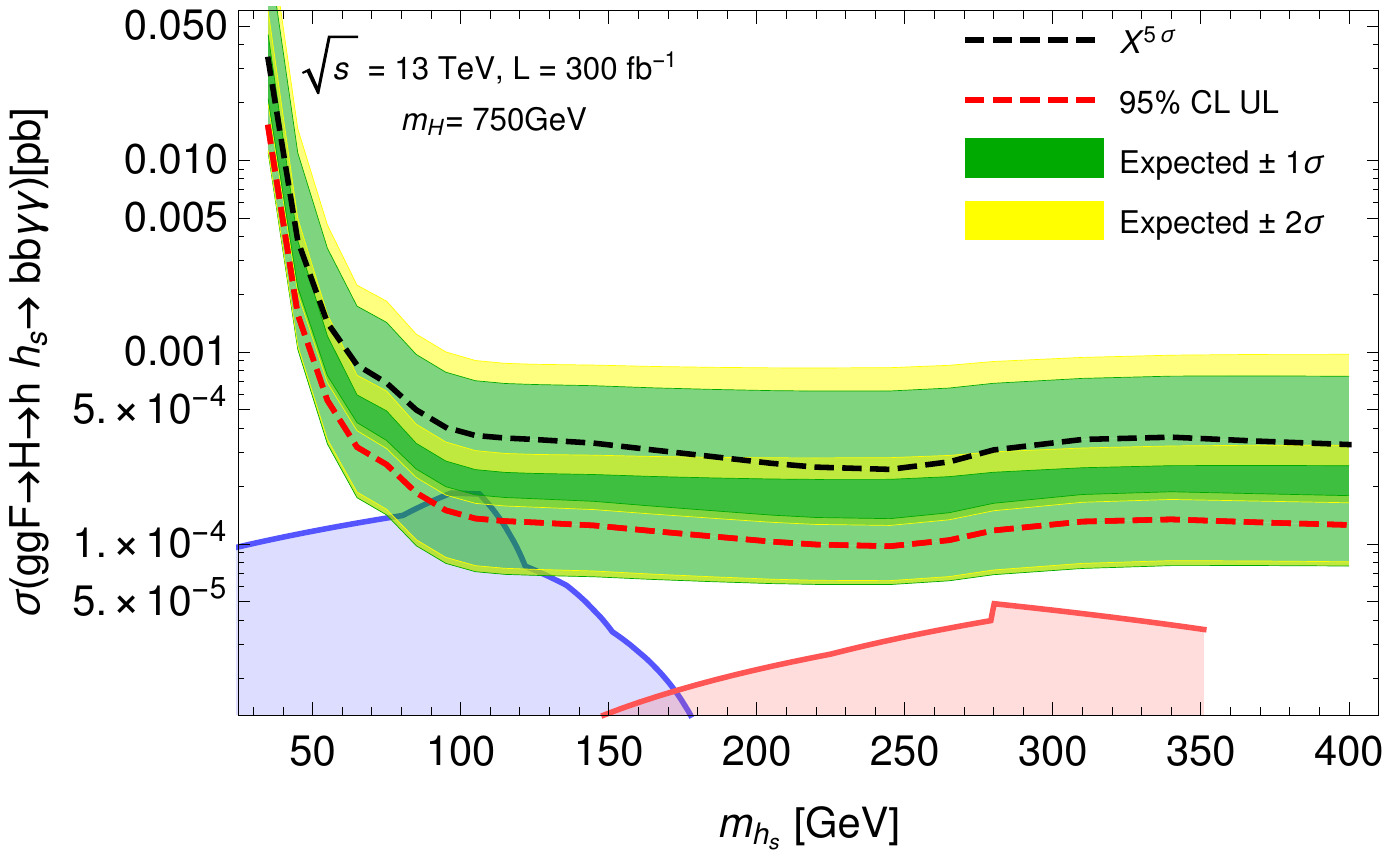}
\caption{95\% CL exclusion limits and $5\,\sigma$ discovery cross sections
for $H_{125}\to b\bar{b}$ and $H_S\to \gamma\gamma$
as function of $M_{H_S}$ for 300~fb$^{-1}$ integrated luminosity and
$M_H=425$~GeV (upper left), $M_H=500$~GeV (upper right), $M_H=625$~GeV (lower left),
$M_H=750$~GeV (lower right). The colored regions are explained in the
caption of Fig.~\ref{fig:sensbbbb_300}.}
\label{fig:SensULPoints_bbaaMX_300}
\end{figure}

\begin{figure}[!h]
\centering
\includegraphics[scale=0.50]{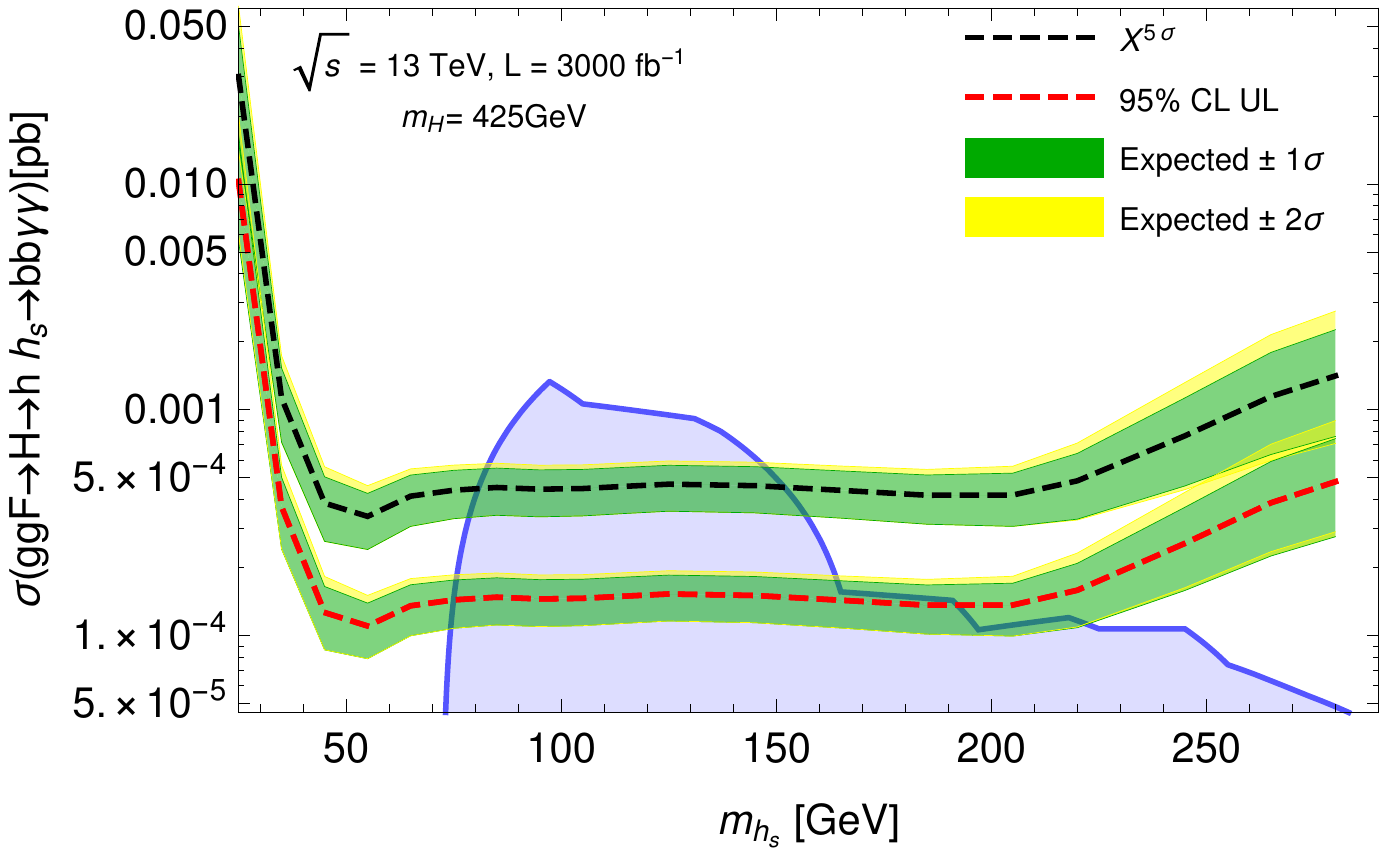}\hfill
\includegraphics[scale=0.50]{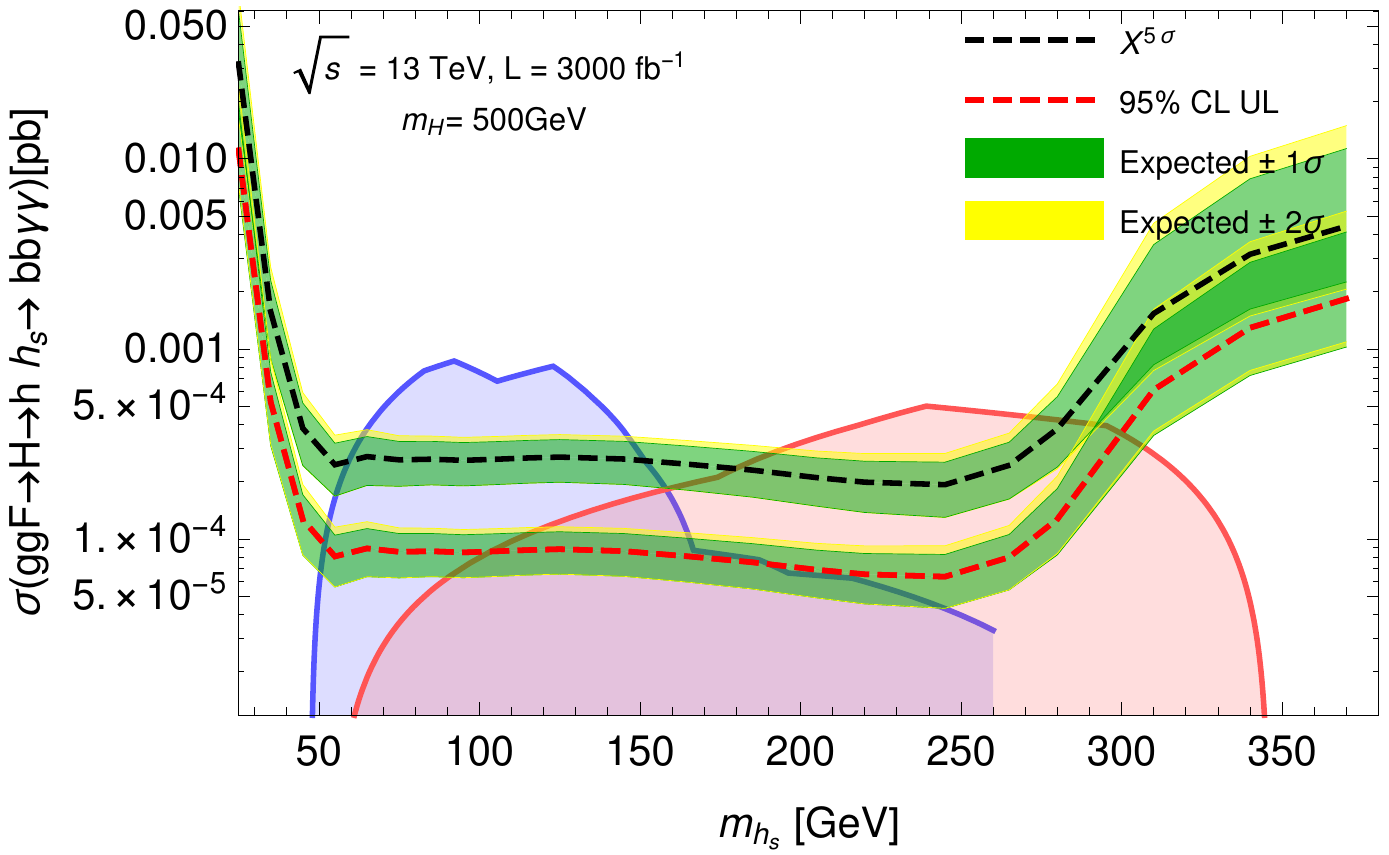}\\
\includegraphics[scale=0.50]{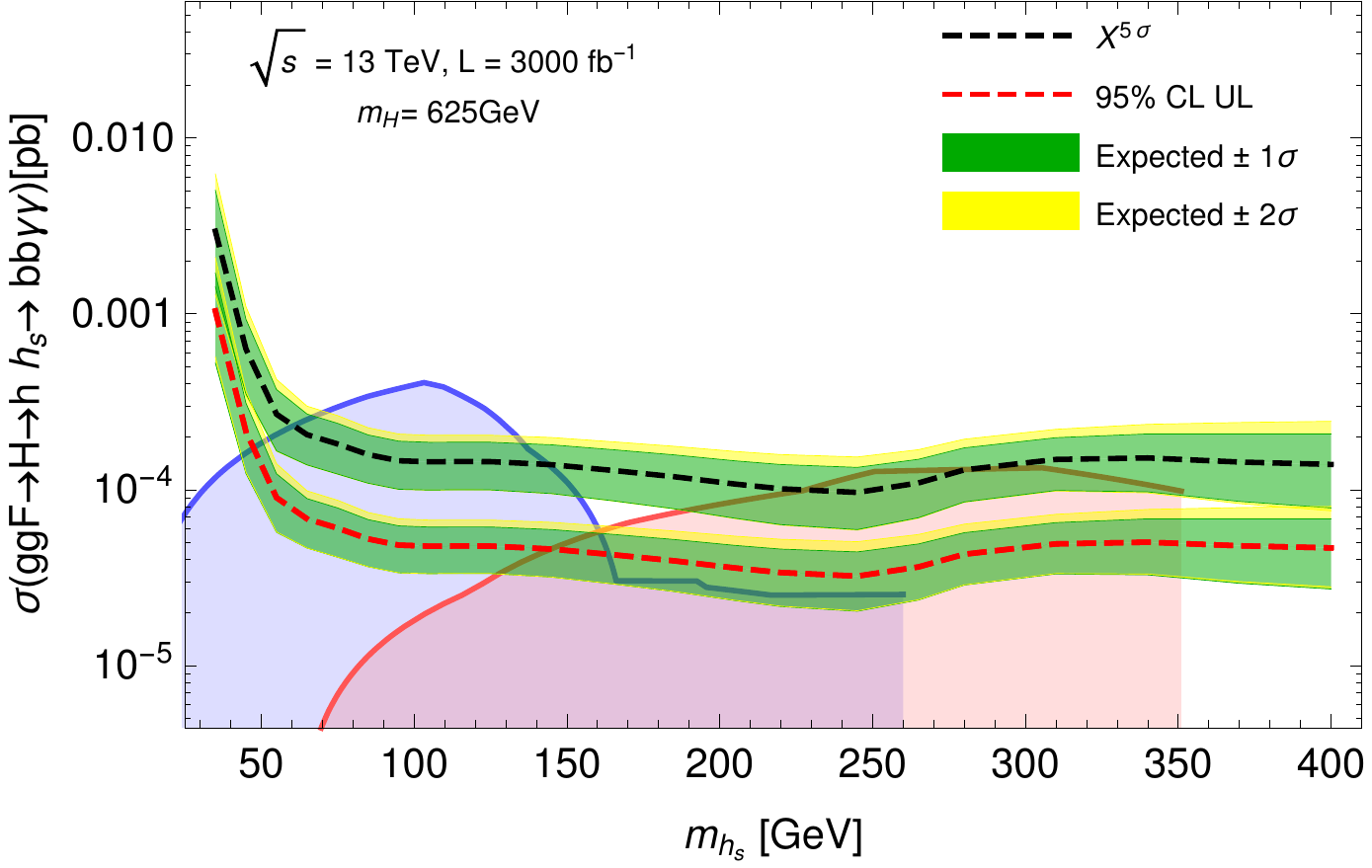}\hfill
\includegraphics[scale=0.50]{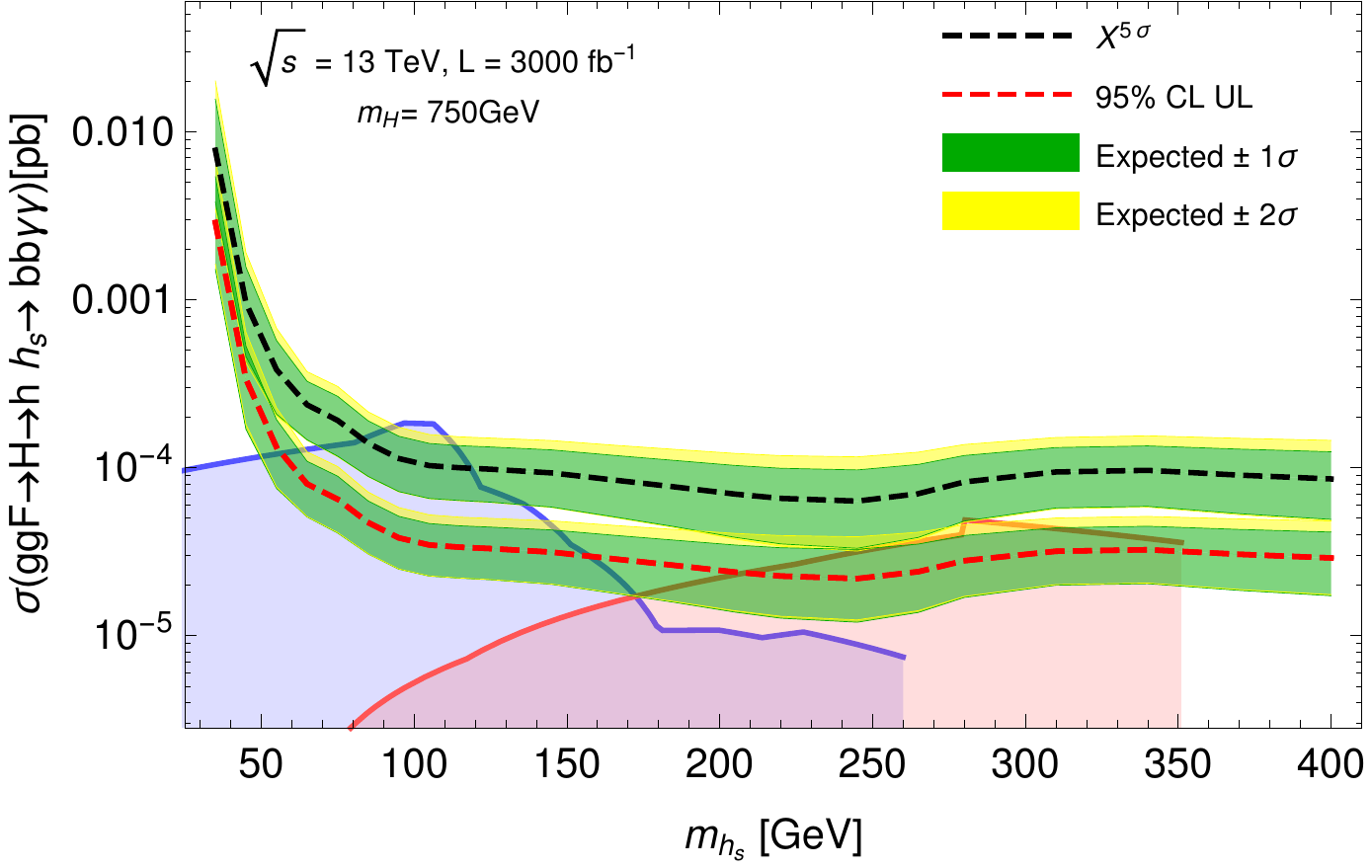}
\caption{95\% CL exclusion limits and $5\,\sigma$ discovery cross sections
for $H_{125}\to b\bar{b}$ and $H_S\to \gamma\gamma$
as function of $M_{H_S}$ for 3000~fb$^{-1}$ integrated luminosity and
$M_H=425$~GeV (upper left), $M_H=500$~GeV (upper right), $M_H=625$~GeV (lower left),
$M_H=750$~GeV (lower right). The colored regions are explained in the
caption of Fig.~\ref{fig:sensbbbb_300}.}
\label{fig:SensULPoints_bbaaMX_3000}
\end{figure}

\begin{figure}[!h]
\centering
\includegraphics[scale=0.50]{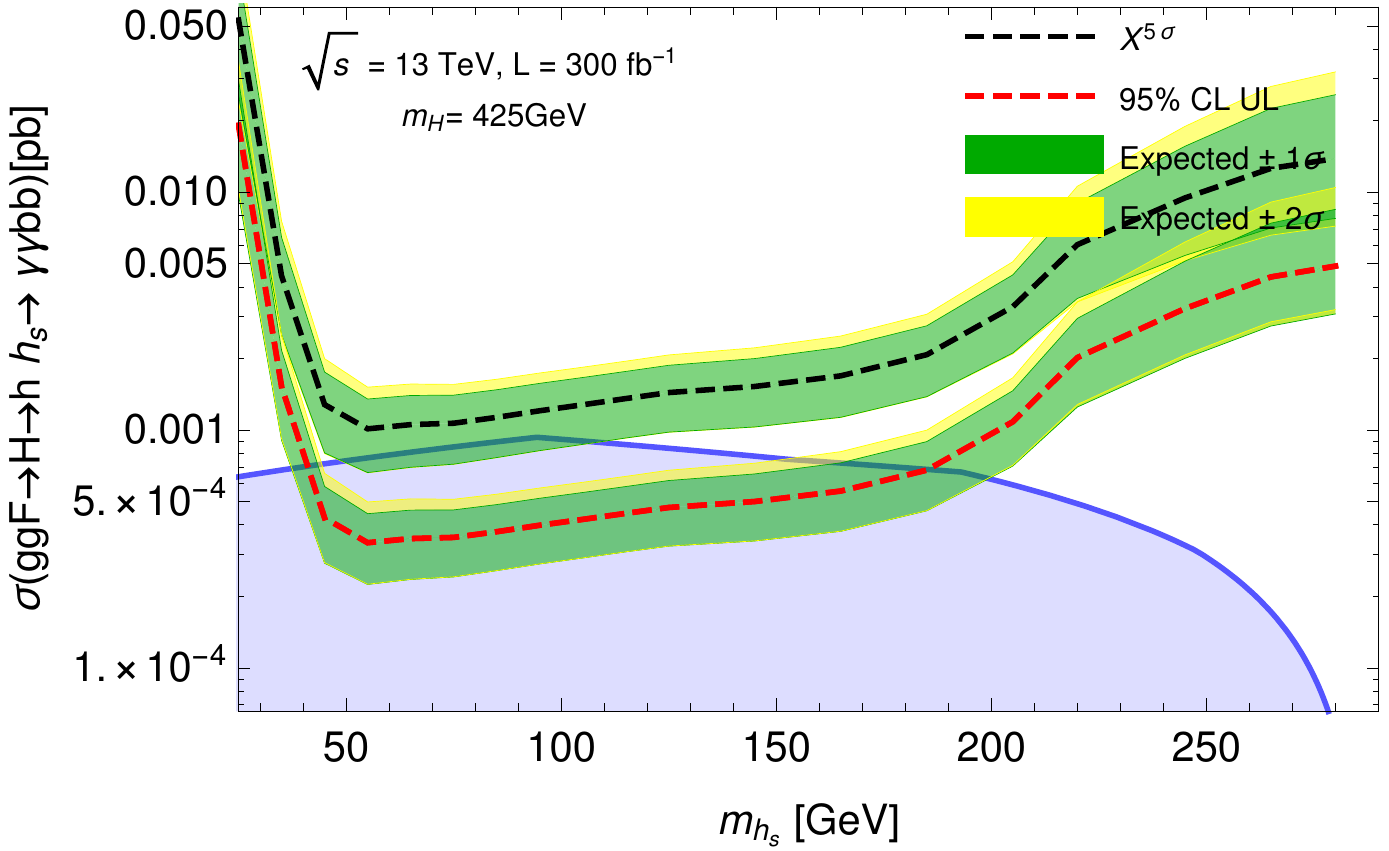}\hfill
\includegraphics[scale=0.50]{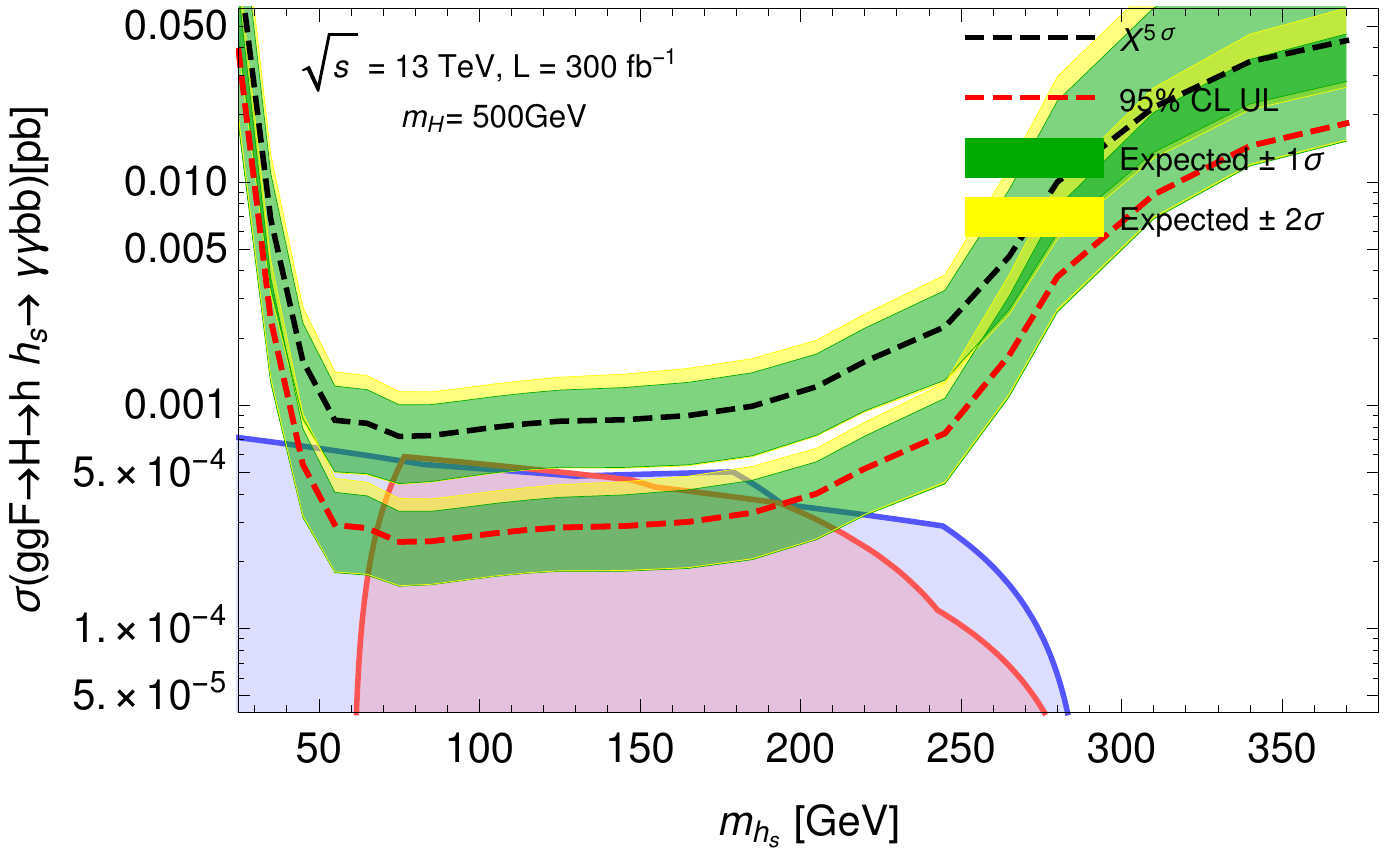}\\
\includegraphics[scale=0.50]{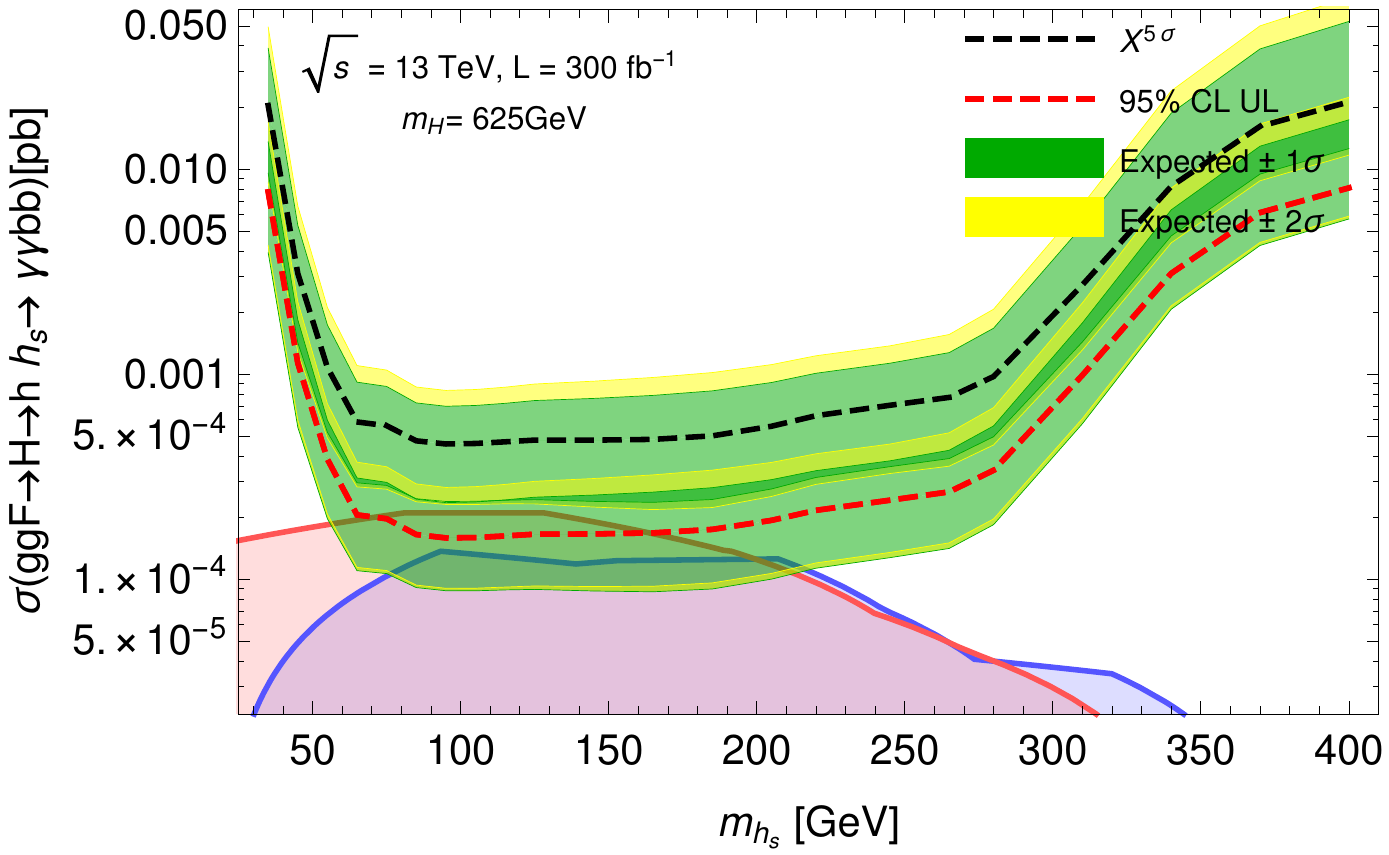}\hfill
\includegraphics[scale=0.50]{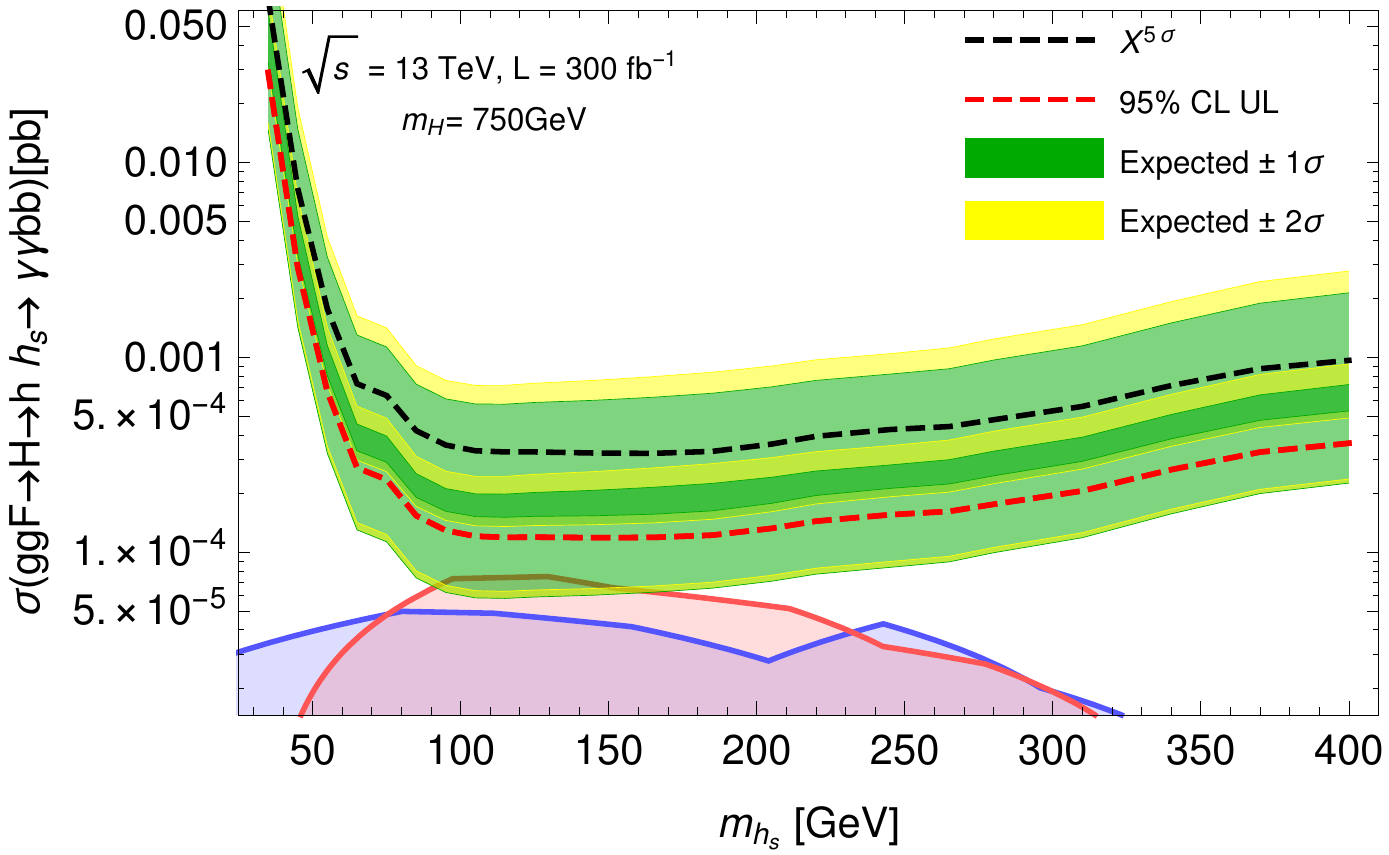}
\caption{95\% CL exclusion limits and $5\,\sigma$ discovery cross sections
for $H_{125}\to \gamma\gamma$ and $H_S\to b\bar{b}$
as function of $M_{H_S}$ for 300~fb$^{-1}$ integrated luminosity and
$M_H=425$~GeV (upper left), $M_H=500$~GeV (upper right), $M_H=625$~GeV (lower left),
$M_H=750$~GeV (lower right). The colored regions are explained in the
caption of Fig.~\ref{fig:sensbbbb_300}.}
\label{fig:SensULPoints_aabbMX_300}
\end{figure}

\begin{figure}[!h]
\centering
\includegraphics[scale=0.50]{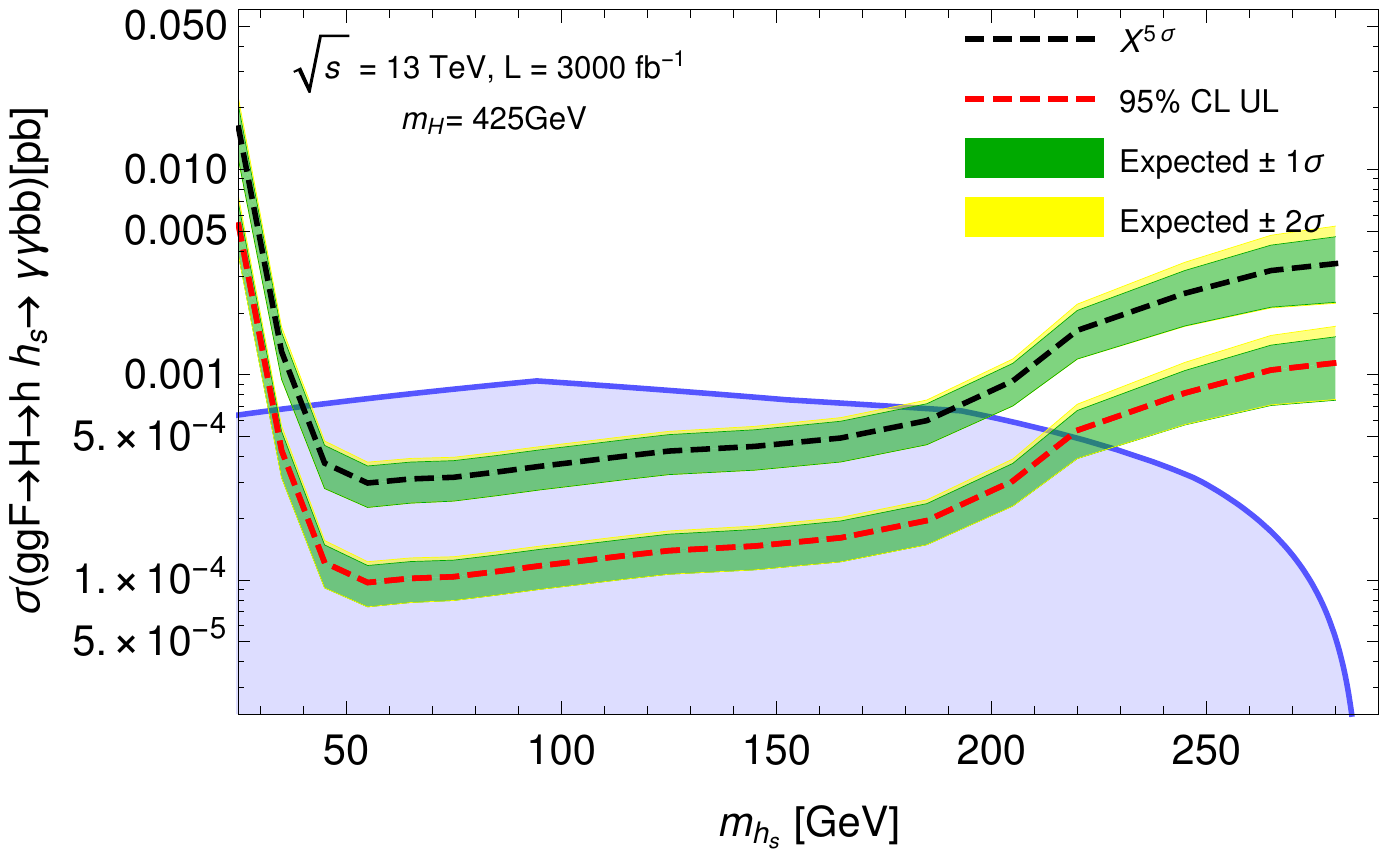}\hfill
\includegraphics[scale=0.50]{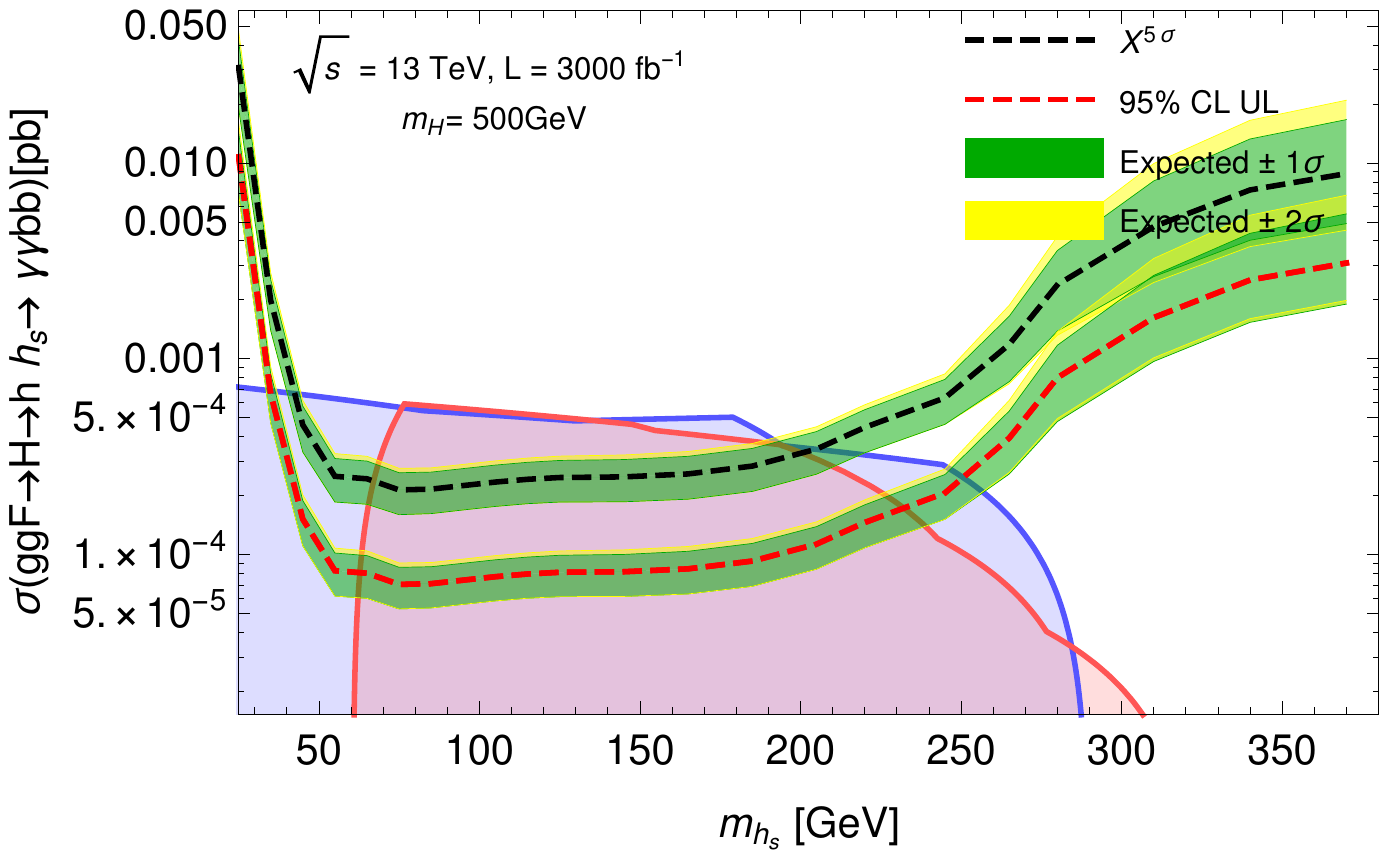}\\
\includegraphics[scale=0.50]{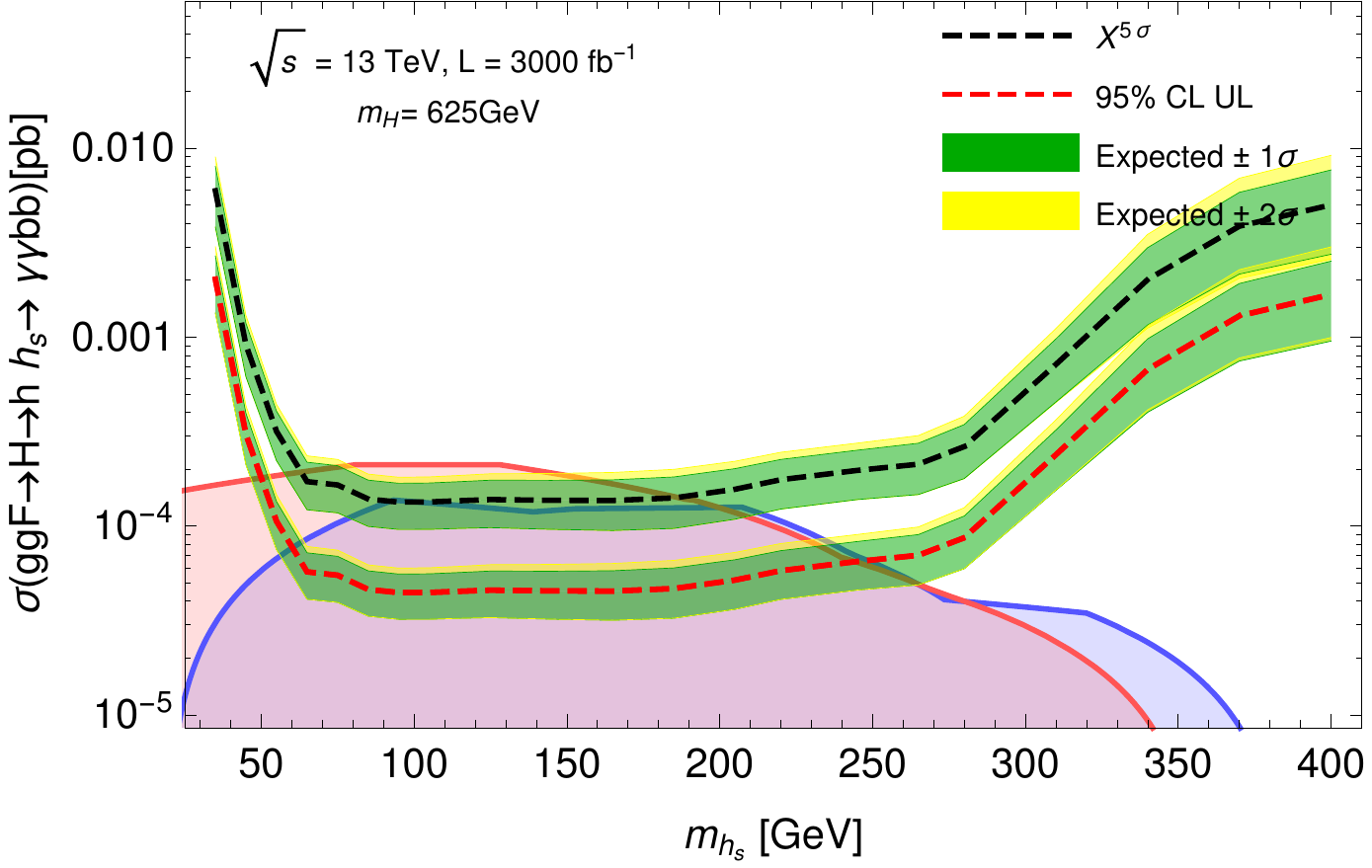}\hfill
\includegraphics[scale=0.50]{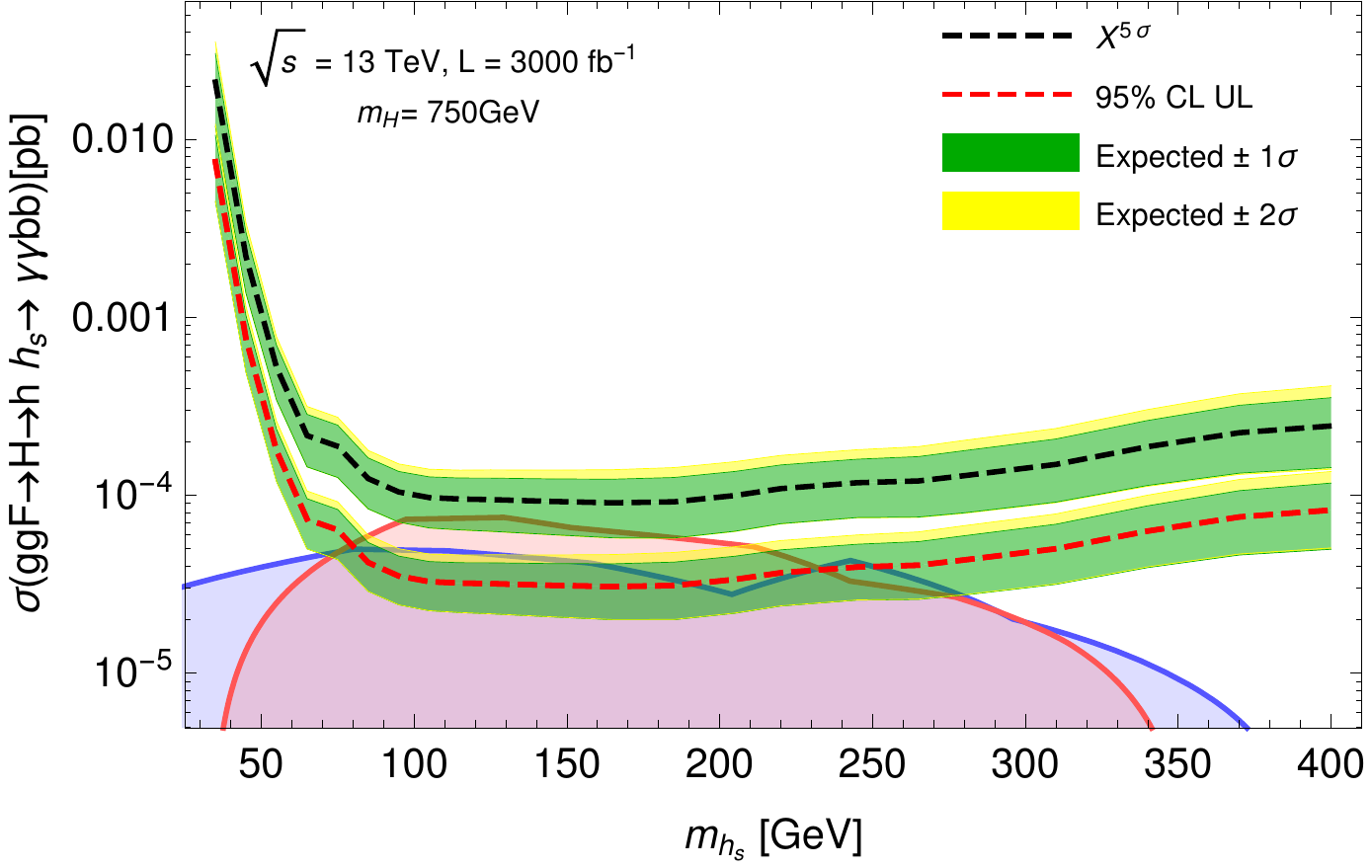}
\caption{95\% CL exclusion limits and $5\,\sigma$ discovery cross sections
for $H_{125}\to \gamma\gamma$ and $H_S\to b\bar{b}$
as function of $M_{H_S}$ for 3000~fb$^{-1}$ integrated luminosity and
$M_H=425$~GeV (upper left), $M_H=500$~GeV (upper right), $M_H=625$~GeV (lower left),
$M_H=750$~GeV (lower right). The colored regions are explained in the
caption of Fig.~\ref{fig:sensbbbb_300}.}
\label{fig:SensULPoints_aabbMX_3000}
\end{figure}

\begin{figure}
\centering
\includegraphics[scale=.7]{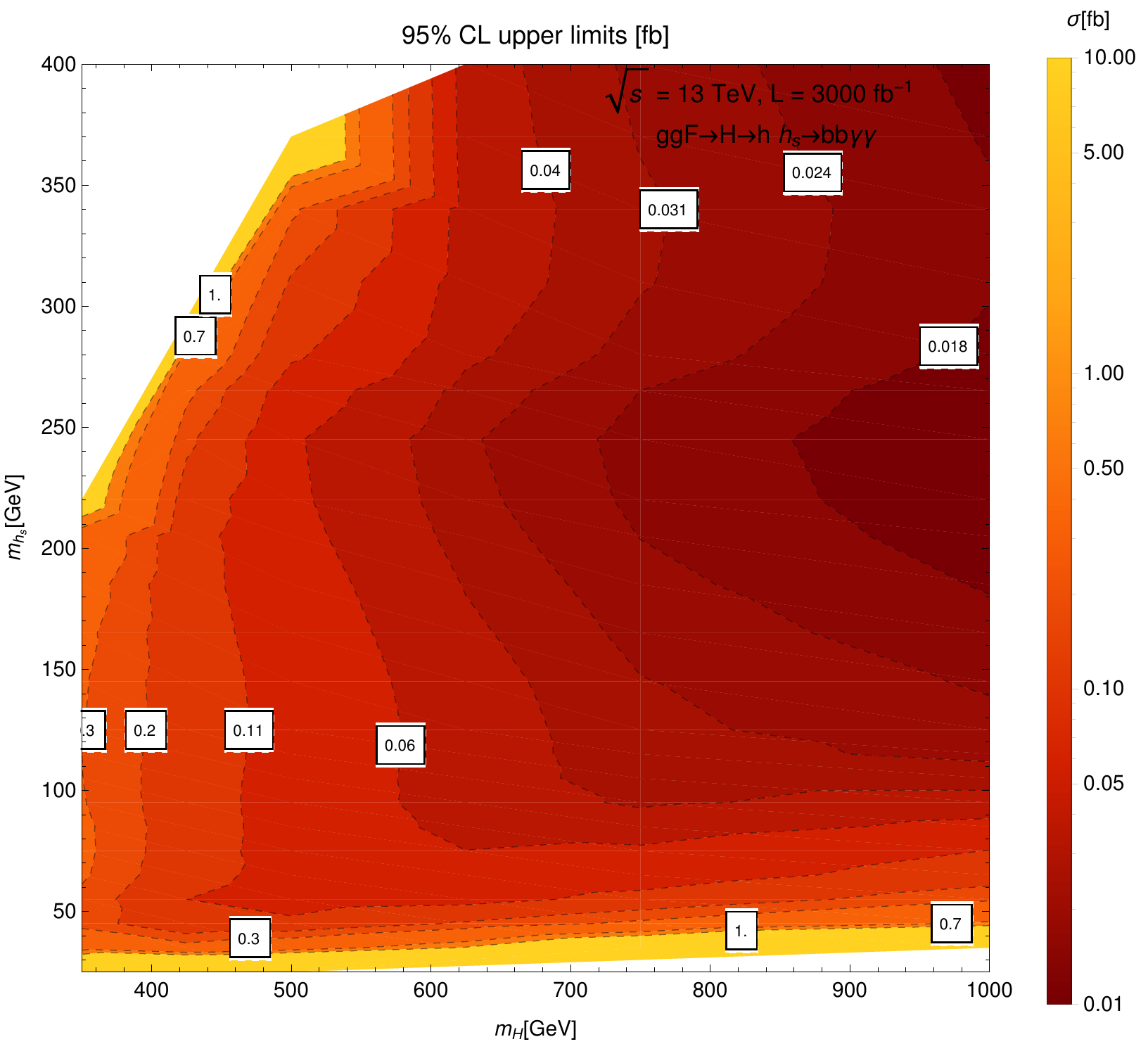}
\includegraphics[scale=.7]{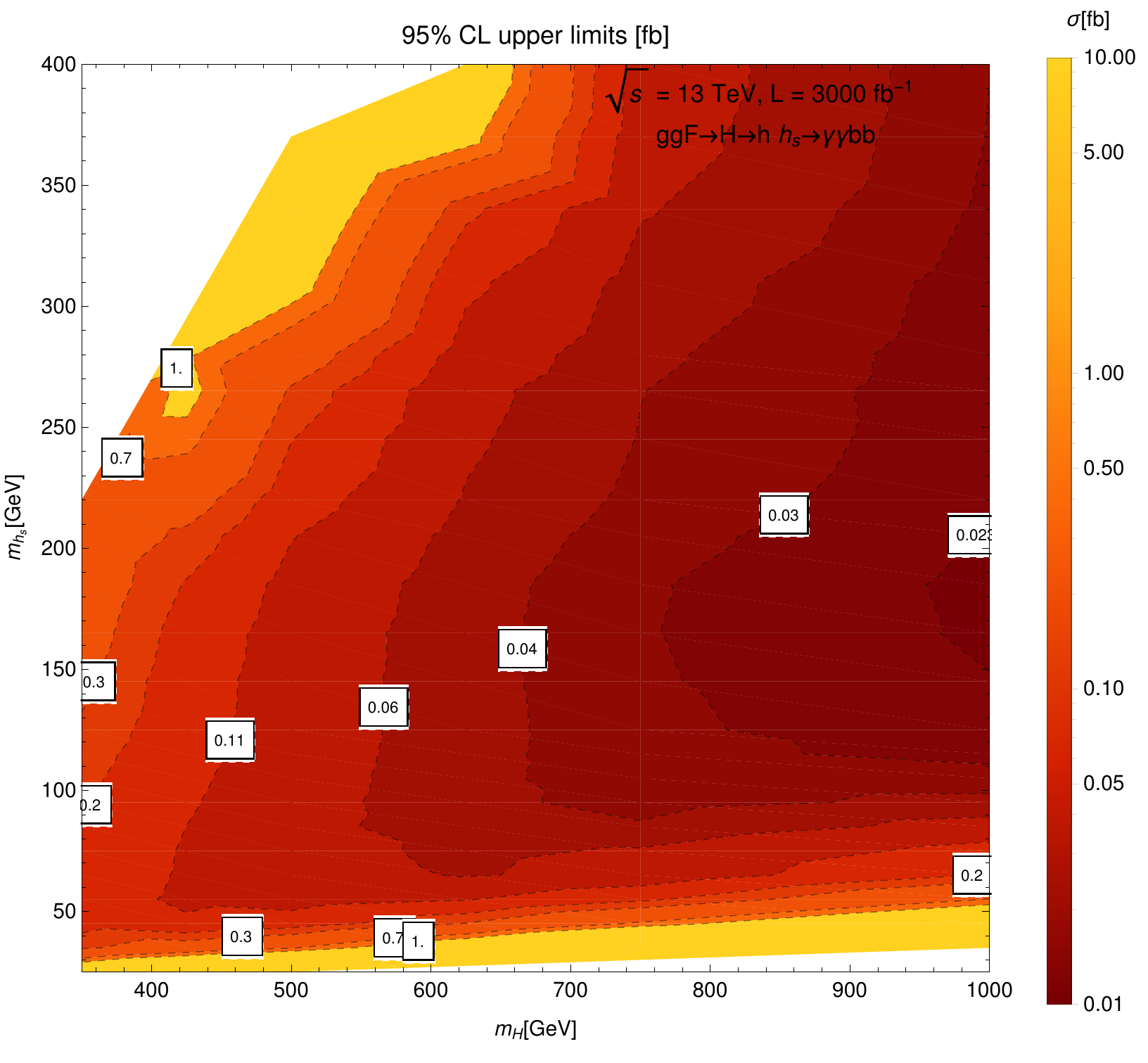}
\caption{95\% CL expected upper limits  for an integrated luminosity L=3000\fb for the process $ggF\rightarrow H\rightarrow H_{125}(b\bar{b}) H_S(\gamma\gamma)$ (up) and  $ggF\rightarrow H \rightarrow H_{125}(\gamma\gamma) H_S(b\bar{b})$ (down), using $m_X$ as the final discriminant. }
\label{fig:expectedbbaaMX3000}
\end{figure}

Our expected 95\%~CL exclusion limits and $5\,\sigma$ discovery cross sections
for $H_{125}\to b\bar{b}$, $H_S\to \gamma\gamma$ as function of $M_{H_S}$
at 300~fb$^{-1}$ integrated luminosity
are shown in Figs.~\ref{fig:SensULPoints_bbaaMX_300} for
four representative values for $M_H=425,\,500,\,625\ \text{and}\ 750$~GeV,
and for 3000~fb$^{-1}$ integrated
luminosity in Figs.~\ref{fig:SensULPoints_bbaaMX_3000}. For $H_{125}\to \gamma\gamma$,
$H_S\to b\bar{b}$ at 300~fb$^{-1}$ integrated luminosity
these are shown in Figs.~\ref{fig:SensULPoints_aabbMX_300}, for 3000~fb$^{-1}$ integrated
luminosity in Figs.~\ref{fig:SensULPoints_aabbMX_3000}.
As in the $b\bar{b}b\bar{b}$ case, we also present the results for the
expected 95\% C.L. upper limits in the $M_{H_{125}}$ vs $M_{H_{S}}$ plane in
fig. \ref{fig:expectedbbaaMX3000}.


%
%
%

The uncertainties include
statistical uncertainties and, added linearly, a factor $\pm 0.35$ considered as
systematic uncertainty originating from the normalisation of the background.

 As before viable NMSSM
regions for scalar production are shown in shaded blue, for pseudoscalar
production in shaded red in case they potentially exceed the ones for
scalar production. Again a sizeable region in the NMSSM
parameter space can be tested in this final state provided $M_H$ is not
too large.
It is, however, \emph{not} the same region potentially visible in the
$b\bar{b}b\bar{b}$ final state: The branching fraction of $H_S$ into
$\gamma\gamma$ can vary in the $0.2\% \pm 0.1\%$ range, and is anticorrelated
with its branching fraction into $b\bar{b}$. Moreover, as it is visible
from the shaded red regions in Figs.~\ref{fig:SensULPoints_bbaaMX_300} and
\ref{fig:SensULPoints_bbaaMX_3000}, the signal rates
$\sigma(ggF\to A\to H_{125}+A_S\to b\bar{b}\gamma\gamma)$ can be relatively
large. These correspond to very singlet-like pseudoscalars $A_S$ with very
suppressed couplings to quarks and leptons, but sizeable coupling $\sim
\lambda$ to higgsinos. Then the charged higgsino-loop
induced coupling to diphotons can dominate, leading to a large
$BR(A_S\to \gamma\gamma)$. The coupling $A-A_S-H_{125}$ is {\emph not}
suppressed in this case, leading to potentially large signal rates.

On the experimental side, the comparison of the upper limits on 
$H_{125}\to b\bar{b}$ and $H_S/A_S\to \gamma\gamma$ versus
$H_{125}\to \gamma\gamma$ and $H_S/A_S\to b\bar{b}$ has a simple answer
depending on $M_{H_S/A_S}$: For $M_{H_S/A_S}<125$~GeV the search for 
$H_S/A_S\to \gamma\gamma$ is sensitive to smaller signal rates, whereas for
$M_{H_S/A_S}>125$~GeV the search for $H_{125}\to \gamma\gamma$,
$H_S/A_S\to b\bar{b}$ is typically sensitive to smaller signal rates.
However, different regions in the parameter space of underlying models
are tested by these searches.


\section{Conclusions and Outlook}

Searches for resonant SM Higgs pair production are performed with considerable
effort by ATLAS and CMS. As explained in the introduction searches for
$ggF\to \Phi\to H_1 + H_2$ can be more promising where either $H_1$ or $H_2$
can be SM-like, and the other state being possibly CP-odd (which does not affect the
search methods). 

This scenario is manifest in the NMSSM where the r\^ole of $\Phi$ is played
by the MSSM-like heavy doublet, but the argument is more general. In the
present paper we have studied the prospects for corresponding searches in the
$b\bar{b}b\bar{b}$, $b\bar{b}\tau\tau$ and $b\bar{b}\gamma\gamma$ final states,
including SM backgrounds. We found that significant regions in the NMSSM
parameter space can be tested by these searches:

The NMSSM specific parameters testable by $b\bar{b}b\bar{b}$ are typically in the region
$\lambda \sim 0.50 - 0.70$ (the conservative upper bound from the absence of
a Landau singularity below $M_{GUT}$), $\kappa \sim 0.4 - 0.7$, $\tan\beta \sim 2 - 3.5$,
$\mu_{eff} \sim 180\;\text{GeV} - 290\;\text{GeV}$,
$A_{\lambda} \sim 150\;\text{GeV} - 550\;\text{GeV}$,
$A_{\kappa} \sim -830\;\text{GeV} - -210\;\text{GeV}$. An exception is
the case $M_{H_S}\sim 85\;\text{GeV} - 110\;\text{GeV}$ where LEP constraints
on the coupling of $H_S$ to the $Z$ boson are somewhat weaker; here values
of $\lambda$ down to 0.16 and $\mu$ down to 100~GeV (together with $\tan\beta$ up to 4.5)
can lead to testable points. These ranges of potentially testable parameters depend little on the total
integrated luminosity, but higher luminosity increases of course the number
testable parameters within these ranges.
Most of the parameters testable by $b\bar{b}\gamma\gamma$ are in the same region
except for $\kappa \sim 0.08 - 0.3$, $A_{\kappa} \sim -50\;\text{GeV} - 10\;\text{GeV}$
which indicates that this region is non-overlapping with the one testable by $b\bar{b}b\bar{b}$.

We are convinced that the here proposed search methods can still be refined,
and that the estimated sensitivities to cross sections times branching fractions
presented here are conservative. This becomes clear from a comparison to the recent CMS search
for resonant SM-Higgs pair production \cite{CMS-PAS-HIG-17-008} in the $b\bar{b}\gamma\gamma$
final state (and actually also from a comparison to the recent CMS search
\cite{Sirunyan:2017djm} in the $b\bar{b}\tau\tau$ final state). Thus we hope
that such promising searches will be performed in the future at the LHC.


\section*{Acknowledgements}

This project has received support from the European Union's Horizon 2020 research and
innovation programs 
ITN Higgs\-Tools (PITN-GA-2012-316704),
ITN Elusives (Marie Sklodowska-Curie grant agreement No 674896),
RISE InvisiblesPlus (Marie Sklodowska-Curie grant agreement No 690575),
RISE NonMinimalHiggs (Marie Sklodowska-Curie grant agreement No 645722),
the ERC advanced grant Higgs@LHC,
and the D\'efi InPhyNiTi project N2P2M-SF.
M.R.V. thanks Marius Wiesemann, Fady Bishara, Dirk E. Zerwas,
Nikola Makovec, Sophie Henrot,  Alberto Escalante and Davide
Napoletano for fruitful discussions and help, and the hospitality received
at the IPPP at the University of Durham, where part of this work has been
done.


\renewcommand{\theequation}{\Alph{section}.\arabic{equation}}
\section*{Appendix A: Functions to Fit Background Distributions}

\setcounter{section}{1}

In this appendix we define the functions used to parametrize the
total invariant mass (or $M_X$) distributions of the backgrounds to
the various final states. The best choice among the functions and the values of the
corresponding parameters have been obtained by maximum likelihood estimates.

\subsection*{Four~Parameter Gamma distribution:}

\bea
\label{eq:gammadist}
f(M_X;\alpha,\beta,\gamma,\mu)=&\frac{\gamma}{\beta\Gamma(\alpha)}
e^{-\left(\frac{(M_X-\mu)}{\beta}\right)^\gamma}
\left(\frac{M_X-\mu}{\beta}\right)^{\alpha\gamma-1}
\ &\text{for}\ M_X\geq \mu\;,\nn \\
f(M_X;\alpha,\beta,\gamma,\mu)=&0\ &\text{for}\   M_X \leq \mu\;,
\eea

\subsection*{Frechet distribution:}

\bea
\label{eq:frechet}
f(M_X;\alpha,\beta,\mu)=&\frac{\alpha}{\beta}
e^{-\left(\frac{(M_X-\mu)}{\beta}\right)^{-\alpha}}
\left(\frac{M_X-\mu}{\beta}\right)^{-\alpha-1}
\ &\text{for}\ M_X\geq \mu\;,\nn \\
f(M_X;\alpha,\beta,\mu)=&0\ &\text{for}\   M_X \leq \mu\;,
\eea

\subsection*{GaussExp function:}

A Gaussian with an exponential tail:

\bea
\label{eq:gaussexp}
f(M_X;\mu,\sigma,k)=&
e^{-\frac{(M_X-\mu)^2}{2\sigma^2}}
\ &\text{for}\ \frac{M_X-\mu}{\sigma} \leq k\;,\nn \\
f(M_X;\mu,\sigma,k)=&
e^{\frac{k^2}{2}-\frac{k(M_X-\mu)}{\sigma}}
\ &\text{for}\  \frac{M_X-\mu}{\sigma} > k\;.
\eea

\subsection*{Landau Distribution}

\beq
\label{eq:landau}
f(M_X;\mu,\sigma)=
\int_0^\infty \sin(2t)
e^{-t\frac{(M_X-\mu)}{\sigma}-\frac{2t}{\pi}\log(t)}dt
\eeq

\section*{Appendix B: $5\,\sigma$ Discovery and 95\%~CL Exclusion Limits}
\setcounter{section}{2}
\setcounter{equation}{0}

In this appendix we sketch the computations of $5\,\sigma$ discovery and 95\%~CL exclusion
limits based on the $M_X$ distributions of the background, and the different $M_X$
signal distributions (depending on $M_{H_S}$), following \cite{Cowan:2010js}.

As shown in Fig.~\ref{fig:MXvsM4} the $M_X$ distribution of the signal, after
event selection and cuts, corresponds to a certain number $s_i$ of expected signal
events per bin. (The bin size in $M_X$ is 20~GeV, and we have checked that the final
results do not vary with this size.) From the MC simulation we know how this event
number depends on the total signal cross section~$\sigma_{sig}$, the integrated
luminosity ${\cal L}_{int}$ and the acceptance $A$ times efficiency $\epsilon$:
\beq
s_i \sim {\cal L}_{int}\cdot \sigma_s\cdot \left<A\cdot \epsilon\right>_s
\eeq
where $\sim$ indicates a bin-dependent proportionality factor $< 1$.
The backgrounds after event selection and cuts have been fitted by continuous
functions of $f(M_X)$ normalized to~1. Thus
the number $b_i$ of expected background events per bin per integrated luminosity is
\beq
b_i = {\cal L}_{int}\cdot \sigma_b\cdot \left<A\cdot \epsilon\right>_b\cdot \int_{bin\ i} f(M_X) dM_X\; . 
\eeq
Due to the large number of simulated events the statistical uncertainties around the
median values $\left<A\cdot \epsilon\right>_s$ and $\left<A\cdot \epsilon\right>_b$
are negligibly small, $\lsim 1\%$ in practically all cases.

Subsequently we use likelihood functions
\beq
L(\sigma_{a},\sigma_b) = \prod_{i}\frac{(b_i+\sigma_a s_i)^{b_i + \sigma_{b} s_i}}
{(b_i + \sigma_{b} s_i)!} e^{-(b_i+\sigma_a s_i)}
\eeq
with $(b_i + \sigma_{b} s_i)!$ interpolated by the $\Gamma$ function for
non integer $(b_i + \sigma_{b} s_i)$.

For $5\,\sigma$ discovery limits on $\sigma_{sig}$ we look for
the value of $\sigma_{sig}$ for which the background
only hypothesis is rejected at the $5\,\sigma$ level. The observed number of events
per bin would be $b_i + \sigma_{sig} s_i$, and the likelihood function
corresponding to the background only hypothesis is $L(0,\sigma_{sig})$.
As function of the number of events per bin it has its maximum at
$L(\sigma_{sig},\sigma_{sig})$. Hence the test statistics $t_{disc}$ for
discovery is
\beq
t_{disc} = -2\ln\frac{L(0,\sigma_{sig})}{L(\sigma_{sig},\sigma_{sig})}\; .
\eeq
Following \cite{Cowan:2010js} the significance $Z_{disc}$ is then
\beq
Z_{disc}=\sqrt{t_{disc}}\; ,
\eeq
and for a $5\,\sigma$ discovery we determine $\sigma_{sig}$ such that
$Z_{disc}=5$.

For 95\%~CL exclusion limits on $\sigma_{sig}$ we look for
the value of $\sigma_{sig}$ for which the signal
hypothesis is rejected at 95\%~CL. The observed number of events
per bin would be $b_i$, and the likelihood function
corresponding to the signal hypothesis is $L(\sigma_{sig},0)$.
As function of the number of events per bin it has its maximum at
$L(0,0)$. Hence the test statistics $t_{excl}$ for
exclusion is
\beq
t_{excl} = -2\ln\frac{L(\sigma_{sig},0)}{L(0,0)}\; .
\eeq
For exclusion at 95\%~CL we determine $\sigma_{sig}$ such that
$\sqrt{t_{excl}}=1.64$ since we consider only positive signal contributions
to the number of events.

Uncertainties from the background are estimated as follows:
$\pm (1-2)\,\sigma$ statistical uncertainties can be obtained bin by bin. To
these we add linearly (to be conservative) the estimated $(1-2)\,\sigma$ systematic
uncertainties from the overall normalisation of the background. Then the
above likelihoods are recomputed with correspondingly larger and smaller
values for all $b_i$ from which we deduce the $\pm (1-2)\,\sigma$ uncertainties
of $\sigma_{sig}$ for
the $5\,\sigma$ discovery limits and 95\%~CL exclusion limits.


\end{document}